\documentclass[referee]{sn-jnl}
\usepackage{times}
\usepackage{amsmath,amsthm, amssymb}
\usepackage{bm}
\usepackage{widetext}
\usepackage[authoryear]{natbib}
\usepackage[normalem]{ulem}

\def\be{\begin{equation}}
\def\ee{\end{equation}}
\def\bea{\begin{eqnarray}}
\def\eea{\end{eqnarray}}

\def\lg{\lambda_g}
\newcommand\Ol{\mathcal O}
\def\Lm{\mathcal{L}_m}
\def\L{\mathcal{L}}

\begin{document}

\title[Tests of gravity with planetary ephemerides]{{Testing Theories of Gravity with Planetary Ephemerides}}

\author*[1,2]{\fnm{Agn\`es} \sur{Fienga}}\email{agnes.fienga@oca.eu}
\author[3,4]{\fnm{Olivier} \sur{Minazzoli}}


\affil[1]{\orgdiv{Geoazur}, \orgname{Observatoire de la Côte d'Azur}, \orgaddress{\street{Av. A. Einstein}, \city{Sophia-Antipolis}, \postcode{06560},  \country{France}}}

\affil[2]{\orgdiv{IMCCE}, \orgname{Observatoire de Paris}, \orgaddress{\street{Av. Denfert-Rocheau}, \city{Paris}, \postcode{75014}, \country{France}}}

\affil[3]{\orgdiv{ARTEMIS}, \orgname{Observatoire de la Côte d'Azur}, \orgaddress{\street{Av. Observatoire}, \city{Nice}, \postcode{06560},  \country{France}}}

\affil[4]{{\orgdiv{Bureau des Affaires Spatiales}, \orgaddress{\street{2 rue du Gabian}, \city{Monaco}, \postcode{98000},  \country{Monaco}}}}


\abstract{
We describe here how  planetary ephemerides are built in the framework of General Relativity and how they can be used to test alternative theories. We focus on the definition of the reference frame (space and time) in which the planetary ephemeris is described, the equations of motion that govern the orbits of solar system bodies and {electromagnetic waves}. After a review on the existing planetary and lunar ephemerides, we summarize the results obtained considering full modifications of the ephemeris framework with direct comparisons with the observations of planetary systems, with a specific attention for the PPN formalism. We then discuss other formalisms such as Einstein-dilaton theories, the massless graviton and MOND. The paper finally concludes on some comments and recommendations regarding misinterpreted measurements of the advance of perihelia.

}


\maketitle

\tableofcontents

\section{Introduction}
\label{intro}
\label{sec:2}

The current definition of what is an ephemeris is  {\it{a table giving the future positions of a planet, comet, or satellite}}. By extension, it also includes the dynamical framework from which the planetary positions and velocities are estimated. This framework includes not only the dynamical modeling and the reference system with which the motion is described, but also the sample of observations used for adjusting the constants of the model. In this review, we will mainly focused on ephemerides of planets, but there are also ephemerides of natural and artificial satellites, small bodies (comets, asteroids) and pulsars. 

With the XIXth century  and the golden age of the big refractors \citep{1990hias.book.....V},  the astrometry of planets has known a significant improvement{,} leading to an increased accuracy of the dynamical theories describing their motions. Together with these improvements, came the inconsistencies between very accurate observed positions of planets{, such as Uranus,} and the classical Newtonian modeling of gravity. In the case of Uranus, a significant---even at this epoch---difference of several seconds of arc between its observed positions and the positions proposed by the models, based on Newton's laws, opens the door to a major controversy of the century.  Why the Uranus orbit does not match with the observations when, for instances, comet ephemerides are able to predict accurately their return in the visible sky ? Since 1821, the explanation of an unknown planet perturbing the orbit of Uranus has been proposed.  A hunt for the unseen planet started with an intensification in 1844 and the hidden celestial object, called Neptune, was finally discovered in 1846 by Galle in Berlin at the location that was predicted by Le Verrier and Addams. This episode signs the paramount of classical celestial mechanics of the XIXth century. However, an other problem remains : the orbit of Mercury. At the beginning of the XXth century,  catalogs of planetary observations have been built with a worldwide 
effort and provided angular positions for almost one century with accuracies below the level of few seconds of arc.  
During this period, when no direct measure of the planetary distance was possible, the constraints on the size of the planetary orbits were obtained thanks to the long interval of observations and precise determination of the mean motions, and consequently, with the third Kepler law, of the semi-major axis. 

In the case of Mercury, the observed variation of the Mercury orbit was far too important to be left unexplained in comparison to the observational accuracy of the end of the XIX century and the beginning of the XXth century.  Following the example of Neptune and Uranus, people proposed the existence of {another} hidden object, called Vulcain, perturbing the orbit of Mercury the same way that Neptune perturbs the one of Uranus. The problem was that no one was able to observe Vulcain. Some proposed that the planet could be always located on the other side of the sun relative to the earth, making it impossible to be seen from earth. Others proposed a modification of Newton's potential equation. However, it was Einstein who ultimately explained Mercury orbit, positing that gravity was not a force, but a result of the spacetime curvature caused by the mass of celestial bodies \citep{einstein1915}.


Decades later, thanks to the remote exploration of the solar system, planetary ephemerides have evolved with high accurate astrometric observations obtained for planet and natural satellites  thanks to the navigation tracking of spacecraft (s/c) orbiting these systems.
Table \ref{tab:3gene} gives a good comparison between three generations of planetary ephemerides developed at key stages of their evolution: from the end of the classical Newtonian celestial mechanics \citep{1913AnPar..31....1G}, to the beginning of the space regular exploration \citep{DE102} and up to the present time \citep{2021AJ....161..105P, 2019NSTIM.109.....V}.

Since the beginning of the space exploration, general relativity has been continuously tested with planetary orbits, for instance, using the first direct radar measurements on the telluric planet surfaces  \citep{1964PhRvL..13..789S}. These measurements were the first direct estimations of the Earth-Venus, Mercury or Mars distances used as direct constraints in the construction of planetary ephemerides \citep{1967AJ.....72..338A, 1976jden.book.....S}.
With the Apollo and Lunakhod missions on the Moon and the installation of light reflecting corner cubes, a new step is reached as centimetric measures of the Earth-Moon distances are obtained {with Lunar Laser Ranging (LLR)} since then, with regular improvement of the measurement accuracy \citep{2008PASP..120...20M, 2013ExA....36..105T, 2017A&A...602A..90C}. 
 In 1983, the first tracking data from the Viking landers on Mars have been included in the numerical integration of the Mars orbit at JPL \citep{1983A&A...125..150N}.  With these measurements and those of the first Pioneer and Voyager flybys of the Jovian and Saturnian systems, the accuracy of the planetary ephemerides enters into the age of {kilometer} accuracies for the planets and few tens of {centimeter} for the Moon thanks to LLR. 
 At this level of accuracy, the modelisation of the planetary dynamics was upgraded including asteroid perturbations of the 5 biggest objects, Mars rotation improvement but also figure-figure effect for the modelling of the Earth-Moon tidal deformations \citep{Standish2001}. 
 Since 2010,  the s/c tracking measurements are still improving  thanks to the installation of more efficient transpondeurs. It is now possible to study the orbit of Mars with an accuracy of less than a meter  and since 2016 to have a monitoring of Jupiter and Saturn orbits with an accuracy of tens of meters.  
 
 With these accuracies, the complexity of the dynamical model increases and it is clear that the solar system becomes a more and more interesting tool for testing general relativity or alternative theories of gravity. First tests of general relativity with modern planetary ephemerides start in 1978 \citep{1978AcAau...5...43A} with regular improvements since then. 
 


 In this review, we will focus on planetary ephemerides, setting aside the topic of gravity tests using the Earth-Moon system due to its complexity. Our incomplete understanding of the Moon internal structure significantly limits, indeed, the testing of alternative theories. But in order to introduce these limitations, a detailed presentation of the complex mechanisms between the internal structure of the Moon, its deformations, general relativity and their impacts on LLR observations (at very close frequencies) has to be done, and we think that it is out of the scope of the present work.
 
Here, we first explain how a planetary ephemeris is built from its native relativistic framework. We give the dynamical model and the data sets used for its fits as well as the problematics related to the fit itself. We try, at this stage, to give an introduction  to some of the concepts hidden behind the construction of planetary ephemerides.  We then review classic results obtained in the general relativity and post-Newtonian limits for the modern period. In a third part, we discuss state-of-the-art approaches for directly confronting alternative theories of gravity with planetary observations in the most consistent {possible way}. 
 Finally we give a quick overview of other types of tests, deduced by interpreting planetary ephemeris accuracy at the light of different gravitational frameworks. We will try to convince the readers that such results have to be considered with a lot of {caution} because of the lack of consistency between the framework with which the planetary ephemeris is built and the one proposed by the authors.
 
Finally, we acknowledge that this review may have limitations. We apologize in advance for any inadvertent omissions in the references and welcome feedback from our readers.

 \section{Basic concepts behind planetary ephemerides}


\subsection{Prerequisites: A Brief Primer on general relativity}
\label{sec:introGR}


In classical Newtonian mechanics, space and time are treated as absolute concepts. For instance, the time span is assumed to be uniform throughout the universe.
But the main lesson from general relativity is that not only space and time are relative notions rather than absolute ones, but also that the structure formed by space and time is curved by the energy of matter. This notably means that the very notion of distances in space and time depends on the observer, and also that observation may be impacted by the curvature of spacetime. This implies a wide range of subtleties when one describes motions at a level at which Newton's theory is no longer accurate enough. In what follows, we provide a brief summary of what is required to understand those subtleties.

\subsubsection{Newton's theory}
\label{sec:newton}

According to the first law of motion in Newton's theory, an inertial motion is an uniform motion in a straight line. As a consequence, free fall motions are not inertial in Newton's theory, but are accelerated by the gravitational force $\bm F$. This force acts between every massive objects and reads
\be
{\bm F_{AB}} =- {\bm F_{BA}} =  \frac{G M_A M_B}{{r^2_{AB}}} \frac{{{\bm r}_{AB}}}{\|{{\bm r}_{AB}} \|}, \label{eq:Fgrav}
\ee
where $M$ is the equivalent to the charge in Coulomb's law, but for gravity and G the constant of gravitation.\footnote{Indeed, Coulomb's law between two charges reads {$\lvert \bm F_{12} \rvert = k_C \lvert q_1\rvert \lvert q_2\rvert/r^2_{12}$, where $k_C$ is Coulomb's constant}.} $M$ is usually named \textit{gravitational mass}, since it is related to the gravitational force.
According to the second law of motion, the acceleration that follows from this force is
\be
m_A \bm a_A =  {\bm F_{AB}} ,\label{eq:agrav}
\ee
where $m_A$ is the \textit{inertial mass} of the body $A$. Namely, to accelerate a body $A$ with acceleration $\bm a_A$, a force $\bm F_{AB}$ needs to be applied, regardless of the type of force in question—it could be gravitational or electrostatic etc. Just as one would not expect a relationship between the charge $q$ in the electrostatic force and the inertial mass, one should not anticipate a relationship between the gravitational ``charge'' $M$ and the inertial mass $m$. However, observations indicate otherwise, compelling Newton to postulate the equivalence of gravitational and inertial mass. This is known as the \textit{equivalence principle}.

A common way to derive the equation of motion in classical mechanics is through the definition of a Lagrangian of motion---of the general form $L = K - V$, where $K$ is the kinetic energy and $V$ the potential energy. The Lagrangian of motion in the theory of Newton reads 
\be
L_N = \frac{m v^2}{2} + m U,
\label{eq:lagrnewt}
\ee
{where $v$ is the velocity of the moving object and $U$ is the Newtonian potential that satisfies the Poisson equation}
\be
\Delta U = - 4 \pi G \rho_m, \label{eq:Poisson}
\ee
where $\rho_m$ is the mass density of the gravitational sources.\footnote{Note that the potential energy with this definition of the gravitational potential indeed reads $V=-mU$, such that one indeed verifies that $L_N = K - V$ in Eq. (\ref{eq:lagrnewt}).} For instance,  for a sum of {spherical bodies A}, 
\be
U(x^i,t) = \sum_A \frac{G m_A}{\lvert x^i-x^i_A(t) \rvert}, \label{eq:potN}
\ee
is solution of Eq. (\ref{eq:Poisson}),{with $x^i$ and $x^i_A(t)$ the positions of the moving object and of the gravitational sources A, respectively}, where
\be
m_A = \int_A \rho_m d^3 r.
\ee 
Indeed, applying the Euler-Lagrange equation 
\be
\frac{\partial L}{\partial \bm{x}} - \frac{d}{dt} \frac{\partial L}{\partial \bm{v}}=0,\label{eq:EulLag}
\ee
on $L_N$ leads to Newton's equation of motion
\be
\bm{a} = \bm{\nabla} U, \label{eq:Nacc}
\ee
which reproduces Eq. (\ref{eq:agrav}) with Eq. (\ref{eq:Fgrav}) for $M=m$.
In the Newtonian framework, the speed of light, $c$, and the constant of gravitation, $G
$, are constants. Time and space are universal and space is flat. 

\subsubsection{Proper time in {special relativity}}

\begin{figure}
\centering
\includegraphics[scale=0.5]{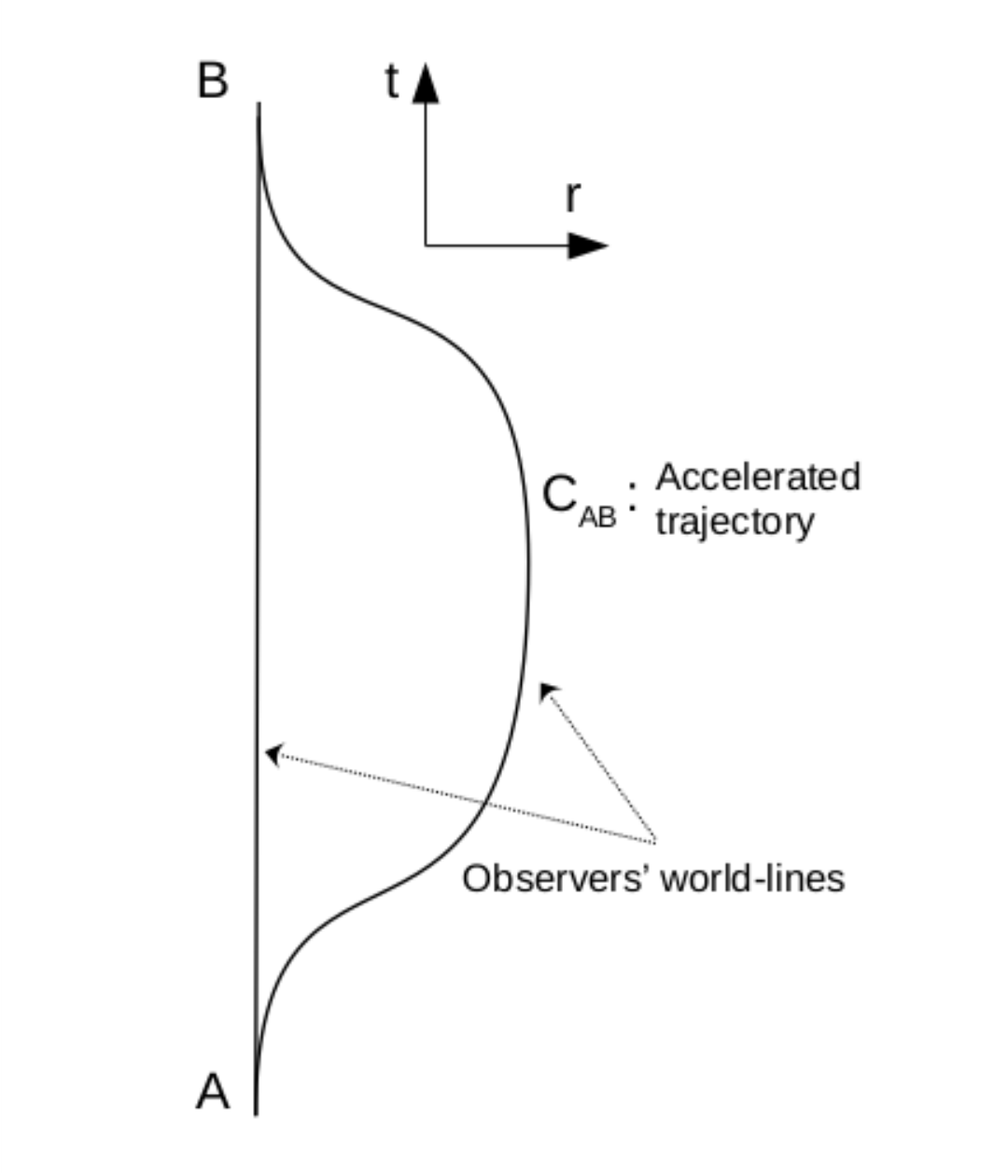}
\caption{{Schematic illustration of the trajectories (world-lines) of two observers, one at rest and the other accelerated. The accelerated trajectory correspond to the curve $C_{AB}$.}}
\label{fig:langevin}
\end{figure}

At the onset of the XXth century, a group of physicists understood that time and space were not separate concepts, but instead, composed a singular entity known as spacetime. In this new understanding, time and space became relative notions, and the structure of spacetime, according to Minkowski, possessed a Lorentzian nature---which means that the variation of the proper time of an observer follows 
{
\be
c^2 d\tau^2 = -\eta_{\alpha \beta} dx^\alpha dx^\beta = c^2 dt^2 - dx^2 - dy^2 - dz^2, \label{eq:Minkowski}
\ee}
where $\eta_{\alpha \beta}$ is named the Minkowski metric, and $\{t,\bm x \}$ inertial and non-rotating coordinate systems.\footnote{Non-rotating with respect to what will be the subject of a discussion in Sec. \ref{sec:EIHeqNa}.} {Throughout the text, we use the metric signature (-,+,+,+) and Einstein's summation notational convention---which is such that $A_\sigma B^\sigma := \sum_\sigma A_\sigma B^\sigma$}. The variable $c$ will be identified in Sec. \ref{sec:EMwaves} as the speed of light, but more fundamentaly, it fixes the causal structure of the flat spacetime equation Eq. (\ref{eq:Minkowski}). Lorentz transformations of special relativity are simply the coordinate transformations that leave the structure of the metric Eq. (\ref{eq:Minkowski}) unchanged. They read
\be
T = \gamma \left(t - \frac{\bm v \cdot \bm x}{c^2} \right),~~\bm X = \bm x + \gamma \bm v t + (\gamma -1) \frac{\bm v (\bm v \cdot \bm r)}{v^2}, \label{eq:Lorentz}
\ee
where 
$v^2 = \bm v \cdot \bm v$ 
and where $\bm v$ is the velocity between the two inertial reference frames, and with the Lorentz factor $\gamma$ defined as
\be
\gamma := \frac{1}{\sqrt{1-\frac{v^2}{c^2}}},
\ee
Contrary to popular wisdom, it is Eq. (\ref{eq:Minkowski})---not the Lorentz transformations equation Eq. (\ref{eq:Lorentz})---that is needed to derive the difference in time elapsed between two twins in Langevin's twin paradox.\footnote{Langevin's twin paradox presents a scenario in which one twin travels to space at a high speed and returns to find the other twin has aged more. This highlights the relativistic effect of time dilation, as described in the theory of special relativity. The paradox's essence is rooted in a perceived contradiction: both twins should ostensibly observe the other with a similar relative motion, which would therefore result in the same `anomalous' aging effect for both when solely using Lorentz's transformations Eqs. (\ref{eq:Lorentz}) in the derivation.} Let us assume a pair of observers that are at the same location at the spacetime events $A$ and $B$, but one of the observer stays at rest while the other accelerates to leave and return to the other observer (see Fig. \ref{fig:langevin}). 
In terms of the reference frame of the observer at rest $\{\tau^r, \bm x\}$, the accelerated observer's proper time $\tau^a$ between the two events $A$ and $B$ reads
\be
\tau^a_{AB} = \int_{C_{AB}} d\tau^a = \int_{C_{AB}} \sqrt{1-\frac{v^2}{c^2}}~ d\tau^r, \label{eq:twinptime}
\ee
according to Eq. (\ref{eq:Minkowski}), with $\bm v = d \bm x /d\tau^r$, and where $C_{AB}$ is the accelerated trajectory of the accelerated observer in the reference frame of the observer at rest. Given a trajectory $C_{AB}$, one can derive the proper time of the accelerated observer elapsed between the events $A$ and $B$. It is worth noting that Eq. (\ref{eq:twinptime}) would not maintain the same form if one attempts to derive the proper time elapsed for the rest observer from the coordinates of the accelerated observer. This is because the Minkowski metric (Eq. \ref{eq:Minkowski}) does not remain invariant for an accelerated reference frame $\{\tau^a, \bm X\}$. In the proper reference frame of an accelerated (but non-rotating) observer, the metric Eq. (\ref{eq:Minkowski}) would instead read \citep{ni:1978pr}
\be
c^2 d \tau^2 = \left(1+\frac{\bm a \cdot \bm X}{c^2} \right)^2 c^2 (d\tau^a)^2 - dX^2 - dY^2 - dZ^2,
\ee
where $\bm a$ is the acceleration of the proper reference frame of an accelerated observer. This fact alone dispels any paradox, as the situations of the two observers are distinct both physically and mathematically. From Eq. (\ref{eq:twinptime}), one can infer that the elapsed proper time of an observer who has been accelerated between two arbitrary spacetime events $A$ and $B$ is always smaller than the elapsed proper time of an observer at rest $\tau^a_{AB} < \tau^r_{AB}$. In other words, this means that for an inertial observer the following action 
{
\be
S  = - mc^2 \int d\tau = -mc^2 \int \sqrt{1-\frac{v^2}{c^2}} dt= \int L_{SR}(x^\alpha,v^\alpha) dt,
\ee
is maximal, where $L_{SR}$ is, therefore, the Lagrangian of motion of special relativity:} 
\be
L_{SR} = 	-mc^2 \sqrt{1-\frac{v^2}{c^2}}. \label{eq:LSR}
\ee

\subsubsection{Free fall and proper time in general relativity}

In general relativity, the free fall of a point mass $m$ aslo maximizes the proper time $\tau$, but which is now defined by
\begin{equation}
\tau_{AB} = \int_{C_{AB}} \sqrt{-g_{\alpha \beta} dx^\alpha dx^\beta}, \label{eq:properT}
\end{equation}
where $C_{AB}$ means that the integration is taken along a given path between the spacetime positions\footnote{Also named ``events''} $A$ and $B$ of the observer. \footnote{Let us note that we use the mostly plus convention for the signature of the metric ($-+++$), and that we use the Einstein summation convention, such that $X_\sigma Y^\sigma := \sum_{\sigma=0}^3 X_\sigma Y^\sigma$. $x^\alpha =(ct,x^i)=(ct, \vec x)=(ct, \bm x)$, where $i=1,2,3$.} In cartesian coordinates for instance, one has $x^i = (x,y,z)$. $g_{\alpha \beta}$ is the spacetime metric, which depends on the position, and which represents the curvature of spacetime. As we will see below, the metric is solution of the Einstein-Hilbert equation of general relativity.

The definition of the proper time for an observer in Eq. (\ref{eq:properT}) is true whether or not the motion of the observer is accelerated. It is important to keep in mind that unlike in Newton's theory, a free falling observer in general relativity has an inertial motion---that is, it is not accelerated. In general relativity, inertial bodies follow geodesics of spacetime that correspond to free fall trajectories. This is forced upon the theory by the \textit{equivalence principle}, which states that the inertial mass is indeed equivalent to the gravitational mass.

But since the free fall of an observer maximizes its proper time, it means that Eq. (\ref{eq:properT}) has to be an extremum in the case of a free fall motion. It means that the path $C_{AB}$ between the events $A$ and $B$ must be such that it extremizes Eq. (\ref{eq:properT}). As a consequence, it must follow the \textit{Least Action Principle}\footnote{The Least Action Principle might be more appropriately termed the Extremum Action Principle, since it primarily requires that the variation of the action remains null: $\delta S = 0$. This implies that the action is at an extremum.} of Lagrangian mechanics that demandes the extremization of an action defined upon a Lagrangian. Indeed, from Eq. (\ref{eq:properT}), one can define the following action
\begin{equation}
S = mc^2 \tau = \int L(x^\alpha,v^\alpha) dt,
\label{eq:defLagrangian0}
\end{equation}
where $v^\alpha = c~ dx^\alpha / dx^0 =  dx^\alpha / dt$ is a velocity expressed in terms of the coordinates $\{x^{\alpha}\}$ and where the Lagrangian reads
\begin{equation}
L = mc^2 \sqrt{-g_{\alpha \beta}(x^\sigma) v^\alpha v^\beta }. \label{eq:defLagrangian}
\end{equation}
The  \textit{Least Action Principle} demands that $\delta S =0$. Given a specific spacetime metric $g_{\alpha \beta}$, one can therefore compute the equations of motion by applying the Euler-Lagrange equation on Eq. (\ref{eq:defLagrangian}). The equation can also be applied on the formal definition of equation Eq. (\ref{eq:defLagrangian})  and one then gets the general form---that is, for all spacetime metrics---of the equations of motion (for massive point particles) that reads
\begin{equation}
\frac{d^2 x^\alpha}{dt^2} = \left(\Gamma^\alpha_{\sigma \epsilon} -\Gamma^0_{\sigma \epsilon}  \frac{v^\alpha}{c} \right) v^\sigma v^\epsilon,
\end{equation}
where
\be
\Gamma^\gamma_{\alpha \beta} := \frac{1}{2}~ g^{\gamma \sigma} (\partial_\alpha g_{\sigma \beta} - \partial_\beta g_{\alpha \sigma} - \partial_\sigma g_{\alpha \beta}), \label{eq:Christoffel}
\ee
is named the Christoffel connection with $g^{\gamma \sigma}$, {the contravariant metric---which is such that $g_{\alpha \sigma} g^{\sigma \beta}=\delta_{\alpha \beta}$, where $\delta$ is the Kronecker symbol}. Besides, an accelerated trajectory would read
\begin{equation}
\frac{d^2 x^\alpha}{dt^2} -\left(\Gamma^\alpha_{\sigma \epsilon} -\Gamma^0_{\sigma \epsilon}  \frac{v^\alpha}{c} \right) v^\sigma v^\epsilon=F^\alpha,
\end{equation}
where $F^\alpha$ is the force responsible for the acceleration.

\subsubsection{Einstein-Hilbert equation and the Newtonian approximation}
\label{sec:EIHeqNa}

The spacetime metric is curved by the energy of matter according to the Einstein-Hilbert equation that reads
\be
R_{\alpha \beta} - \frac{1}{2} g_{\alpha \beta} R = \frac{8 \pi G}{c^4} T_{\alpha \beta}, \label{eq:EHeq}
\ee
where $R_{\alpha \beta}$ is the Ricci tensor (see sec \ref{sec:tensor}) defined as
\begin{equation}
R_{\alpha \beta}= \partial_\mu \Gamma_{\alpha \beta}^\mu-\partial_\beta \Gamma_{\alpha \mu}^\mu+\Gamma_{\sigma \mu}^\mu \Gamma_{\alpha \beta}^\sigma-\Gamma_{\sigma \beta}^\mu \Gamma_{\alpha \mu}^\sigma , \label{eq:EIeq}
\end{equation}
and
\be
R = g^{\sigma \epsilon} R_{\sigma \epsilon} \label{eq:Ricci}
\ee 
is the Ricci scalar, and $T_{\alpha \beta}$ is the stress-energy tensor of the material fields---which we will define in Sec. \ref{sec:L4fields}. The stress-energy tensor represents the energy content of the matter fields that generate the curvature of spacetime. Otherwise, we will see below that $G$ corresponds to the constant of Newton, also known as the gravitational constant.

\paragraph{The Newtonian approximation}

In the Newtonian formalism given in Sec. \ref{sec:newton}, we assume that the speed of the considered bodies is small with respect to the speed of light $v^2/c^2 \ll 1$. Let us add the constant $-mc^2$ to this Lagrangian, which does not impact the equation of motion, but which is such that one reproduces the Lagrangian of motion of special relativity equation Eq. (\ref{eq:LSR}) at leading order when $U=0$, that is
\be
L_{SR} = 	-mc^2 \sqrt{1-\frac{v^2}{c^2}} = -mc^2 \left( 1-\frac{v^2}{2 c^2} \right) + \Ol\left(\frac{v^4}{c^{4}}\right).
\ee
Now, the Lagrangian of motion for the theory of Newton reads
\be
L_N = -mc^2 \left(1- \frac{v^2}{2 c^2}  - \frac{U}{c^2}\right). \label{eq:LNewt}
\ee
Let us stress that for an orbit, one has $v^2/c^2 \sim U/c^2 \ll 1$ in the theory of Newton. {This approximation} ought to remain valid at leading order in general relativity in the weak field limit. As a consequence, for now on, we will use the notation $\Ol(c^{-2n})$ to refer to both $\Ol(v^{2n}/c^{2n})$ and $\Ol(U^n/c^{2n})$.
Since the theory of Newton is already  accurate in the solar system, we expect the theory of general relativity to reproduce the trajectories of the Newtonian theory at leading order. Therefore, one expects that there exists a spacetime metric that reproduces the Lagrangian equation Eq. (\ref{eq:LNewt}) at leading order. One can check that injecting the following metric in Eq. (\ref{eq:defLagrangian})
\be
-c^2 d\tau^2 = ds^2 = - \left(1- \frac{2 U}{c^2} \right) c^2 dt^2 + dx^2 + dy^2 + dz^2, \label{eq:metrNewt}
\ee
where $ds$ is the spacetime line element defined as $ds^2 = g_{\alpha \beta} dx^\alpha dx^\beta$, gives back the Lagrangian of motion of the theory of Newton equation Eq. (\ref{eq:LNewt}), up to corrections of order $\Ol  (c^{-4})$. Therefore, one expects the metric in Eq. (\ref{eq:metrNewt}) to be solution of general relativity at leading order for the kind of weakly gravitating sources that we have in the solar system.\footnote{As opposed to a neutron star or a black-hole.} It can be indeed verified that by injecting the metric Eq. (\ref{eq:metrNewt}) in the Einstein-Hilbert equation Eq. (\ref{eq:EIeq}), the resulting differential equation reads
\be
\Delta U = - 4 \pi G \rho_m + \Ol(c^{-2}), \label{eq:Newtoneq}
\ee
where $\rho_m$ is the mass density of the gravitational sources defined as $\rho_m = T^{00}/c^2 + \Ol(c^{-2})$. It corresponds to Newton's \textit{universal law of gravitation} as long as $G$ is indeed the gravitational constant. One has therefore identified $G$ in Eq. (\ref{eq:EHeq}) to Newton's constant from the weak-field and slowly moving limit of the theory that leads to the same field equation as Newton's theory at leading order. 

Therefore, general relativity  produces the same trajectories at leading order as Newton's theory. This is called the \textit{Newtonian approximation} of general relativity.

However, let us stress that even at this level of approximation, the two theories differ drastically---in a way that can be tested at the experimental level already. Indeed, in general relativity, the variation of an observer proper time $d\tau$ (Eq. \ref{eq:metrNewt}) with respect to the proper time of another observer depends explicitly on their different positions in a gravitational potential $U$. This means that two observers at different locations in the gravitational potential will not agree on the evolution of time. This effect, although minute, can be tested if one has accurate enough clocks. In other words,  {had we} developed atomic clocks with sufficient precision prior to our ability to observe the motions of celestial bodies in the solar system, we could have confirmed the superiority of general relativity over Newton's theory. This could have been achieved by comparing the frequencies of two clocks located at different positions relative to the geoid. However, due to our atmosphere's transparency, astronomers identified the limitations of Newton's theory through the anomalous advance of Mercury's perihelion---see Table \ref{tab:3gene} for the numerical value of this advance---before quantum physicists could create sufficiently accurate atomic clocks. We will further discuss the advance of the perihelion of Mercury in Sec. \ref{sec:EIHDLeqm}.

Finally, one can see in Eq. (\ref{eq:metrNewt}) that space it flat\footnote{Space is Euclidean in Eq. (\ref{eq:metrNewt}), such that it satisfies the Pythagorean theorem.} at leading order, such that free fall trajectories in the solar system are essentially the consequence of the curvature of the temporal dimension, not the spatial one. 

\paragraph{Gauge invariance of Newton's equation and the definition of coordinate times}

Newton's equation Eq. (\ref{eq:Newtoneq}) is notably invariant under the following change of the potential:
\be
U \rightarrow U + a_k(t) x^k + b(t), \label{eq:gaugeN}
\ee
where $a_k(t)$ and $b(t)$ are arbitrary three-vector and scalar that depend on time. \footnote{Adding any term $\delta U$ that is such that $\Delta \delta U =0$ leaves Newton's equation invariant.} One often says that the Eq. (\ref{eq:Newtoneq}) is invariant under the change of gauge defined by Eq. (\ref{eq:gaugeN}).

The change of gauge in Eq. (\ref{eq:gaugeN}) corresponds to a change of the coordinate time being used to describe a motion. This is a major difference with respect to Newton's theory where there is a unique time.
A coordinate time is simply a mathematical construction of a time that can be related to the proper time of an observer, but not necessarily. For instance, it can simply be defined according to purely mathematical criteria, such as the simplicity of the field equations. 

In order to illustrate the dependence of the Newtonian potential to the coordinate time being used, let us define a coordinate time that would correspond to the proper time of an imaginary observer that is so remote from the barycenter of the solar system that it can be at rest with respect to the solar system because the potential of the solar system and its gradient at their location is negligible. Assuming that this imaginary observer is indeed at rest with respect to the solar system, this means that the coordinate time $t$ corresponds to the (fictional) proper time $\tau$ such that $dt = d\tau$ at the location of this (fictional) observer. The Newtonian potential in this gauge---or, equivalently, with this coordinate time---reads
\be
U(x^i,t) = \sum_A \frac{G m_A}{\lvert x^i-x^i_A(t) \rvert}, \label{eq:potN1}
\ee
where $m_A$ and $x^i_A(t)$ are the masses and the positions of the bodies $A$. Alternatively, one could define a coordinate time that corresponds to a (fictional) observer that would be at rest at the center of the coordinate system $x^i = 0$. In this gauge, the Newtonian potential would instead read
\be
U(x^i,t) = \sum_A G m_A\left[ \frac{1}{\lvert x^i-x^i_A(t) \rvert} - \frac{1}{\lvert x^i_A(t) \rvert} - \frac{x^j x_A^j(t)}{{\lvert x^i_A(t) \rvert}^3} \right],\label{eq:potN2}
\ee
because having $dt = d\tau$ in Eq. (\ref{eq:metrNewt}) at the center imposes $U(x^i=0)=0$, and being at rest imposes $\vec \nabla U \rvert_{x^i=0}=0$. The transformation from the potential expressed in the coordinate time of a remote observer to the potential expressed in the coordinate time of an observer at the center of the coordinate system takes the form of Eq. (\ref{eq:gaugeN}) with 
\be
a^k(t) = - \sum_A G m_A \frac{x_A^k(t)}{{\lvert x^i_A(t) \rvert}^3}\qquad \textrm{, and }\qquad b(t) =  - \sum_A \frac{G m_A}{\lvert x^i_A(t) \rvert}. 
\ee
This notably means that the leading order of the equation of general relativity is invariant under specific changes of time coordinate. As we will see in Sec. \ref{sec:covariance}, this is a leftover of the invariance of general relativity through a change of coordinate system. It is really important to notice that while the field equation Eq. (\ref{eq:Newtoneq}) is invariant under some changes of the coordinate time, the actual solution of the potential is not, as one can check from Eqs. (\ref{eq:potN1}) and (\ref{eq:potN2}).

In the coordinate time defined as the proper time of the fictional remote observer $t=\tau_r$, the proper time of fictional observer at the center of the coordinate system $\tau_o$ reads
\be
\tau_o = t - \frac{1}{2 c^2} \int \sum_A \frac{G m_A}{\lvert x^i_A(t) \rvert} dt, \label{eq:TtransfoU}
\ee
provided that they are indeed at rest with one another. {See Sec. \ref{sec:timescale} for a discussion on the coordinate times that are used by the community, following the IAU recommendations \citep{Soffel2003}.}

\paragraph{Dependence of a coordinate time to the trajectories of the gravitational bodies}

Eq. (\ref{eq:TtransfoU}) teaches us something very important. Depending on its definition, a coordinate time can depend explicitly on the trajectories of the gravitational bodies {${\bf x}_A(t)$}. As a consequence, in order to define such a coordinate time in practice, it is necessary to estimate the positions of the celestial bodies over time with enough accuracy.

But why would one want to define such a coordinate time if it means that one must already have an accurate knowledge of the trajectories of the gravitational bodies in order to define it? Simply because one has to define such a coordinate time in order to compare different observations made at various locations by several observers. The proper time of each observer is indeed different and the differences between the various proper times depend on the observers' relative trajectories in the gravitational potential. Therefore, in order to compare the various observations, one first has to define a coordinate time that will be used to transform the proper time of every observer to this coordinate time, such that after the transformation, all the observations are expressed in a common time. We will see in Sec. \ref{sec:pN} that at the post-Newtonian level---or, the next-to-leading order---similar considerations must be taken into account for the definition of space.

Any definition of a coordinate time could be used in theory. However, in practice, a coordinate time is usually defined as being the proper time of a fictional observer at some convenient location in the solar system, such as at its barycenter or at the geoid of the Earth. Such coordinate times are defined by the International Astronomical Union, as we will further discuss in Sec. \ref{sec:timescale}.

But since one needs to know the trajectories of the gravitational bodies in the solar system in order to construct this coordinate time, it means that one needs to use planetary ephemerides to construct such a time. Therefore, planetary ephemerides actually deliver the coordinate times defined by the International Astronomical Union to the community, as discussed more in details in Sec. \ref{sec:timescale}.

\paragraph{Rotating reference frames}

The Newtonian approximation of the spacetime metric in general relativity assumes the form of Eq. (\ref{eq:metrNewt}) only within a specific category of coordinate systems, typically referred to as \textit{inertial frames}. Notably, the coordinate system must be kinematically non-rotating relative to distant celestial objects such as quasars. 
This suggests that the rotation (or absence thereof) of a local reference frame must be defined in relation to exceedingly distant objects. In support of his theory, Newton posited this coincidence as evidence of an absolute space, serving as the stage for dynamic events and providing a reference for defining rotation. In contrast, figures such as Leibniz and later Mach, contended that this apparent coincidence reveals that inertia is relative, depending more on the universe matter content than on an absolute space. Heavily influenced by these perspectives, Einstein proposed the \textit{principle of inertia relativity}, later referred to as \textit{Mach's principle} \citep{einstein:1918an}. This principle stipulates that spacetime is wholly determined by its matter content. Considering general relativity accommodates vacuum solutions, whether Einstein's theory satisfies the principle of inertia relativity remains debated \citep{book_mach_principle}. However, practically speaking, the solar system asymptotic metric must be integrated into the larger spacetime metric effectively generated by distant sources. This requirement accounts for why inertial reference frames are non-rotating relative to very remote sources in general relativity---assuming they can be approximated as stationary due to their minimal angular velocity in the sky over certain timescales.

\subsubsection{Invariance of the laws through a change of coordinate system}
\label{sec:covariance}
\paragraph{Tensors}
\label{sec:tensor}
The tensorial nature of the Einstein-Hilbert equation Eq. (\ref{eq:EHeq}) enforces the \textit{covariance principle},\footnote{Or, \textit{principle of relativity} \citep{einstein:1918an}.} which demands that the laws of Nature do not depend on the choice of the coordinate system. Indeed, tensorial equations---such as Eq. (\ref{eq:EHeq})---are invariant under change of coordinates. A tensor is defined by the way it transforms under a change of coordinates. A tensor $W$, with $n$ contravariant indices and $m$ covariant indices transform as

\be
W^{\alpha_1 ... \alpha_n}_{\beta_1 ... \beta_m}\{x^\sigma\} = \left(\prod_{\alpha_k} \prod_{\beta_k}  \frac{\partial x^{\alpha_k}}{\partial X^{\mu_k}} \frac{\partial X^{\nu_k}}{\partial x^{\beta_k}}\right) W^{\mu_1 ... \mu_n}_{\nu_1 ... \nu_m}\{X^\sigma\}. \label{eq:chgtCoord}
\ee
Tensorial equations are manifestly invariant under coordinate change, that is:
\begin{equation}
A_{\alpha_1 \alpha_2 \ldots \alpha_n}^{\sigma_1 \sigma_2 \ldots \sigma_p} \{x^\sigma\}=B_{\alpha_1 \alpha_2 \ldots \alpha_n}^{\sigma_1 \sigma_2 \ldots \sigma_p} \{x^\sigma\}\Leftrightarrow A_{\beta_1 \beta_2 \ldots \beta_n}^{\omega_1 \omega_2 \ldots \omega_p} \{X^\sigma\}=B_{\beta_1 \beta_2 \ldots \beta_n}^{\omega_1 \omega_2 \ldots \omega_p} \{X^\sigma\}.
\end{equation}
Because the metric field is a tensor, one can notably verify with Eq. (\ref{eq:chgtCoord}) that the line-element
\be
ds^2 = - c^2 d\tau^2 = g_{\alpha \beta}\{x^\sigma\} dx^\alpha dx^\beta = g_{\alpha \beta}\{X^\sigma\} dX^\alpha dX^\beta \label{eq:lineE}
\ee
is invariant under coordinate change. Indeed, the proper time of an observer cannot depend on any coordinate system as it is a measured quantity. Let us note however, that according to this definition, the Christoffel connection defined in Eq. (\ref{eq:Christoffel}) is not a tensor as it transforms as follow:
\begin{equation}
\Gamma_{\mu \nu}^{\alpha}  \{x^\sigma\}=\frac{\partial x^{\alpha}}{\partial X^\beta} \frac{\partial X^\sigma}{\partial x^{\mu}} \frac{\partial X^\rho}{\partial x^{\nu}} \Gamma_{\sigma \rho}^\beta \{X^\sigma\}-\frac{\partial^2 x^{\alpha}}{\partial X^\sigma \partial X^\rho} \frac{\partial X^\sigma}{\partial x^{\mu}} \frac{\partial X^\rho}{\partial x^{\nu}} .
\end{equation}
One can also check that the difference between two connections is a tensor. 
What the covariance principle means in particular is that the laws of physics are independent with respect to the observer, whether the observer is inertial or not. While this is very satisfying at the fundamental level, it also implies an intrinsic ambiguity about the coordinates that one can use, as they are all equivalent in principle.

\paragraph{Observables}

Certain coordinate systems can result in what are known as \textit{spurious effects}. These are essentially false effects arising solely from the coordinate system and do not reflect reality.
Determining whether an effect is genuine or spurious can be challenging unless all calculations are made with respect to actual observables. Parameters referred to as {\it{observables}} are independent of the coordinate system, and therefore, can correspond to quantities that an observer can directly measure. These include values such as the proper time elapsed for an observer between two events, or, as we will discuss shortly, the angle between two light cones at the observer's location. On the contrary, it is very important to keep in mind that positions, trajectories, or velocities, are not observables, but have to be reconstructed from observations after assuming a specific coordinate system---as well, as we will see, as a model for the dynamics of bodies and light. 
\subparagraph{Observables Are Scalar Quantitites}

According to Eq. (\ref{eq:chgtCoord}), the only type of tensors that do not depend on the coordinate system are scalars, which are such that
\be
S\{x^\sigma\} = S\{X^\sigma\}.
\ee
Tensors, if not scalars, therefore cannot represent observable quantities. 

\subparagraph{Positions in two dimensions: Projection on the Celestial Sphere}

\begin{figure}
\centering
\includegraphics[scale=0.5]{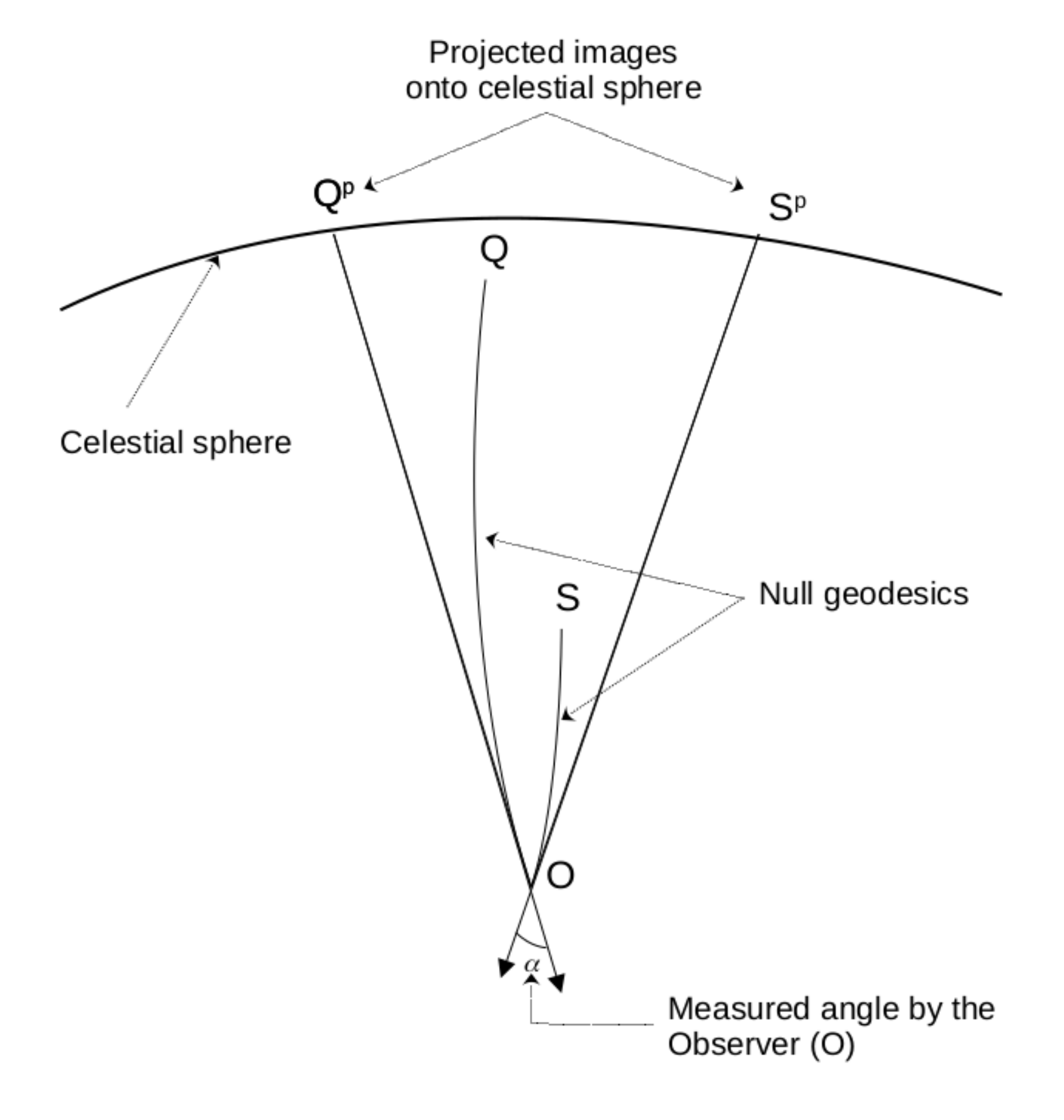}
\caption{{Schematic illustration of the angle ($\alpha$) determined by an observer (O) at their location, between two \textit{null-geodesics}---see Sec. \ref{sec:EMwaves}---that link the observer to two sources (Q and S) of electromagnetic waves. Each null-geodesic represents a segment of the \textit{light-cone} centered on an emitter.}}
\label{fig:lightcones}
\end{figure}

From an observer's perspective, the only tangible element related to the spatial distribution of any distant object is the relative angular separation in the sky between the images of two objects. This measurement can be made without reference to any specific coordinate system. This angular separation corresponds to angles between the light cones connecting the distant objects to the observer at the observer's location, as shown in Figure \ref{fig:lightcones}. 

However, observers typically project at their location the vector of the light cones, which join the distant objects to the observer's location, onto {a coordinate-dependent representation of} their celestial spheres. This process is where coordinate systems become to be used. Importantly, it must be noted that the relative position (angle) of the same distant objects as viewed by two different observers can vary. This is because in general relativity, both time and space are relative to the observer and to its gravitational environment. Hence, there is no such thing as absolute positions. In simpler terms, the celestial spheres for different observers differ in general. Although this effect is minute for observers in the solar system, it nonetheless needs to be accounted for in modern astrometry.


In practical terms, an object location on the celestial sphere is determined by observing the angular separation in the sky between the object and reference points like quasars, from the perspective of the observer. Indeed, due to their extremely remote location, the apparent movement of these reference points is negligible at the present level of accuracy for the astrometric measurements for these objects (see Sec. \ref{sec:ICRF}), and they can therefore indeed serve as static reference points (or astrometric candle light).

{However, there} is another significant effect to consider, which is also illustrated in Fig. \ref{fig:lightcones}: the propagation of light is not generally linear. This means that if an observer sees a remote object at a particular location on his celestial sphere, it doesn't imply that the object is actually at that position on the sphere. This effect is referred to as gravitational lensing (or deflection of light in the solar system) and is due to the fact that light propagates on null-geodesics, see Sec. \ref{sec:EMwaves}---which are not straight lines in general. Although its impact is relatively weak for objects within the solar system, it highlights the inherent difficulty of attributing positions on the celestial sphere in general relativity.


Hence, assigning a two-dimensional position to an object in general relativity not only necessitates the definition of a coordinate system, but also requires a model of the gravitational field through which light has propagated. Consequently, any position in general relativity is model-dependent and is essentially a reconstruction from observations.

\subparagraph{The Third Dimension: Range}


There are various methods to infer the distance between a remote object and an observer. For s/c, the most accurate technique is typically the measure of the time delay between the emission of a signal in the direction of the spacecraft and the reception of the returned signal by the probe (ranging). Essentially, this involves measuring the round-trip time of an electromagnetic signal emitted at a given frequency between an observer and a spacecraft. Given that the signal propagates at a constant speed in vacuum, this propagation time can be converted into a distance. However, while  the signal indeed travels at the speed of light in a vacuum locally, its propagation time measured by a given observer can be affected by the gravitational field localised between the source of emission and this observer, along {the trajectory of the electromagnetic signal}. As a result, even when a the signal follows a straight line between two points, the propagation time would not correspond to the Euclidean distance between these points divided by the speed of light. This effect is known as the \textit{Shapiro delay}---See Sec. \ref{sec:EMwaves}. 

{Therefore, determining} distances through ranging techniques { not only depends on the specific choice of the coordinate system considered but also depends on the knowledge one has on the gravitational potential in the solar system. Consequently, distances are both model and theory dependent. Indeed, the potential in the solar system is reconstructed by solving the field equations of a given theory---e.g. Eq. (\ref{eq:metrNewt})---after assuming a specific model for the solar system that ``sources'' the field equations---e.g. the right hand side of Eq. (\ref{eq:metrNewt}).}

{Let us stress in particular that a} modification of the field equations {of general relativity therefore} implies {the need for} a new analysis of the measures, leading to a new determination of the distances. 



\subparagraph{The Need of Conventions for Coordinate Systems}

Because space and time are relative, for various observers to compare their respective observations and agree on the positions of bodies, they must define a set of {coordinates} that allows to convert their respective observations in terms of positions in space and time. This is the role of the IAU recommendations that we will discuss in Sec. \ref{sec:modeleRF}.

\subsubsection{The action of general relativity}
\label{sec:L4fields}
Just as the equations of motion can be derived from the Euler-Lagrange equation, the field equation of general relativity can be derived from a Lagrangian density that is defined based on the metric field and the various material fields. The key difference lies in the type of dynamical variables used in the Lagrangian: instead of using position and velocity as dynamical variables, one considers the fields and their derivatives as dynamical variables in the case of a Lagrangian density.\footnote{Note that mathematicians typically refer to $\sqrt{-g} \L$ as a \textit{density} because $\L$ is a scalar and a scalar multiplied by the square root of the determinant corresponds to a mathematical object referred to as a density of weight $-1$; however, physicists typically refer to $\L$ as a density, as it carries the dimension of an energy density and is frequently associated with the definitions of kinetic and potential energy densities.} One can define the action of general relativity as follows
\be
S_{GR} = \int \sqrt{-g}d^4x \left[ \frac{R}{2\kappa} + \Lm \right], \label{eq:GRaction}
\ee
where $g$ is the metric determinant, $\Lm$ the Lagrangian density of matter fields{, and $\kappa = 8 \pi G / c^4$ the coupling constant between matter and curvature}. In order to derive the Euler-Lagragian equation, $R$ is treated as a functional constructed upon the metric field and its derivative---see Eq. (\ref{eq:Ricci}). $\Lm$ is then a functional constructed upon the matter fields and their derivative. For instance, for an electromagnetic field one has 
\be
\Lm =\frac{1}{4\mu_0} g^{\alpha \sigma} g^{\beta \epsilon} (\partial_\alpha A_\beta - \partial_\beta A_\alpha)(\partial_\sigma A_\epsilon - \partial_\epsilon A_\sigma) \label{eq:LEM}
\ee
where $\mu_0$ is the magnetic permeability of vacuum, and $A^\alpha$ is the electromagnetic four-vector. Let us note that one has 
$A^\alpha = (\phi/c,\bm A)$ where $\phi$ is the scalar potential and $\bm A$ the vector potential of classical electromagnetism, form which one can compute the electric and magnetic fields \citep{jackson:1998bk}. 

Defining a theory from its action rather than from its equations is convenient because it allows one to be sure that the theory possesses conservation laws for the considered theory, which derive from Noether's theorem {that states that every symmetry of a Lagrangian implies the existence of a conservation law \citep{noether:1918mk,wald1984bk}}. 
Therefore, most modern theories are defined from an action, although there are some exceptions---such as MOND, see Sec. \ref{sec:mond}.

Applying the principle of least action on the action (\ref{eq:GRaction}), one recovers the Einstein-Hilbert equation of general relativity Eq. (\ref{eq:EHeq}), with
\be
T_{\alpha \beta} = -\frac{2}{\sqrt{-g}} \frac{\delta(\sqrt{-g} \Lm)}{\delta g^{\alpha \beta}},
\ee
where $\delta$ stands for a variational derivative. {It is worth noting that from Noether's theorem, the diffeomorphism invariance---that is, the invariance under change of coordinates---of the matter Lagrangian $\L_m$ implies the (covariant) conservation of the stress-energy tensor}
\be
\nabla_\sigma T^{\mu \sigma}  =0, \label{eq:consSET}
\ee
{where $\nabla_\sigma$ is the covariant derivative defined as followed for a tensor $X$ with $k$ contravariant indices and $l$ covariant indices:
\bea \label{eq:covnabla}
\nabla_\sigma X^{\mu_1 ... \mu_k}_{\nu_1...\nu_l}& =& \partial_\sigma X^{\mu_1 ... \mu_k}_{\nu_1...\nu_l}\\
&& +\Gamma^{\mu_1}_{\sigma \lambda} X^{\lambda ... \mu_k}_{\nu_1...\nu_l}+ ... + \Gamma^{\mu_k}_{\sigma \lambda} X^{\mu_1 ... \lambda}_{\nu_1...\nu_l} \nonumber\\
&& - \Gamma^\lambda_{\sigma \nu_1} X^{\mu_1 ... \mu_k}_{\lambda...\nu_l} - ... - \Gamma^\lambda_{\sigma \nu_l} X^{\mu_1 ... \mu_k}_{\nu_1...\lambda}. \nonumber
\eea}
This is, of course, consistent with the geometrical fact that $\nabla_\sigma G^{\mu \sigma}=0$, with $G^{\mu \nu}= R^{\mu \nu} - 1/2~ g^{\mu \nu} R$, the Einstein tensor, {where $R^{\mu \nu}$ is the contravariant Ricci tensor defined from the covariant Ricci tensor $R_{\mu \nu}$ and the contravariant metric tensor $g^{\mu \nu}$ as $R^{\mu \nu} = g^{\mu \alpha} g^{\nu \beta} R_{\alpha \beta}$.} 

\subsubsection{The post-Newtonian approximation of general relativity and gauge invariance}
\label{sec:pN}

First, let us recall that the metric of general relativity at the Newtonian level, for a reference frame that is not kinematically rotating with respect to distant celestial bodies,\footnote{Beyond rotating coordinate systems, it's noteworthy that one could also define and utilize coordinate systems with shear, instead of pseudo-cartesian coordinate systems. However, such systems are rarely used because they would notably complicate the calculations, and likely also hinder understanding.} reads
\bea
g_{00} = -1 + \frac{2 U}{c^2} + \Ol(c^{-3}), \label{eq:N00}\\
g_{0i} = \Ol(c^{-2}),\\
g_{ij} = \delta_{ij}+  \Ol(c^{-1}), \label{eq:Nij}
\eea
such that one indeed has
\be
ds^2 = - \left(1- \frac{2 U}{c^2} \right) c^2 dt^2 + dx^2 + dy^2 + dz^2,
\ee
at leading order. The truncation between leading and next-to-leading orders in Eqs. (\ref{eq:N00}-\ref{eq:Nij}) are made with respect to the equation of motion that derives from the Lagrangian of motion Eq. (\ref{eq:defLagrangian}). Indeed, the Lagrangian of motion with Eqs. (\ref{eq:N00}-\ref{eq:Nij}) reads
\be
L \propto1 + \frac{1}{c^2} \left(\frac{v^2}{2} - U  \right) + \Ol(c^{-3}) .
\ee
One can show that a set of coordinate systems exists in general relativity  that are such that, at next-to-leading order, the metric reads \citep{damour1990prd}
\bea
g_{00} = -1 + \frac{2 w}{c^2} - \frac{2 w^2}{c^4}+ \Ol(c^{-5}), \label{eq:pN00}\\
g_{0i} = - \frac{4 w^i}{c^3}\Ol(c^{-4}),\\
g_{ij} = \delta_{ij} \left(1+ \frac{2 w}{c^2} \right)+  \Ol(c^{-3}), \label{eq:pNij}
\eea
{where $w$ and $w^i$ are gravitational potentials.
This is the so-called post-Newtonian metric of general relativity, in conformally cartesian coordinates.\footnote{The name \textit{conformally cartesian} stems from the fact that one has $g_{00}g_{ij} = - \delta_{ij}+\Ol(c^{-4})$---that is, the space-space component of metric $g_{ij}$ multiplied by a conformal factor (here, the time-time component of the metric $g_{00}$) is flat up to $c^{-4}$ corrections. It's worth noting that such coordinate systems generally do not exist in general in theories other than general relativity---at least when the metric considered still defines the proper time $d\tau^2 = - g_{\alpha \beta} dx^\alpha dx^\beta$ and not a conformal frame.} Assuming this metric, the field equation of general relativity reduces to
\bea
\Delta w+\frac{3}{c^2} \partial_{t t}^2 w+\frac{4}{c^2} \partial_{t j}^2 w_j=-4 \pi G \sigma+O\left(c^{-3}\right), \label{eq:pNf00}\\
\Delta w_i-\partial_{i j}^2 w_j-\partial_{t i}^2 w=-4 \pi G \sigma^i+O\left(c^{-2}\right),\label{eq:pNfij}
\eea
where $\sigma = c^{-2} (T^{00}+T^{kk})$ and $\sigma^i = c^{-1} T^{0i}$, such that $\sigma$ and $\sigma^i$ are zeroth order quantities---because $T^{\mu \nu} = \Ol(c^2,c,c^0)$ in the weak-field and slowly moving approximation.\footnote{One can verify that with the stress energy tensor of dust, which reads $T^{\alpha \beta} = - \rho_m u^\alpha u^\beta$, where $\rho_m$ is the rest mass density and $u^\alpha = dx^\alpha/d\tau$, such that $u^0 \sim c$.}

\paragraph{Post-Newtonian gauge invariance}

One can check that the equations Eqs. (\ref{eq:pNf00}-\ref{eq:pNfij}) are invariant under the following transformations
\bea
w^{\prime}=w-\frac{1}{c^2} \partial_t \lambda, \\
w_i^{\prime}=w_i+\frac{1}{4} \partial_i \lambda ,
\eea
where $\lambda$ is an arbitrary differential function. Indeed, while the form of the metric in Eqs. (\ref{eq:pN00}-\ref{eq:pNij}) completely fixes the type of spatial coordinates being considered, it leaves a freedom at the level of the time coordinate. This gauge invariance indeed corresponds to a shift of the time variable 
\be
\delta t = c^{-4} \lambda(x^\alpha). \label{eq:deltatgauge}
\ee
After having imposed a non-kinematically-rotating frame with respect to distant celestial objects (such as quasars), a central world-line and the use of conformally cartesian coordinates, Eq. (\ref{eq:deltatgauge}) is a remaining freedom of our coordinate system.

\paragraph{Harmonic gauge}

The International Astronomical Union recommends \citep{Soffel2003} the use of harmonic coordinates that are such that
\be
g^{\alpha \beta} \Gamma^\sigma_{\alpha \beta}=\Ol(c^{-5},c^{-4}).
\ee
However, because of the use of conformally cartesian coordinates, the space component of this condition $g^{\alpha \beta} \Gamma^i_{\alpha \beta}=\Ol(c^{-4})$ is already satisfied, and one is left with the time component of this condition that, considering the metric Eqs. (\ref{eq:pN00}-\ref{eq:pNij}), reduces to
\be
\partial_t w + \partial_j w_j = \Ol(c^{-2}). \label{eq:redHG}
\ee
With this condition, Eqs. (\ref{eq:pNf00}-\ref{eq:pNfij}) reduce to
\bea
\square_m w=-4 \pi G \sigma+O\left(c^{-3}\right), \\
\triangle w_i=-4 \pi G \sigma^i+O\left(c^{-2}\right),
\eea
where $\square_m$ and $\triangle w_i$ are respectively the d'Alembertian and Laplacian of the usual flat Minkowski spacetime: 

\be \label{eq:MinDalemb}
\square_m = - c^{-2} \partial^2_{tt} + \triangle,
\ee
where, for instance, $\triangle = \partial^2_{xx}+\partial^2_{yy}+\partial^2_{zz}$ in Cartesian coordinates.
\subsubsection{The Einstein-Infeld-Hoffman-Droste-Lorentz equation of motion} \label{sec:EIHDLeqm}

Up to corrections that can be taken into account at a later stage, celestial bodies in the solar system can be approximated as being non-rotating point particles. This approximation has been explored for the first time by Lorentz and Droste \citep{ld19171,ld19172}---translated in \citep{lorentz1937}---and re-derived later  by Einstein, Infeld and Hoffman \citep{eih1938}. The stress-energy tensor for point particles is simply the stress-energy tensor of a dust fluid
\be
T^{\mu \nu} =  \rho_m u^\mu u^\nu,
\ee
{where $u^\alpha = dx^\alpha / d\tau$ is the \textit{proper} four-velocity of the fluid.}
In terms of conserved mass along the fluid geodesics $dm_A/d\tau = 0$, it reads
{
\be
T^{\mu \nu} = \sum_A \frac{m_A~ u^0/c}{\sqrt{-g}}  v_A^\mu v_A^\nu \delta^{(3)}(\bm x - \bm x_A(t)), \label{eq:SETEIH}
\ee
}
where $\delta^{(3)}$ is the 3-dimensional delta function{, and $v^\alpha = dx^\alpha/dt$ the \textit{coordinate} four-velocity of the fluid, such that $u^\alpha = u^0 v^\alpha / c$}. Eq. (\ref{eq:SETEIH}) used the fact that $u_\mu \nabla_\sigma T^{\mu \sigma} =0$ ---which follows from Eq. (\ref{eq:consSET})---such that one has the usual Newtonian conservation of the mass density {$\partial_\sigma (\rho^* v^\sigma)=\partial_t \rho^* + \partial_i (\rho^* v^i)=0$} for the density {$\rho^* := \sqrt{-g}  \rho_m u^0/c$}. 
Solving the Einstein-Hilbert equation Eq. (\ref{eq:EHeq}) with this approximation, and in the harmonic gauge Eq. (\ref{eq:redHG}), leads to the metric in Eqs. (\ref{eq:pN00}-\ref{eq:pNij}) with 
\bea
&&w=w_{0}-\frac{1}{c^2}\Delta+O(c^{-4}) \label{eq_wet}, \\
&&w^i=\sum_A\frac{G m_{A}}{r_A}v_A^i+O(c^{-2}), \label{eq_wiet}	
\eea
where
\be
w_0=\sum_A\frac{G m_{A}}{r_A},
\ee
where $\bm{r}_A=\bm{x}-\bm{x}_A(t)$, $r_A= \lvert\bm{r}_A\rvert$, and
\be
\Delta= \sum_A \frac{G m_A}{r_A}\left\{-2 v_a^2+\sum_{B \neq A} \frac{G m_B}{r_{B A}}+\frac{1}{2}\left[\frac{\left({\bm r_A} \bm{v_A}\right)^2}{r_A^2}+{\bm r_A} {\bm a_A}\right]\right\},
\ee
with $r_{AB}= \lvert \bm{x}_B - \bm{x}_A \rvert$.
We should note that this corresponds to the metric recommended by the International Astronomical Union \citep{Soffel2003}, subject to corrections accounting for the fact that celestial bodies are not point-like but extended objects, and that these bodies possess angular momentum relative to the frame that is fixed with respect to distant objects, such as quasars. Fortunately, these corrections are numerically small, such that they can safely be added a posteriori at the leading order, without impacting the calculation of the next-to-leading order.

Now, injecting this metric into the definition of the Lagrangian of motion Eq. (\ref{eq:defLagrangian}), one gets for a test particle $B$ the following Lagrangian of motion
\bea
L_{B}&&=-m_{B}c^2+m_{B}\frac{v_B^2}{2}+\sum_{A\ne B}\frac{G m_{A}m_{B}}{r_{AB}}+\frac{m_{B}v_{B}^4}{8c^2} \nonumber\\
&&+\frac{1}{c^2}\sum_{A\ne B}\frac{G m_{A}m_{B}}{r_{AB}}\left[-4 \bm{v}_A\cdot\bm{v}_B +\frac{3}{2}v_B^2+2 v_A^2 -\frac{(\bm{v}_A\cdot\bm{r}_{AB})^2}{2r_{AB}^2}-\frac{\bm{r}_{AB}\cdot\bm{a}_A}{2} \right]\nonumber\\
&& -\frac{1}{c^2}\sum_{A\ne B}\frac{G m_{A} m_{B}}{r_{AB}}\left[ 
\sum_{D\ne A}\frac{G m_{D}}{r_{AD}}+\sum_{D\ne B}\frac{1}{2} \frac{G m_{D}}{r_{DB}}\right]+\Ol(c^{-4}).\nonumber
\eea
Injecting this Lagrangian into the Euler-Lagrange equation Eq. (\ref{eq:EulLag}), one finally gets the Einstein-Infeld-Hoffman-Droste-Lorentz (EIHDL) equation of motion given in Eq. (\ref{eq:accgeneral relativity}) in Sec. \ref{sec:dynam}.
\paragraph{The advance of perihelion of Mercury}

When focusing solely on the two-body problem, one can examine the secular changes of the orbital elements by treating the $c^{-4}$ term in the equation of motion Eq. (\ref{eq:accgeneral relativity}) as perturbations to Keplerian orbits. For instance, the secular advance of a perihelion in the two-body problem can be expressed as follows
\be
\Delta\dot\varpi = 6 \pi \frac{G (m_1+m_2)}{a (1-e^2)c^2},
\ee
where $a$ is the Keplerian semi-major axis, $e$ the eccentricity and $m_i$ the mass of the body $i$. 

\subsubsection{The propagation of light in general relativity}
\label{sec:EMwaves}


From the Lagrangian of a free electromagnetic field Eq. (\ref{eq:LEM}), one derives the following equation for the free electromagnetic field in a curved spacetime
\be
\nabla_\sigma F^{\alpha \sigma}= \partial_\sigma F^{\alpha \sigma} + \Gamma^\alpha_{\sigma \epsilon} F^{\epsilon \sigma} + \Gamma^\sigma_{\sigma \epsilon} F^{\alpha \epsilon} =0, \label{eq:EMfreefield}
\ee
where $\nabla_\sigma$ defines the covariant derivatives associated to the Christoffel connection {Eq. (\ref{eq:covnabla})}. Using the definition of the electromagnetic tensor in Sec. \ref{sec:L4fields}, and translating the 4-vector potential in terms of electric and magnetic fields \citep{jackson:1998bk}, one recovers the equations of Maxwell for free electromagnetic fields if spacetime is flat---that is $\Box_m \vec E=0$ and $\Box_m \vec B=0$, where $\Box_m$ is the usual D'Alembertian of flat Minkowski spacetimes defined in Eq. (\ref{eq:MinDalemb}). Considering the Lorenz gauge $\nabla_\sigma A^\sigma = 0$, Eq. (\ref{eq:EMfreefield}) reduces to
\be
\Box A^\alpha - g^{\alpha \sigma} R_{\sigma \epsilon} A^\epsilon = 0.,
\ee
where one defines the covariant D'Alembertian as $\Box = g^{\sigma \epsilon} \nabla_\sigma \nabla_\epsilon$. Now, let us expand the 4-vector potential as follows \citep{MTW}
\begin{equation}
A^\mu=\mathfrak{R}\left\{\left(a^\mu+\epsilon b^\mu+O\left(\epsilon^2\right)\right) \exp^{i \theta / \epsilon}\right\},
\end{equation}
{where $\mathfrak{R}\{X\}$ means the real part of $X$.} The leading order {in the $\epsilon$ expansion} corresponds to the \textit{geometric optics approximation}. It induces that\be
k^\sigma \nabla_\sigma k^\alpha = 0,
\label{eq:72}
\ee
where {the wave-vector is defined as} $k^\alpha := g^{\alpha \sigma} \partial_\sigma \theta$ and
{
\be
k_\sigma k^\sigma = g_{\sigma \epsilon} k^\sigma k^\epsilon = 0. \label{eq:nullC}
\ee}
Eq. (\ref{eq:72}) means that in the geometric optics limit, electromagnetic waves follow spacetime geodesics
\begin{equation}
\frac{d k^\alpha}{d \lambda}+\Gamma_{\mu \nu}^\alpha k^\mu k^\nu=0, \label{eq:geodlight}
\end{equation}
where $\lambda$ is an affine parameter of the geodesics, such that $k^\alpha = dx^\alpha / d\lambda$. Eq. (\ref{eq:nullC}) means that those geodesics are such that the line element is null ($ds^2 =0$) along the trajectories of the electromagnetic waves in the geometric optics approximation. One therefore generically says that light follows \textit{null-geodesics}, although this is in fact correct only in the geometric optics approximation. The fact that $ds^2=0$ along the geodesics of light implies that the trajectory lies on the null spacetime cones, which define the causal structure of spacetime. In other words, the speed of light is indeed equivalent to the speed $c$ appearing in the definition of the line element Eq. (\ref{eq:lineE}). But it also means that there is no notion of proper time for light since the line element $ds^2 = - c^2 d\tau^2$ is null along their geodesics. 

\paragraph{{Null-geodesics and astrometric observables}}

Because electromagnetic waves follow null-geodesics of a curved spacetime in the geometric optic approximation, the trajectories of electromagnetic waves are curved in general, notably leading to the deflection mentioned in Sec. \ref{sec:covariance} and represented in Fig. \ref{fig:lightcones}.

{One side of astrometry is about determining the projection of the positions of celestial bodies on the celestial sphere as they are seen by an observer, based on their angular measurements. Indeed, as detailed in Sec. \ref{sec:covariance}, what measures an observer are the angles between the null-cones that link them to the sources of the observed electromagnetic waves.}

\begin{figure}
\centering
\includegraphics[scale=0.4]{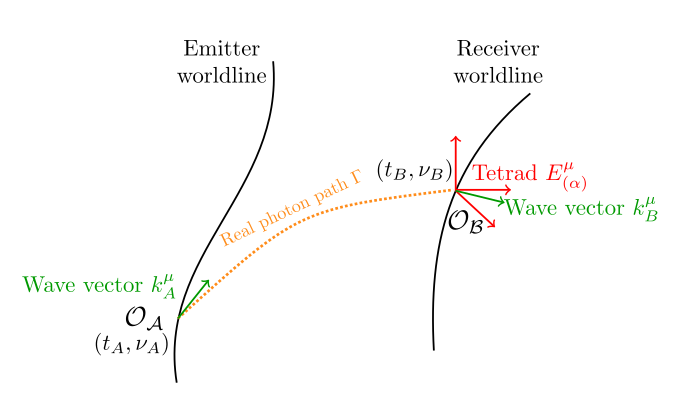}
\caption{{Illustration of an electromagnetic wave with frequency $\nu_A$ emitted by $\mathcal{O}_{\mathcal{A}}$
with a wave four-vector of components $k^\mu_A$ and received by $\mathcal{O}_{\mathcal{B}}$ at a frequency $\nu_B$ and with a wave four-vector of components $k^\mu_A$. Figure taken from \citep{hees:2014pr}, with the authorization of the lead author.}}
\label{fig:astrometry}
\end{figure}

As explained notably in \citep{hees:2014pr}, one way to get a covariant definition\footnote{That is, a definition that is invariant under change of coordinate systems.} of the position of the electromagnetic sources in the celestial sphere {as it appears for} an observer, is to use the tetrad formalism \citep{MTW,klioner:1992aj}, by giving the direction of observation of an incoming electromagnetic wave in a tetrad $E$ comoving with the observer $\mathcal{O}_{\mathcal{B}}$, as one can see in Fig. \ref{fig:astrometry}. We write $E^\mu_{\langle \alpha \rangle}$ the components of this tetrad, where $\langle \alpha \rangle$ corresponds to the tetrad index and $\mu$ is a normal tensor index that can be lowered and raised by use of the metric. The tetrad is assumed to be
orthonormal so that
\begin{equation}
g_{\mu \nu} E_{\langle\alpha\rangle}^\mu E_{\langle\beta\rangle}^\nu=\eta_{\langle\alpha\rangle\langle\beta\rangle},
\end{equation}
$\eta$ being the flat Minkowski metric, and the vector $E^\mu_{\langle 0 \rangle}$ is chosen to be timelike, such that $E^\mu_{\langle i \rangle}$ is spacelike. Then, the wave-vector $k_\mu = g_{\mu \nu} k^\nu$ becomes in the tetrad frame associated to the observer
\begin{equation}
k_{\langle\alpha\rangle}=E_{\langle\alpha\rangle}^\mu k_\mu.
\end{equation}
The incident direction of the wave in the tetrad frame is given by the following normalization
\begin{equation}
n^{\langle i\rangle}=\frac{k^{\langle i\rangle}}{\sqrt{\delta_{j k} k^{\langle j\rangle} k^{\langle k\rangle}}}=\frac{k^{\langle i\rangle}}{k^{\langle 0\rangle}}=-\frac{k_{\langle i\rangle}}{k_{\langle 0\rangle}}.
\end{equation}
This quantity is the actual astrometric observable at the location of the observer. For more details, we refer the reader to \citep{MTW,klioner:1992aj,hees:2014pr}.

{The other side of astrometry is about determining the distance of remote objects, which leads us to the concept of the Shapiro delay.}

\paragraph{Shapiro delay}

In the solar system, most of the time, the trajectory of light can be approximated as being straight lines at leading order.\footnote{This approximation holds true when one is sufficiently distant from the {\textit{gravitational lensing regime}}, but {starts to fail} near this {regime}, resulting in the emergence of the so-called \textit{enhanced terms} during the derivation of light trajectory in geometric configurations where this approximation is only marginally accurate \citep{ashby:2010cq,linet:2016pr}. For a brief discussion on this matter, we refer to Sec. \ref{sec:shapiro}.} That is, one has $x^i(t) = n^i c (t-t_e) + x^i_e + \Ol(c^{-2})$---where $e$ stands for the emission{, and $n^i$ is a constant normalized vector}. One can use this information in order to compute the coordinate time elapsed between the emission and the reception of light, without the need to actually solve the geodesic equation Eq. (\ref{eq:geodlight}). From the null condition Eq. (\ref{eq:nullC}), one has, indeed, that $ds^2 = 0$ between the emission and the reception of the electromagnetic wave. This means that {$g_{\mu \nu} dx^\nu dx^\mu=0$, where $g_{\mu \nu}$ is given by Eqs. (\ref{eq:pN00}-\ref{eq:pNij}). However, since light travels at the speed of light, $v^i/c$ is not a negligible quantity, and it is necessary to maintain the same order of development in terms of $c^{-n}$ in the metric to account for all the relevant terms at a given order for the propagation of light. As a result, when considering deviations from the trajectory of light relative to special relativity up to the order $\Ol(c^{-3})$, the relevant metric is}
\be
ds^2 = - \left(1- \frac{2 U}{c^2} \right) c^2 dt^2 + \left(1+ \frac{2 U}{c^2} \right) (dx^2 + dy^2 + dz^2)+ \Ol(c^{-3}) =0,
\ee
{where $U = w + \Ol(c^{-2})$---see Eqs. (\ref{eq:Newtoneq}) and (\ref{eq:pNf00})}.
Integrating over this equation, one gets the coordinate time elapsed between the emission and the reception $T_{er} = t_r-t_e$ 

\be
c~ T_{er} = \lvert x^i_r - x^i_e \rvert + \frac{2}{c^2} \int_e^r U d l + \Ol(c^{-3}),\label{eq:orShapiro}
\ee
where the integration is taken along the straight line $x^i(t) = n^i (t-t_e) + x^i_e$ that connects the emission and the reception events. It appears as if light experiences a delay due to the presence of a gravitational field. This delay is known as the \textit{Shapiro delay}, named after Irwin Shapiro who was the first to predict this effect \citep{1964PhRvL..13..789S}. More on this delay in Sec. \ref{sec:shapiro}.


Hence, the Shapiro delay  has to be taken into account in order to recover the distance---in terms of a given coordinate system---from the measured round-trip propagation time. This means that distances constructed from observations not only depend on the coordinate system being used, but also depend on the model for the gravitational potential $U$ {along the trajectory of the electromagnetic wave}.


It is crucial to have in mind that what is typically probed by solar system experiments{---such as in \citep{bertotti2003nat}---}is not so much the delay in Eq. (\ref{eq:orShapiro}), but rather its variation as the observed electromagnetic signal traverses different sections of the gravitational potential $U$.\footnote{The mathematical expression of the delay as given in Eq. (\ref{eq:orShapiro}) is gauge-dependent, and thus does not represent an observable quantity by itself. For detailed discussions on this, we refer readers to \cite{gao:2000cq,minazzoli:2019pr}.} More specifically, these experiments usually probe the delay variation with respect to the the minimum distance between the electromagnetic signal trajectory and the gravitating body (also called impact parameter). For a comprehensive discussion on this topic, we direct readers to Chapter 6.3 in \cite{wald1984bk}.

\subsubsection{Alternative gravitational theories}

All the aspects of the aforementioned content may be subject to modifications in an alternative theory to general relativity---see Sec. \ref{sec:fulltest}. Hence, one must exercise caution when considering alternative theories of gravity, as the introduced modifications can impact the {entire modeling process}---from the definition of the coordinate system, {to} the equations of motion of light and {massive} bodies{---}and consequently the analysis of the observations.

 \subsection{Basics on Ephemeris}
 \label{sec:basisephem}
 An ephemeris is a table of positions and velocities given at different time steps. One can compute an ephemeris for artificial satellites, planetary bodies (planets, natural satellites, asteroids, comets...) but also pulsars.
 In order to provide to the user accurate estimations of the dynamical states of the considered body, several ingredients are necessary.\\
 
 First, one needs to agree upon a set of both time and space coordinates given in a properly defined frame. This frame will be preferably inertial and the coordinate system will give an easily understandable representation of the body motion. This coordinate system will also be used to confront the dynamical modeling of the motion to the observations (see in step 3). The selection of the frame---in which the motion and the observations will be described---and its characterisation in space and time (metrics) will constitute the step 1 of the ephemeris construction (red boxes in Fig. \ref{fig:ephem}).
 

 

 Second, we shall identify and develop the appropriate dynamical {model} for describing the most accurately {the motion of the bodies}. This step (black box in Fig. \ref{fig:ephem}) requires the writing of an equation of motion according to the frame defined in the step 1. In the dynamical modeling, one must include all the gravitational perturbations expected for the considered system. A numerical integration of the equations of motion is also usually performed in order to provide to the users positions and velocities for discrete times. Analytical resolutions were also proposed up to the beginning of the XXI century but stopped when the planetary observations became too accurate in regards to the size of the analytical series (see for example \cite{2005A&A...429..361F}).
 
 At this step, one may consider that the ephemeris is built. However, in particular in the solar system, the localisation of the object of interest has been monitored by observers and a comparison between its modeled and observed dynamical states (positions and velocities) is used for improving the model and then, to continue the process of construction of the ephemeris. 
 
 It is important to stress that the transformation from computed positions and velocities to observed quantities (direct radar range to the planetary surface, s/c navigation range and frequency shift, angular positions, pulsar time of arrival) requires some hypothesis on the space and time coordinate system in play and on how the observations have been obtained (see for example the Shapiro delay in Sec. \ref{sec:mass_shapiro} or Sec. \ref{sec:fit}). The data used for building planetary ephemerides are then not independent from the framework used for describing gravity. 
 
 In a third step, we consider the information on the positions and/or velocities obtained by the observers. 
 As explained previously, and as it will be described in details in Sec. \ref{sec:obs} in the case of planetary ephemerides, these {informations are deduced from observations that can be ranges}, frequency shifts, angular positions relative to reference stellar or quasi-stellar objects (part of astrometric catalogues or not) in different wavelengths (from optical to radio), in most of the cases centered on Earth or on specific locations at its surface. In general, at this step (blue boxes in Fig. \ref{fig:ephem}), the observations are closely analysed in order to remove outliers, and to identify and correct potential biases and systematics. 
 
 The forth step (orange box in Fig. \ref{fig:ephem}) is the confrontation between the observed quantities analysed at step 3 and the same quantities estimated with the dynamical model at step 2. 
 From this comparison, are obtained the residuals, which are the differences between model and observations. In order to have the most accurate modelisation of the observed positions, one looks for minimizing the residuals by considering different causes:
 \begin{itemize}
 \item a mismodeling in the motion of the body. In this case, the step 2 has to be reconsidered and  the dynamical modeling modified for example by adding more perturbations.
 \item parameters used in the model that are not close enough from their {\it{real}} values. The model is then corrected by updating the values of the parameters.
 \item some systematics or bias in the observations were not accounted for properly and still remain in the residuals. A modification of step 4 is then necessary.
 \end{itemize}
 It is of course not easy to isolate the different causes of the residuals. But, nevertheless, in order to reduce the differences obtained at step 4, one systematically starts with correcting the parameters of the model {by} using either a classic least squares method {or} a Bayesian approach (see for example the discussion in Sec. \ref{sec:fit}). 
 Once the model has its parameters {updated} considering the current set of observations, if residuals still present some signatures different from white noise, one can investigate the correctness of the dynamical modeling or the data analysis procedure.
This very generic procedure----that can be used for a wide set of natural or artificial bodies---- can be sketched by Fig. \ref{fig:ephem}. This Figure shows that the definition of the frame in which will be described both the motion and the observations is a crucial step for the prediction of the dynamical status of whatever object, from artificial satellites to quasi-stellar objects.

 \begin{figure}
\centering
\includegraphics[scale=0.4]{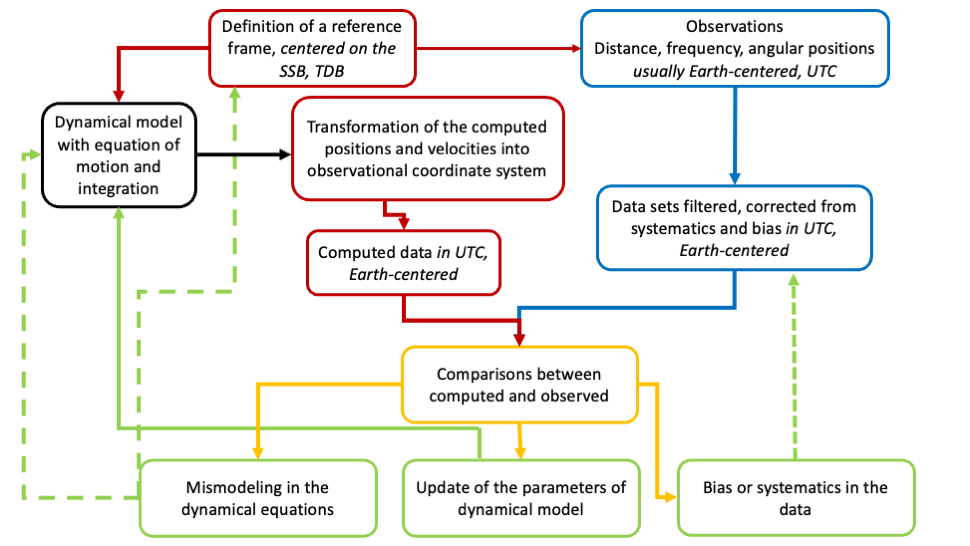}\\
\caption{Schematic representation of the construction of ephemeris applied to the planetary case. SSB stands for solar system barycenter (see Sec. \ref{sec:SSB}), TDB (UTC) stands for Barycentric Dynamimcal Time (Universal Time Coordinate, respectively) (see Sec. \ref{sec:timescale}). The description of the dynamical model is the case of planetary ephemerides is done in Sec. \ref{sec:dynam}, and the presentation of the observational data sets in Sec. \ref{sec:obs}}
\label{fig:ephem}
\end{figure}


 With the improvement of the measurement accuracy on planetary positions and velocities, the Newtonian paradigm of an universe with flat Cartesian coordinate systems and straight photon path failed to explain the observations. As discussed in introduction, the first evidence of the Newtonian failure was the case of the advance of perihelia of Mercury, explained by \cite{einstein1915}. On Table \ref{tab:3gene}, are compared the accuracies reached by three generations of planetary ephemerides and the advance of perihelia as predicted by general relativity. It is immediately visible that even with the Gaillot ephemerides in the late XIX century \citep{1888tamp.book.....G,1913AnPar..31....1G}, after three years of observations, the accumulated advance of perihelia for Mercury (1.29 seconds of arc) is greater than the observational accuracy of this epoch (1 second of arc). This leads to the choice of general relativity as a preferred framework over Newton's laws.

 However, other frameworks can also be proposed for describing the most accurate astrometric observations of planets in our solar system. In most cases, these alternative theories tend asymptotically to general relativity in the context of our weak field solar system. 
 In Section \ref{sec:fulltest}, we review the alternative gravity models for which dedicated planetary ephemerides have been constructed and published in existing literature. We prioritize these models because we believe that only a fully developed ephemeris, as described in Fig. \ref{fig:ephem}, can conclusively constrain, or even rule out, an alternative theory of gravitation. For a detailed discussion with examples, we refer to Section \ref{sec:derivedQ}.

\begin{table}
 \caption{4 generations of planetary ephemerides: accuracies of the observations used for the construction of the ephemerides either with angular measurements (Columns labeled {\it{angle}}) or with direct distance radar and spacecraft tracking observations (Columns labeled {\it{distance Earth-}}). These values give an idea of the expected precision obtained by the ephemerides during the period of observations indicated in the second row of the Table. It does not preclude from a degradation of the quality of the ephemerides out from the period. The last column gives the estimation of the advance of perihelia $\Delta \dot{\varpi}$ as computed with general relativity. For the earth, $\Delta \dot{\varpi}$ is of about 0.1 ".yr$^{-1}$. }
 \begin{tabular}{r | r r | r r | r r | r }
  \hline
Ephemerides  & \multicolumn{2}{c}{{ {Gaillot}} } & \multicolumn{2}{c}{{ {DE102}}} & \multicolumn{2}{c}{{ {DE440/INPOP19a}}} & GR\\ 
& \multicolumn{2}{c}{{\cite{1913AnPar..31....1G}}} & \multicolumn{2}{c}{{ \cite{DE102}}} & \multicolumn{2}{c}{{ {\cite{2021AJ....161..105P}}}} & \\
& \multicolumn{2}{c}{{}} & \multicolumn{2}{c}{{ }} & \multicolumn{2}{c}{{ {\cite{2019NSTIM.109.....V}}}} & \\
Data span  & \multicolumn{2}{c}{{ {1800-1913}} } & \multicolumn{2}{c}{{ {1913-1983}}} &  \multicolumn{2}{c}{{ {1924-2021}}} &\\
  \hline
 &{ angle} & { distance} & { angle} & { distance}  &  { angle} & { distance}  & $\Delta \dot{\varpi}$ \\
 &       & { Earth- }  &       & { Earth-}    &       & { Earth-}    &\\
 \hline
 & { " }    &  { km  }    & { "}     & {  km   }    & { "}      & {  km  }&  { ".yr$^{-1}$}\\
{ Mercury }& { 1 }& { 450}& { 0.050}& { 5}&  { 0.002} & { 0.004 } & 0.43\\
{ Venus }& { 0.5 }& { 100}& { 0.050}& { 2}&  { 0.002} & { 0.006 }& 0.14\\
{ Mars }& { 0.5 }& {{ 150}}& { 0.050}& {{ 0.050}} & { 0.001} & { 0.0015 } & 0.065\\ 
{ Jupiter} & { 0.5 }&{ 1400}& { 0.1}& { 10}& { 0.010} & { 0.020 } &0.019\\
{ Saturn }& { 0.5 }&{ 3000}& { 0.1}& { 600}&  { 0.001} & { 0.020 } &0.010\\
{ Uranus }& { 1 }&{ 12700}& { 0.2}& { 2540}&  { 0.050} & {10 } & 0.005\\
{ Neptune }& { 1 }&{ 22000}& { 0.2}& { 4400}&  { 0.050} & { 50} & 0.0033 \\
{ Pluto }& { 1 }&{ 24000}& { 0.2}& { 4800} &  { 0.050} & { 2400 } & 0.0027\\
\hline
 \end{tabular}
 \label{tab:3gene}
 \end{table}
 
\section{Planetary and lunar ephemerides in General relativity}

\label{sec:modele}

\subsection{State-of-the-art for planetary and lunar ephemerides}
\label{sec:ephemmodele}

The motion of the planets and asteroids in  our solar system can be solved directly by the numerical integration of their equations of motion, or with analytical approximations of their orbits. As it has been shown in \cite{2005A&A...429..361F}, the analytical {models for} the main planet orbits  are not accurate enough (due to the limited number of terms in the series) in comparison with the meter level uncertainties reached by the modern observations of planets. Therefore, in the following, we will only consider the planetary ephemerides in their numerical form. 

Based on the first preliminary  versions of the numerical integration of planetary motions \citep{Devine66, 1967AJ.....72..338A}, the DE96 JPL ephemerides \citep{1976jden.book.....S} was first of the known and widely distributed accurate numerical ephemerides fitted to observations developed  by JPL. These were followed by DE102 \citep{DE102}, DE200 \citep{1990A&A...233..252S}, DE403 \citep{Standish1995} and DE405 \citep{Standish2001}. All these ephemerides are numerically integrated with a variable step-size, variable-order, Adams method.
Their dynamical model includes point-mass interactions between the eight planets and Pluto, the Sun and a diverse number of asteroids, relativistic PPN effects  \citep{Moyer1971,Moyer2000} and lunar librations \citep{1983A&A...125..150N}. 
Since DE96, regular improvements have been added to the DE ephemerides. Ephemerides such as DE421 \citep{DE421}, DE430 \citep{folkner2014in} and DE440 \citep{2021AJ....161..105P} have been constructed and fitted with increasingly dense sets of space mission tracking data. 
Numerical ephemerides have also been developed at the Institute of Applied Astronomy of the Russian Academy of Sciences (EPM) and at the Observatory of Paris and the C\^ote d'Azur Observatory (INPOP). 
They are based on a dynamical model similar to the JPL one but with specific characteristics, in particular regarding the interactions between the main planets and the asteroids. Several possible additional contributions have been included in the EPM ephemerides such as the interactions of Trans-Neptunian Objects (TNOs) by the mean of one or several rings \citep{2018CeMDA.130...57P} and the influence with the Jupiter Trojens \citep{2020AstL...45..855P}. 
The EPM ephemerides, are  fitted to optical, radar and space tracking data and have an accuracy comparable to the JPL ephemerides \citep{1988CeMec..45..219K, 1993CeMDA..55....1K, 2001CeMDA..80..249P, 2005SoSyR..39..176P, 2014CeMDA.119..237P}. They  have been intensively used for estimating PPN parameters and the hypothetical secular variation of the gravitational constant \citep{1993CeMDA..55..313P,2005AstL...31..340P, pitjeva2021}.  
Since 2003, INPOP planetary ephemerides are developed, integrating numerically the  Einstein-Infeld-Hoffmann-Droste-Lorentz (EIHDL) equation{, as} proposed by \cite{Moyer1971,Moyer2000} (see sect. \ref{sec:modele}), and fitting the parameters of the dynamical model to the most accurate planetary observations. The main INPOP releases are INPOP08 \citep{2009AA...507.1675F}, INPOP10a \citep{fienga2011cm}, INPOP17a \citep{viswanathan2018dc}, and INPOP19a \citep{fienga2019inpop}.  For this family of ephemerides, a specific care has been brought on the consistency between the GR framework defined by the IAU and the actual equations of motion and time-scale used in the ephemerides. In particular in 2009, INPOP08 \citep{2009AA...507.1675F} was the first ephemeris built with consistent planetary orbits and  time-scales (see sect. \ref{sec:timescale}). In 2010, INPOP10a \citep{fienga2011cm} was also the first to fit the gravitational mass of the Sun instead of the astronomical unit for {consistency reasons}.\\

These three families of planetary ephemerides differ in the dynamical model (see sect. \ref{sec:aste}) such as the number of point-mass objects (more or less main-belt asteroids, TNOs) to consider in the EIHDL point-mass interaction, additional accelerations to implement (Lense-Thirring acceleration, TNOs rings, Trojan rings{, etc.}) as well as in the  size of the planetary datasets used for the adjustments of the models (see sect \ref{sec:obs}) and by the way this adjustment is performed  (see sect \ref{sec:fit}). Table \ref{tab:param} summarises these distinctions. 
However, it is important to stress that, despite their specific characteristics, the accuracy of these models are very close to each other and involve towards an even closer consistency.

\begin{table}
\caption{Solve-for parameters for recent ephemerides.  
S01 \cite{Standish2001}, F08 \cite{DE421}, F14 \cite{folkner2014in}, P17 \cite{2017AJ....153..121P}, P05 \cite{2005SoSyR..39..176P}, P14 \cite{2014CeMDA.119..237P}, P18 \cite{2018CeMDA.130...57P}, P21  \cite{pitjeva2021},    F09 stands for \cite{2009AA...507.1675F}, F11 for \cite{fienga2011cm}, F19 \cite{fienga2019inpop}, F21 \cite{fienga2022}. Column 3 and 4 indicate the model used for describing the perturbations induced by the main belt asteroids as well as the number of corresponding parameters  where  Columns 5 and 6 give the same informations but for Trans-Neptunian objects (sect. \ref{sec:aste}). Column 7 indicates which ephemeris accounts for the Lense-Thirring acceleration (see sect. \ref{sec:LT}) and Column 8  if the ephemeris was built with the astronomical unit ({\em{AU}}) fitted or with the gravitational mass of the sun fitted ({\em{GM$_{\odot}$}}). It also gives if the integration of the TT-TDB equation was done simultaneously with the equations of motion (see sect. \ref{sec:timescale}). Finally, Column 9 gives the time interval covered by the observations used for the fit.}
\begin{center}
    \begin{tabular}{l l | l l | l l | l l | l}
   & Ref &\multicolumn{2}{c} {Main Belt asteroids} & \multicolumn{2}{c} {TNO} & \multicolumn{2}{c} {Others}  & Period \\
   &  &  Model & Fit & Model & Fit &  LT & &\\
        \hline
{\bf{DE}} & & & & & & & \\
DE405 & S01 & 300 & 3 GM +  & N & N & N  & AU &1924:1998\\
& & & 3 densities &   &   &  &  & \\
DE421 & F08 & 343 & 11 GM + & N & N & N & AU &1924: 2007\\ 
& & & 3 densities &   &   &  &  & \\
DE430 & F14 & 343 & 343 GM & N & N & N & GM$_{\odot}$,TDB &1924:2018 \\
DE440 & P17 & 343  & 343 GM & 36 + 1 ring & 1 & Y & GM$_{\odot}$,TDB &1924:2020 \\
{\bf{EPM}}  && & & & & & &\\
EPM2004 &  P05 & 301 + 1 ring & 6 GM + 
& N & N & N & AU & 1913: 2004\\
& & & 3 densities &   &   &  &  & \\
EPM2011 & P14 & 301 + 1 ring  & 21 GM +  & 21 + 1 ring  & 1 & N & AU,TDB &1913:2011 \\
& & & 3 densities &   &   &  &  & \\
EPM2017 & P18 & 301 + 1 ring & 21 GM +  & 30  + 3 rings & 1 & Y & GM$_{\odot}$,TDB &1913:2016 \\
& & & 3 densities &   &   &  &  & \\
EPM2019 & P21 &301 + 1 ring  +  & 24 GM +  & 30  + 3 rings & 1 & Y & GM$_{\odot}$,TDB & 1913:2017 \\
 &  & 2  Trojan groups &  3 densities &  &  &  &  &\\
 {\bf{INPOP}} & & & & & & & &\\
INPOP08a &  F09 & 300 + 1 ring & 34 GM + & N  & N  & N & AU,TDB &1913:2007  \\
& & & 3 densities &   &   &  &  & \\
INPOP10a & F11 & 289 + 1 ring & 145 GM & N & N & N & GM$_{\odot}$,TDB & 1913:2010\\
INPOP19a & F19 & 343 & 343 GM & 9 + 3 rings & 1 & N & GM$_{\odot}$,TDB &1924:2019 \\
INPOP21a & F21 & 343 & 343 GM & 509 & 1 & Y & GM$_{\odot}$,TDB &1924:2020 \\
     \hline
    \end{tabular}
\label{tab:param}
\end{center}
\end{table}



\subsection{Reference frame theory in general relativity}
\label{sec:modeleRF}

The general relativistic framework of the planetary ephemerides since 2006 is the one summarized by the IAU2000 and IAU2006 conventions \citep{Soffel2003, 2010ITN....36....1P}.

Because general relativity is a covariant theory, an infinite set of coordinate systems could be used in principle in order to describe space-time events---see Sec \ref{sec:introGR}. The International Astronomical Union (IAU) has therefore set the standard coordinate systems that people are recommended to use, notably through the IAU2000 recommendations \citep{Soffel2003}. Two main reference systems have been defined, as well as the transformation between one another: the barycentric celestial reference system (BCRS) and the geocentric celestial reference system (GCRS). Both reference systems are defined at the post-Newtonian level and use the harmonic gauge. {Beyond the harmonic {gauge} condition, t}he freedom in chosing the coordinate systems is further reduced by fixing the form and properties of the metric and the gravitational potentials.

Planetary ephemerides are integrated in the BCRS and are linked to the realization of ICRS, by VLBI observations of s/c orbiting planets (see sect. \ref{sec:ICRF}).

\subsubsection{general relativity Barycentric Metric}
\label{sec:metric}

The BCRS is defined with the coordinates ($c t$,$x^i$), where $t=$TCB (see sect. \ref{sec:timescale}). The metric is taken to be kinematically fixed with respect to distant quasi stellar objects (QSO). The catalog gathering QSO astrometric positions and velocities used as fixed standards for the definition of the kinematically fixed BCRS is the International Celestial Reference Frame (ICRF)  \citep{1998AJ....116..516M, 2015AJ....150...58F, 2020A&A...644A.159C}. The general form of the BCRS metric is taken to be the following \citep{Soffel2003} (see Sec. \ref{sec:pN} for details on its derivation)
\begin{widetext}
		\begin{align}
			g_{00}=&-1 + \frac{2 w}{c^{2}} -  \frac{2 w^{2}}{c^{4}} + \Ol(c^{-5}) \nonumber\\
			g_{0i}=&- \frac{4 w_{i}}{c^{3}}  + \Ol(c^{-4}) \nonumber\\
g_{ij}=& \delta_{ij} \left(1 + \frac{2 w}{c^{2}} \right) + \Ol(c^{-3}) \, ,\label{eq:metric}
		\end{align}
		\end{widetext}
where $w$ and $w_{i}$ are respectively a scalar gravitational potential and a vector potential, $c$ being the speed of light.
The harmonic gauge conditions then imply that the potentials $w$ and $w_i$ satisfy the following equations:
\bea
\left(-\frac{1}{c^{2}} \frac{\partial^{2}}{\partial t^{2}}+\nabla^{2}\right) w=-4 \pi G \sigma+\Ol\left(c^{-4}\right), \label{eq:poissongeneral relativity} \\
\nabla^{2} w^{i}=-4 \pi G \sigma^{i}+\Ol\left(c^{-2}\right),\label{eq:poissongeneral relativityi}
\eea
where $\sigma$ and $\sigma^i$ are the gravitational mass and mass current defined upon the stress-energy tensor:
\begin{equation}
\sigma=\frac{1}{c^{2}}\left(T^{00}+T^{s s}\right), \quad \sigma^{i}=\frac{1}{c} T^{0 i}. \label{eq:defsigma}
\end{equation}
In this definition, the gravitational perturbations induced by other bodies in the vicinity of the solar system (stars, galaxies, dark matter, dark energy) are ignored. The solar system is considered as an isolated system---which is possible in general relativity, thanks to the equivalence principle and the resulting ``effacement of internal degrees of freedom in the global problem and of the external world in the local system'' \citep{damour:1989bk,klioner:2000pd}, but not in general for alternative theories of gravity---e.g. not in MOND \citep{10.1111/j.1365-2966.2009.15302.x,Milgrom:2014}, see Sec. \ref{sec:mond}.
For a solar system composed by {non-rotating} point-mass objects, the previous barycentric potentials are $w=\sum_{A} w_{A}$ and  $w^{i}=\sum_{A} w^{i}_{A}$ with

\begin{widetext}
		\begin{align}
			w_{A} =& \frac{\mu_A}{r_{A}} \left[ 1 + 
			2\frac{v_A^2}{c^2} - \frac{1}{c^2}\sum_{B\neq A} \frac{\mu_B}{r_{BA}}-\frac{1}{c^2} \left( \frac{( \bm r_A. \bm v_A)^{2}}{r^{2}_A} + \bm r_A. \bm a_A \right)\right],  \nonumber\\
			w^{i}_{A} =& \frac{\mu_A}{r_{A}} v^{i}_{A}\, ,\label{eq:metric2}
		\end{align}
		\end{widetext}
where $\mu_A:=G m_A$ is the gravitational parameter of the body $A$, $\bm r_{AT}$ is the relative position of body $T$ with respect to $A$, $r_{AT}=\lvert \bm r_{AT}\rvert $ and $\bm v_A$ is the coordinate velocity of body $A$ while $\bm a_A$ is its coordinate acceleration in the BCRS. 

The same type of framework can be defined for a reference system centered on the Earth  center of mass and leads to the definition of the Geocentric Celestial Reference System (GCRS). The GCRS is suitable {in practice} for the modeling of processes in the vicinity of the Earth{, whereas} the metrics of Eqs. (\ref{eq:metric}) and (\ref{eq:metric2}) will be used for modeling the light propagation and motion of celestial objects in the solar system in the BCRS.

The coordinate transformations between the BCRS and the GCRS involve a complicated set of functions that are defined in the resolution B1.3 of the IAU2000 resolution \citep{Soffel2003}.

\subsubsection{Time-scales in the solar system}
\label{sec:timescale}

\begin{figure}
\centering
\includegraphics[scale=0.5]{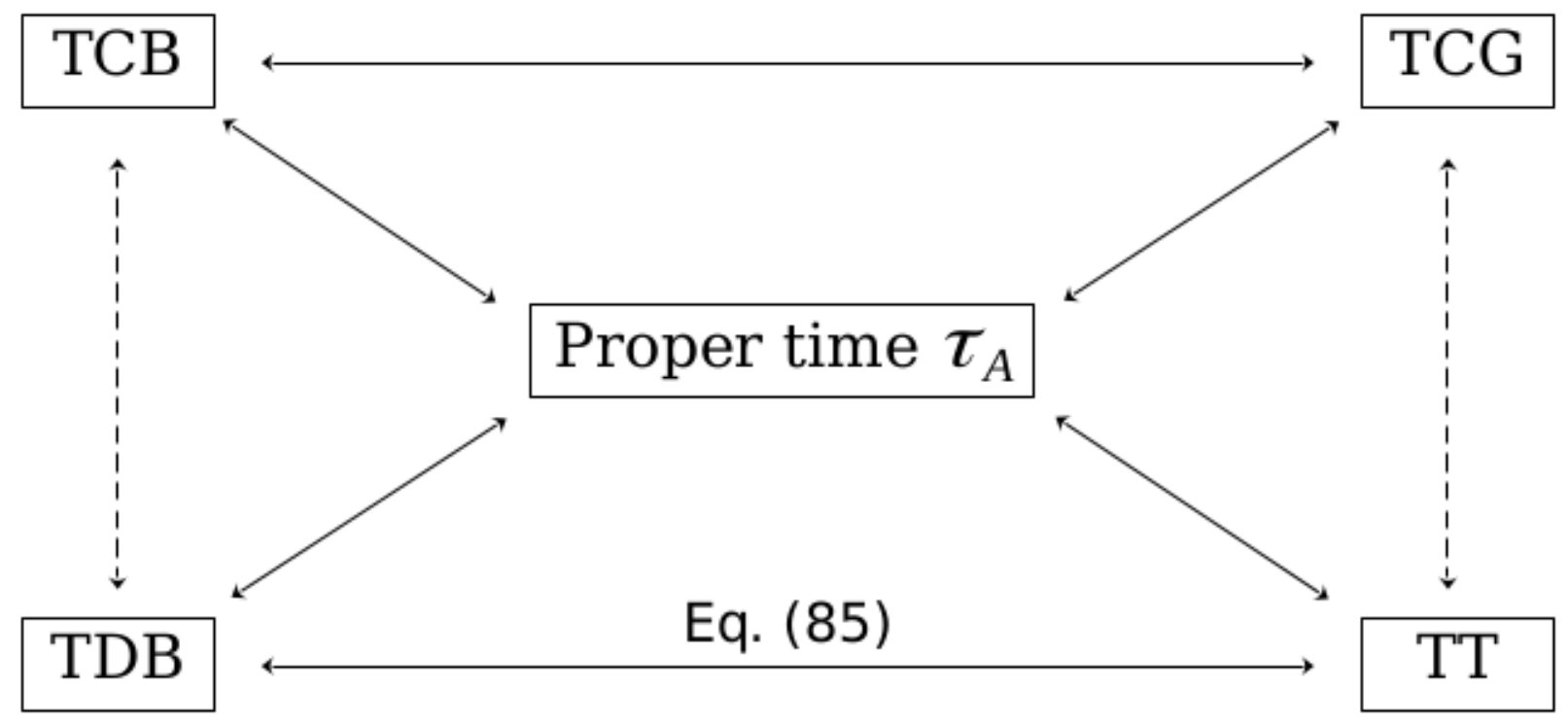}\\
\caption{Various relativistic time-scales and their relations. Each of the coordinate time-scales TCB, TCG, TT and TDB can be related to the proper time $\tau_A$ of an observer $A$, provided that the trajectory of the observer in the BCRS and/or GCRS is known \citep{2010ITN....36....1P}. Dashed lines represent transformations with fixed rates, whereas full lines represent transformations that depend on the metric potentials, following to the IAU recommendations \citep{Soffel2003}.}
\label{fig:times}
\end{figure}

The time-scales in the BCRS and GCRS are denoted by TCB and TCG, respectively \citep{Soffel2003}.
The relation between TCB and TCG are given in \cite{Soffel2003}. 
At the Earth level, the Terrestrial Time (TT) has been defined in order to remain close to the realized atomic time (TAI). It differs from the TCG by a constant rate \citep{Soffel2003}.

Likewise, at the level of the solar system barycenter, the TDB is defined as a linear transformation of the TCB. The relations between the various time scales is shown in Fig. \ref{fig:times}.

The difference between TT and TDB is produced by  planetary ephemerides, {by} integrating {the following equation together with the equations of motion \citep{2008A&A...478..951K,2009AA...507.1675F}}
\begin{widetext}
		\begin{align}
\frac{d(TT-TDB)}{d(TDB)} = & (L_{B}+ \frac{1}{c_{2}}a)(1+L_{B}-L_{G}) - L_{G}+ \frac{1}{c_{4}}b,\label{eq_time1}
		\end{align}
		\end{widetext}
with $L_{B}$ and $L_{G}$ are defining constants for TDB relatively to TCB  and TT relatively to TCG, respectively (see i.e. \cite{2008A&A...478..951K, 2010ITN....36....1P}  for the full definition) and where
\begin{widetext}
		\begin{align}
	a =& -\frac{1}{2} v_T ^{2} - \sum_{A\neq T} \frac{\mu_A}{r_{AT}} \, \nonumber \\
b =& -\frac{1}{8} v_T ^{4} +  \frac{1}{2} \left[ \sum_{A\neq T} \frac{\mu_A}{r_{AT}} \right]^{2} + \sum_{A\neq T} \frac{\mu_A}{r_{AT}} \Bigg\{ 4\bm v_T.\bm v_A - \frac{3}{2} v_T^{2} - 2 v_A^{2}  \nonumber \\
 & + \frac{1}{2}\bm a_A.\bm r_{AT} + \frac{1}{2} \left(\frac{\bm v_A.\bm r_{AT}}{r_{AT}} \right)^{2} + \sum_{B\neq A} \frac{\mu_B}{r_{BA}} \Bigg\},\label{eq_time2}
		\end{align}
		\end{widetext}
following the notations of Eq. (\ref{eq:metric2}).
In \cite{2021AJ....161..105P},  terms related to  the perturbations induced by the oblateness of the Sun were also added in the DE440 (TT-TDB) computation.
\\
Finally, planetary observations are also related to time. Following the IERS conventions \citep{2010ITN....36....1P}, those observations obtained on the ground are given in UTC, related to TDB by TT and TAI.
The observations obtained in other planetary systems (i.e. positions of a spacecraft orbiting Mercury) are, up to now, given also in UTC as the differences between the coordinate time defined for the corresponding planetary system (in the previous example, Mercury) and TDB or TT, are not significant at the present day accuracy. 
However, with missions such as Bepi-Colombo, it will be necessary to account for the local gravitational potential in the definition of the observational time scales (i.e \cite{2013PhRvD..87b4020T, Nelson2006RelativisticTF, 2002PhRvD..66h2001M}).
In conclusion, it is important to keep in mind that the various measured proper times are converted beforehands into coordinate times---according to the laws of general relativity.
%


\subsubsection{Inertial Frame}
\label{sec:ICRF}
The coordinate frame of the planetary and lunar ephemerides is linked to the International Celestial Reference System (ICRS) by its current realization  achieved by VLBI measurements of the positions of extragalactic radio sources (i.e., quasars) defined in the ICRF \citep{1998HiA....11..310S}. The planetary orbits are tied to ICRF because the observations used for their adjustment have been obtained in the ICRF. For the inner planets, the VLBI observations of Venus and Mars-orbiting missions give a tie with an accuracy better than the milliarcsecond \citep{fienga2011cm,folkner2014in}. For the outer planets, the link is maintained with the same level of uncertainty thanks the VLBI observations of  Juno and Cassini \citep{2019AAS...23330201J, fienga2011cm, folkner2014in} orbiting Jupiter and Saturn, respectively. 
Some other methods have been investigated for enhancing the link between planetary ephemerides and ICRF. In particular, one can notice the use of Lunar Laser Ranging observations \citep{2020JGeod..94....5P} and the use of GAIA DR2 asteroid positions \citep{2022CeMDA.134...32D}. The tie between the planetary ephemerides frame and ICRF is confirmed to be sub-mas level accuracy in both approaches.

\subsubsection{Definition of the solar system barycenter (SSB) }
\label{sec:SSB}

The definition of the solar system barycenter at the origin of the time of integration is based on the hypothesis of the conservation of the energy of the dynamical system composed by the planets, the Moon and the asteroids (including Trans-Neptunian Objects) in the solar system. The position $R$ and the velocity $V$ of the center of mass are then invariant. {Using} the notation of DE430 \citep{2014IPNPR.196C...1F}, it comes
\bea
\begin{cases}
    R = \frac{\sum_{i} \mu^*_i \bm r_i}{\sum_{i} \mu^*_i} = 0 \\
    V = \frac{dR}{dt}= 0 \\
\end{cases}
\label{eq:SSB0}
\eea
where  $\bm r_i$ and $\bm v_i$ being respectively its barycentric position and velocity vectors of the planet $i$. $\mu^*$  is given by
\begin{equation}
\begin{cases}
    \mu^*_i = \mu_i \left( 1 + \frac{v^2_i}{2c^2} - \frac{1}{2c^2} \sum_{j \ne i} \frac{\mu_j}{r_{ij}} \right) \\
    \\
    \dot{\mu}^*_i =  \frac{\mu_i}{2c^2} \left( \sum_{j \ne i} \mu_j \frac{(\bm r_j - \bm r_i).(\bm v_j + \bm v_i)}{r^{3}_{ij}} \right) \\
\end{cases}
\label{eq:SSB2}
\end{equation}
with $\mu_i$ the product of the gravitational constant G with the inertial mass of the body $i$. The term $\dot{\mu}^*_i$ is not included in DE430 nor DE440 ephemerides \citep{folkner:2014in,2021AJ....161..105P} but is accounted for in the INPOP ephemerides \citep{fienga:2008aa}.
The positions and velocities of the SSB are then obtained by integration of the following equations and up to the c$^{-2}$ order, 
\bea
\begin{cases}
    \sum_{i} \mu^*_i \bm r_i = 0 \\
    \sum_{i} \mu^*_i \bm v_i + \dot{\mu}^*_i \bm r_i = 0 \\
\end{cases}
\label{eq:SSB}
\eea
In the INPOP formalism \citep{fienga:2008aa}, Eqs. (\ref{eq:SSB}) and (\ref{eq:SSB2}) are solved only at the initial step of the planetary integration and a constant vector in positions and velocities is subtracted to all the body positions and velocities for having $R$ and $V$ to 0. Once the frame is centered on the SSB defined by Eq. (\ref{eq:SSB}) at J2000 (inital date of integration for INPOP epehemerides), the equations of motion of the solar system bodies and of the Sun are integrated in this fixed frame. The method described here is used by the INPOP planetary ephemerides since INPOP08 \citep{fienga:2008aa}. In JPL DE430 ephemerides \citep{folkner2014in}, the Sun initial coordinates are set up such as $R$ and $V$ are 0, the equations of motion of the Sun, the Moon, planets, and asteroids being then integrated in the fixed frame.  In the former JPL DE versions, such as DE421 \citep{DE421}, the Sun coordinates were set up at each step of integration for maintaining $R$ and $V$ to 0 all along the integration process. A description of the successive SSB implementations for older JPL versions can be found in \cite{folkner2014in} and discussions concerning possible uncertainties are presented in \cite{fienga:2008aa} and \cite{folkner2014in}. It is interesting to note that, for the tests of the Equivalence Principle (sect. \ref{sec:ep}), $\mu_i$ of Eq. (\ref{eq:SSB}) and (\ref{eq:SSB2}) will {still} correspond to the product of G with the inertial mass of the body. It will then differ from the {gravitational} mass used for computing the planetary acceleration as in Eq. (\ref{eq:accgeneral relativity}).


\subsection{The dynamical model}
\label{sec:dynam}
\subsubsection{Point-mass interactions}

In the context of  \cite{Soffel2003} and of the metric defined in Eq. (\ref{eq:metric}) and (\ref{eq:metric2}), one can then, in the mass-monopole approximation, write the equations of motion of bodies as well as the conservation laws satisfied by the SSB as given in \cite{1995PhRvD..52.4455D}.
In BCRS, the equation of motion for the point-mass interaction is given by

\begin{widetext}
		\begin{align}
			\bm {a}_T=&-\sum_{A\neq T} \frac{\mu_A}{r_{AT}^3}\bm r_{AT}  \nonumber\\
			&-\sum_{A\neq T} \frac{\mu_A}{r_{AT}^3c^2}\bm r_{AT}\Bigg\{v_T^2+2v_A^2-4\bm v_A.\bm v_T  -\frac{3}{2}\left(\frac{\bm r_{AT}.\bm v_A}{r_{AT}}\right)^2\nonumber \\
			&\hspace{3cm}-\frac{1}{2}\bm r_{AT}.\bm a_A -4\sum_{B\neq T}\frac{\mu_B}{r_{TB}}-\sum_{B\neq A}\frac{\mu_B}{r_{AB}}\Bigg\} \nonumber \\
			&+\sum_{A\neq T}\frac{\mu_A}{c^2r_{AT}^3}\left[4\bm r_{AT}.\bm v_T-3\bm r_{AT}.\bm v_A\right](\bm v_T-\bm v_A)  \nonumber \\
			&+ \frac{7}{2}\sum_{A\neq T} \frac{\mu_A}{c^2r_{AT}}\bm a_A \, ,\label{eq:accgeneral relativity}
		\end{align}
		\end{widetext}

This equation is known as  the Einstein-Infeld-Hoffmann-Droste-Lorentz (EIHDL) equation of motion, and it is numerically integrated for obtaining the modern planetary ephemerides in general relativity at the first post-Newtonian c$^{-2}$ level. Additional accelerations induced by the oblateness of the Sun (sect \ref{sec:j2}) or frame dragging effects (sect \ref{sec:LT}) are also accounted for in order to describe the motion of the planets at the level of accuracy required by the modern space mission observations. Other multipole moments than the Sun  quandrupole moment are negligeable at the current level of accuracy.

\subsubsection{Asteroid perturbations}
\label{sec:aste}

Differences in the number of point-mass perturbations exist between planetary ephemerides. Table \ref{tab:param} summarizes them {by} splitting the case of the Main-Belt asteroids (MBA) {on the} one hand, and of the Trans-Neptunian objetcs (TNO) {on} the other hand. The problem of the modelisation of the MBA perturbations on the inner planet orbits has been addressed since DE405 \citep{Standish2001}. The issue is that there is a considerable number of objects (at least 250,000), for which the masses are unknown, that can potentially interact gravitationally with the inner planets (mainly Mars).  With DE405 \citep{DE405}, 5 asteroids (Ceres, Pallas, Vesta , Iris and Bamberga) have been identified as the main perturbers associated with an averaging of the effect induced by the rest of the MBA based on taxonomic classifications. But this modelisation shows its limits with the improvement of the accuracy of the Mars-orbiter tracking data \citep{2002A&A...384..322S}. \cite{Williams1984} proposed a list of 343 individual objects among the MBA gathering the most perturbing asteroids to consider for the construction of accurate planetary ephemerides. That list has been confirmed by \cite{2013Icar..222..243K} and since DE430  this is the \cite{Williams1984} objects that are taken into account in DE and INPOP planetary ephemerides, with their masses individually fitted to observations together with the initial conditions of planet orbits, the mass of the Sun and the Earth-Moon mass ratio \citep{folkner2014in, fienga2019inpop}. A detailed description of the complexity of such an adjustment is out of the scope of this review but more comments on the accuracy obtained for these mass determinations can be found in \cite{2013Icar..222..243K} and  \cite{2020MNRAS.492..589F}.  In order to overcome this problem, but also to reduce the time cost of integrating an important number of objects, it has been proposed to substitute or to complement the MBA individual point-mass contribution by a global ring potential. 

The first ring modelisation was proposed by \cite{2002Icar..158...98K} as a static ring fixed at 2.8 AU from the SSB. An updated model maintaining the total linear and angular momenta of the system has been implemented in \cite{2009AA...507.1675F} and \cite{2010A&A...514A..96K} as a way to account for additional MBA contribution associated with a shorter list of individual point-mass perturbers. This approach has not be retained for MBA in DE \citep{folkner2014in} and INPOP (since INPOP13c \citep{INPOP13c}) ephemerides but kept in EPM ones associated with individual mass and density determinations for 3 classes of MBA \citep{2018CeMDA.130...57P}.

After the introduction of the Juno and Cassini tracking data, it became clear that the perturbations induced by the TNO have to be included in the model. In \cite{2018CeMDA.130...57P, 2020A&A...640A...7D} and \cite{2021AJ....161..105P}, individual point-mass perturbations of the most massive and binary Trans-Neptunian objects  have been added together with a circular ring for accounting for the average effect of the rest of the TNO. However as the TNO orbits are more eccentric than the MBA ones, \cite{fienga2022} had introduced, instead of a circular ring, 500 individual point-mass TNO perturbers with observed orbits\footnote{extracted from the astord database \citep{2022A&C....4100661M}} but of equal mass. This latest assumption simplifies the adjustment as only one parameter (the total mass spread over 500 objects) is used and fitted for characterising the TNO perturbations.

The full 1PN general relativity equation of motion is only taken into account for the first biggest MBA (Ceres, Pallas, Vesta , Iris and Bamberga). For the rest of the MBA and the TNO point-mass interactions, the perturbations on the planet orbits, and on the other asteroid orbits, are estimated at the Newtonian level only. This is done in order to reduce the computational cost, but it is consistent with the overall PN perturbative approach---given that those perturbations are already very small at the Newtonian level, such that their contributions at the post-Newtonian level are negligible.

\subsubsection{Lense-Thirring acceleration}
\label{sec:LT}
The Lense-Thirring (LT)  acceleration induced by the Sun rotation has been recently added in modern planetary ephemerides \citep{2018CeMDA.130...57P, 2021AJ....161..105P, inpop21a} . Its new introduction is justified by the accuracy reached by space missions, especially those close to the Sun and Mercury such as Messenger and Bepi-Colombo \citep{genova2018nc}. It also has been shown that it helps for disentangling the PPN parameters {for} their contribution to the planetary motions \citep{fienga2022}.

In 2011, \cite{2011Ap&SS.331..351I} has also shown that this effect contributes to about 10$\%$  of the dynamical acceleration induced by the shape of the Sun in General relativity (see sect. \ref{sec:j2}).  The acceleration induced by the Lense-Thirring effect generated by a central body (here the Sun) at the first post-Newtonian approximation  is given by
\begin{equation}
\bm{a}_{LT} = \frac{2 G S}{c^{2}r^{3}} \left[  \frac{3 \bm{k} . \bm{r}}{r^{2}} (\bm{r} \wedge \bm{v}) -  (\bm{k} \wedge \bm{v}) \right]
\label{eq:LT}
\end{equation}
where $G$ is the gravitational constant, $c$ the speed of light, $\bm{S}$ is the Sun angular momentum such as $\bm{S} = S \, \bm{k}$ where $\bm{k}$ is the direction of the Sun rotation pole defined according to the IAU right ascension and declination \citep{2018CeMDA.130...22A}, $\bm{r}$ and $\bm{v}$ are the position and velocity vectors of the planet relative to the Sun and $\gamma$ is the PPN parameter for the light deflection, equal to 1 in general relativity.
The Lense-Thirring effect hence depends on the Sun angular momentum, $\bm{S}$ that can be obtained by considering different models for the Sun rotation including both the rotation of the convective region well constrained by helioseismology and the rotation of the Sun core (see \cite{fienga2022} for discussion).

\subsubsection{Shape of the Sun}
\label{sec:j2}
Among the solar multipole moments, only the degree 2 order 0 term of the spherical harmonic decomposition of the Sun gravity field, J$_{2}^{\odot}$---which is identified to the oblateness of the Sun--- leads to a significant impact on the planetary ephemerides at the current level of accuracy.
The J$_{2}^{\odot}$ induces an acceleration $\bm{a}_{J_{2}^{\odot}}=({a}^x_{J_{2}^{\odot}},{a}^y_{J_{2}^{\odot}},{a}^z_{J_{2}^{\odot}})$ to be added to EIHDL Eq. (\ref{eq:accgeneral relativity}) such as \citep{2016arXiv161002156S,2021CosRe..59..324I} :

\begin{equation}
     \begin{cases}
{a}^x_{J_{2}^{\odot}} = & - \frac{3}{2} \mu_{\odot} J_{2}^{\odot}  R^2_{\odot}  x \frac{(x^2+y^2-4z^2)}{2 r^7},\\
{a}^y_{J_{2}^{\odot}} = & - \frac{3}{2} \mu_{\odot} J_{2}^{\odot}  R^2_{\odot}  y \frac{(x^2+y^2-4z^2)}{2 r^7},\\
{a}^z_{J_{2}^{\odot}}  = & - \frac{3}{2} \mu_{\odot} J_{2}^{\odot}  R^2_{\odot} z \frac{(3x^2+ 3y^2-2z^2)}{2 r^7}
\end{cases}
\label{eq:accJ2}
\end{equation}
where  $\mu_{\odot}$ and  R$_{\odot}$ are the gravitational parameter and the radius of the Sun, respectively, $r$ is the heliocentric distance of the planet, $(x,y,z)$ being its heliocentric coordinates.

It is common to see in the literature dealing with geodesy and geophysics the contribution of J$_{2}^{\odot}$ as time variations of the osculating elements of the perturbed orbit such as \citep{1966tsga.book.....K,2003ASSL..293.....B}:
\begin{equation}
     \begin{cases}
        \dot{a} = \, 0 ;  & \dot{e} = \, 0 ;   \dot{i} = \, 0  \\
        \dot{\omega} =  & \frac{3}{4}J_{2}^{\odot}  \frac{n}{(1-e^{2})^2}\frac{R^2_{\odot}}{a^2} (5\, \cos^2 i - 1)\\
        \dot{\Omega} =  & - \frac{3}{2}J_{2}^{\odot}  \frac{n}{(1-e^{2})^2}\frac{R^2_{\odot}}{a^2} \cos i \\
        \dot{M} =  &  n + \frac{3}{4}J_{2}^{\odot}  \frac{n}{(1-e^{2})^{3/2}}\frac{R^2_{\odot}}{a^2}(3\, \cos^2 i - 1)\\
     \end{cases}
     \label{eq:J2}
\end{equation}
with $a$, $e$, $n$ and $i$ are the semi-major axis, the eccentricity, the mean motion and the inclination of the planetary orbit impacted by the oblateness of the Sun. $M$ is the planet mean anomaly. We give these equations Eq. (\ref{eq:J2}) as a simple illustration of the type of contribution one may get from the Sun quadrupole. It is then clear that the major contribution of this quadrupole term is on the advance of nodes $ \dot{\Omega}$ and perihelia $\dot{\omega}$ of the planetary orbits. This system of equations is also somehow useful to apprehend the potential correlations that may appear between the Sun  quadrupole moment and the semi-major axis and eccentricity.

Fig \ref{fig:j2} shows $J_{2}^{\odot}$ values  deduced from planetary ephemerides  including or not the Lense-Thirring contribution. 
It is clear, from this Figure,  that the introduction of the Lense-Thirring acceleration in the model improves significantly the $J_{2}^{\odot}$ determinations. The values obtained with LT \citep{fienga2022, 2017AJ....153..121P}  are indeed closer to the ones issued from helioseismologic surveys \citep{1998MNRAS.297L..76P,Antia2008} than those obtained before the LT introduction \citep{fienga2019inpop,2011Ap&SS.331..351I,viswanathan2018dc,INPOP13c,folkner2014in,fienga2011cm,2009AA...507.1675F}. In general relativity, the difference in $J_{2}^{\odot}$ with or without LT is of about about 7 to 10 $\%$ .

\begin{figure}
\centering
\includegraphics[scale=0.5]{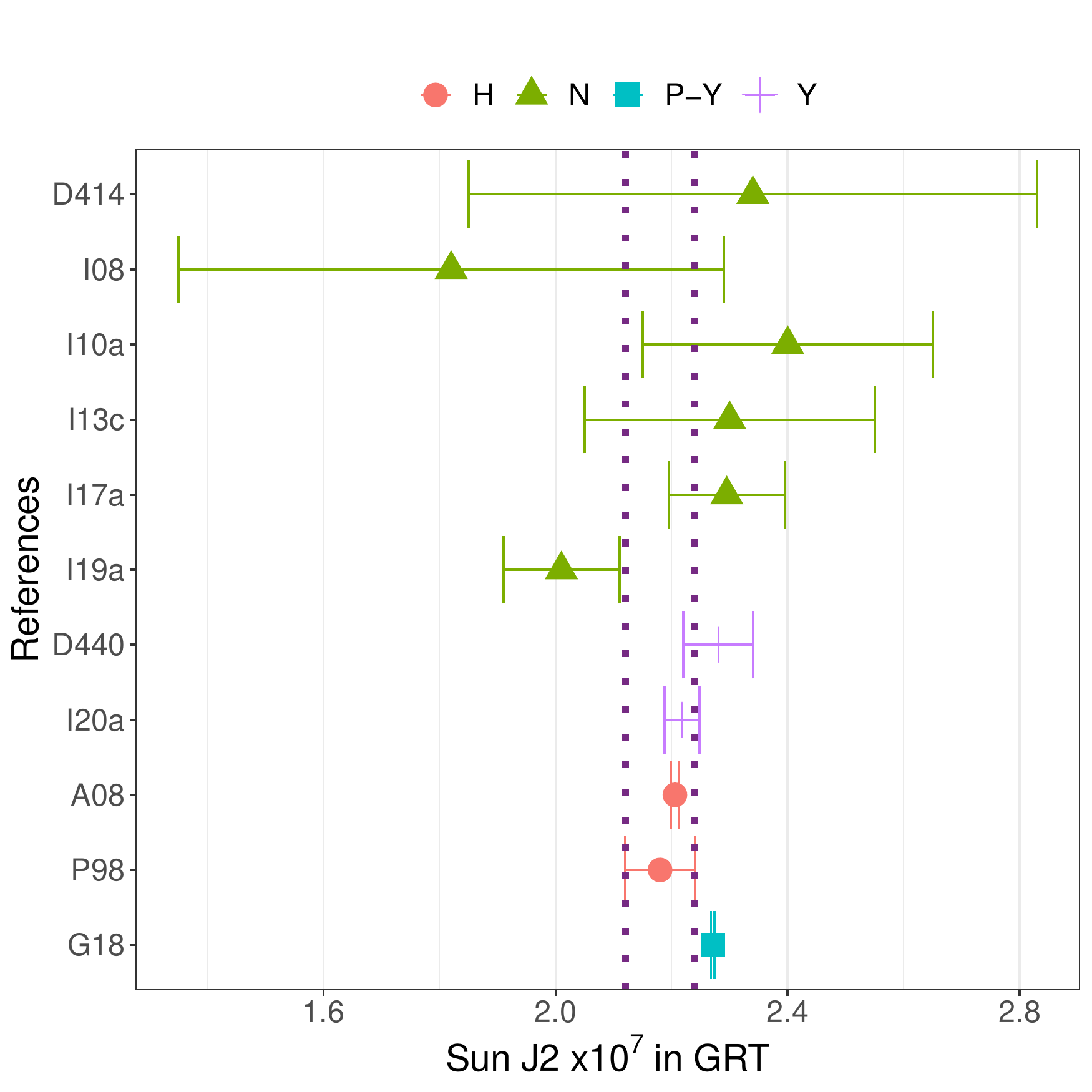}\\
\caption{Values of the Sun oblateness $J_{2}^{\odot}$ obtained with different ephemerides and other methods in the general relativity frame. H stands for values deduced from helioseismology, P points values obtained in a partial fit of planetary orbits ({i.e. }considering only a couple of planets {for the fit}, like Mercury and the Earth), N indicates values obtained before the introduction of the Lense-Thirring acceleration in the dynamical modeling and Y gives the values obtained after the introduction. The dotted lines give the limits of the less constraining helioseismologic value from \cite{1998MNRAS.297L..76P}. In y-axis, are given the references : G18 is \cite{genova2018nc}, P98 \cite{1998MNRAS.297L..76P}, A08 \cite{Antia2008}, I20a \cite{fienga2022}, D440 \cite{2017AJ....153..121P}, I19a \cite{fienga2019inpop}, I17a \cite{viswanathan2018dc}, I13c \cite{INPOP13c}, D414 \cite{2006Icar..182...23K}, I10a \cite{fienga2011cm} and I08 \cite{2009AA...507.1675F}. It is important to note that in \cite{genova2018nc}, only the orbit of Mercury is considered.}
\label{fig:j2}
\end{figure}


As it will be developed in sect. \ref{sec:PPN}, the introduction of gravity theories that are different but close to general relativity induces  modifications of the previous Eqs. (\ref{eq:accgeneral relativity}-\ref{eq:accJ2}). 
In this context,  the value of $J_{2}^{\odot}$ can also be determined with planetary ephemerides in the PPN framework, like given in Table \ref{tab:ppn} . But we will see that the value of $J_{2}^{\odot}$ can be strongly correlated to the value of non-general relativity parameters, and in particular the PPN parameter $\beta$ \citep{2002PhRvD..66h2001M}---as one can get an intuition from Eq. (\ref{eq:ppnom}).
In sect. \ref{sec:PPN}, we will see that the determination of the Sun oblateness when one includes PPN deviations plays a key role in the context of the global adjustment of planetary ephemerides .

\begin{table}
    \centering
       \caption{Detailed example of a planetary ephemeris (here INPOP21a) data sample extracted from \citep{inpop21a}. Column 1 gives the observed planet and information on the type of observations, Column 2 gives the type of data or the name of spacecraft used for producing the data. Columns 3 and 4 give the time interval and the {\it{\textup{a priori}}} uncertainties provided by space agencies or the navigation teams, respectively. In Column 2, $MRO$ for Mars Reconnaissance Orbiter and $MO$ for Mars Orbiter. Flybys stands for average normal points obtained during a flyby of a spacecraft with $U,C,P,V$ gives the normal points obtained after the flybys of  Ulysses, Cassini, Pionneer and Voyager respectively. $radar$ is for direct radar ranging on planetary surfaces, $VLBI$ stands for Delta-DOR observations and $Transit+CCD$ gives the optical angular observations obtained by transit, photographic plates or CCD. $H14$ stands for \cite{2014PhRvD..89j2002H}.}
   \begin{tabular}{l l l l}
    \hline
    Planet / Type [unit] & s/c or method &  Period & Averaged \\ 
    & &  & Accuracy \\
    \hline
   {\bf{Mercury}} && & \\
Direct range [m]&  Surface &  1971.29 : 1997.60 &   900 \\
Radio Science range [m]& Messenger &   2011.23 : 2014.26 &     5   \\
Navigation range [m]& Mariner &  1974.24 : 1976.21 &    100   \\
{\bf{Venus}} & && \\
VLBI [mas]& Magellan, Venus Express  &  1990.70 : 2013.14 &   2.0  \\
Direct range [m]& Surface &  1965.96 : 1990.07 &   1400    \\
Navigation range [m]& Venus Express &   2006.32 : 2011.45 & 7.0   \\
{\bf{Mars}} & && \\
VLBI [mas]& MGS, MRO &  1989.13 : 2013.86 &   0.3   \\
Navigation range [m]& Mars Express &   2005.17 : 2019.37 &  2.0  \\
Radio Science range [m]& MGS &  1999.31 : 2006.70 &   2.0   \\
Radio Science range [m]&MRO/MO &  2002.14 : 2014.00 &  1.2  \\
Navigation range [m] & Viking&   1976.55 : 1982.87 &    20.0 \\ 
{\bf{Jupiter }}& && \\
VLBI [mas]& Galileo &  1996.54 : 1997.94 & 11  \\ 
& Juno &  2016:2020 & 0.5 \\
Optical RA/Dec [arcsec]& Transit+CCD &  1924.34 : 2008.49 &   0.3  \\
Flyby RA/Dec [mas]& U,C, P, V &  1974.92 : 2001.00 &  8.0 \\
Flyby range [m]& U,C, P, V  &   1974.92 : 2001.00 &    2000 \\
Radio Science range [m]& Juno &   2016.65 : 2020.56 &  20   \\
{\bf{Saturn}} & &&   \\
Optical RA/Dec [arcsec]& Transit+CCD &   1924.22 : 2008.34 &    0.3 \\
VLBI RA/Dec [mas]& Cassini &   2004.69 : 2017.9 &   0.6 \\
JPL H14 [m]& Cassini &    2004.41 : 2014.38 &  25.0  \\
Navigation [m] & Cassini &   2006.01 : 2009.83 & 6.0 \\
Radio Science : Titan Flybys  [m] & Cassini  &  2006.01 : 2016.61 & 15.0\\ 
Radio Science : Grand Finale)  [m]& Cassini  &  2017.35 : 2017.55&  1.0 \\
{\bf{Uranus}} && & \\
Optical RA/Dec [arcsec]& Transit+CCD+photo &   1924.62 : 2013.75 &   0.25  \\
Flyby RA/Dec [mas]& Voyager &  1986.07 &   50 \\
Flyby range [m]& Voyager &   1986.07  & 50 \\ 
{\bf{Neptune}} & && \\
Optical RA/Dec [arcsec]& Transit+CCD+photo  &   1924.04 : 2007.88 &   0.3  \\
Flyby RA/Dec [mas]& Voyager &   1989.65  &  15.0 \\
Flyby range [m]& Voyager &   1989.65  &    2 \\
\hline
    \end{tabular}
    \label{tab:res}
\end{table}

\subsection{Planetary Datasets }
\label{sec:obs}

The modern planetary ephemerides are characterised by the role of planet positions deduced from spacecraft missions either by navigation tracking or by the use of radio science data. In both cases, it is the distance between a station on the ground and a spacecraft orbiting of flying over a planet that is used for the construction of planetary ephemerides. The angular position between the same spacecraft and an extra-galactic source (generally a reference beacon for the construction of the inertial frame, ICRF) is also of great importance for the adjustment of the planetary ephemerides as it allows the tie of the planetary planes to the ICRF (see sect. \ref{sec:ICRF}). Optical meridian transit and CCD angular positions are mainly kept for the constraint on the outer planet orbits. 
A full description of the different data sets used for the construction of the planetary ephemerides can be found in \cite{folkner2014in} and \cite{2009AA...507.1675F} and regular updates are presented for every new releases such as for DE440 \citep{2021AJ....161..105P} or INPOP19a \citep{2020A&A...640A...7D}. Table \ref{tab:res} gives the different data sets used for the construction of INPOP21a \citep{inpop21a}, counting a total of about 200,000 observations, this number {of course} varying from one ephemeris to {another}. Very similar datasets have been used by DE440 \citep{2021AJ....161..105P}.

\subsubsection{Time delay of the propagation of light}
\label{sec:shapiro}

The propagation of light is affected by the curvature of spacetime. Although the trajectory of light is bent in gravitational potentials, this bending is negligible for radio science in the solar system\footnote{Because it is a second order effect in the post-Newtonian development of the propagation time.}, and the main effect is known as the Shapiro delay: the propagation time between two points is not simply the Euclidean BCRS distance over the speed of light between an emission point $e$ and a reception point $r$, but it also involves a delay in general relativity that depends on the gravitational potentials as follows \citep{Moyer2000}
	\begin{equation}
			c(t_r-t_e)_{GRT}=R+\sum_A 2\frac{\mu_A}{c^2}\ln\frac{\bm{n}\cdot\bm{r}_{rA}+r_{rA}+\frac{4\mu_A}{c^2}}{\bm{n}\cdot\bm{r}_{eA}+r_{eA}+\frac{4\mu_A}{c^2}} \, ,
\label{eq:shapirogeneral relativity}
		\end{equation}
where the  $R$ is the Euclidean coordinate distance, $\bm{n}=(\bm{r}_r-\bm{r}_e)/\|\bm{r}_r-\bm{r}_e\|$, $\bm{r}_{iA}=\bm{r}_i-\bm{r}_A$ and $r_{iA}=\|\bm{r}_{iA}\|$ with $i=e$ or $r$. 
		Note that Eq. (\ref{eq:shapirogeneral relativity}) slightly differs from the actual Shapiro equation through the $c^{-2}$ terms in the logarithm function. It has lately been realized that this version of the Shapiro delay---rather than the usual equation without the additional $c^{-2}$ terms in the logarithm---allows to account for the so-called second order time propagation enhanced terms, which can become numerically significant for particular conjunction situations despite formally being next-to-leading-order (or $c^{-4}$) terms \citep{ashby:2010cq,linet:2016pr}. In practice, the difference between the two versions of the Shapiro delay is irrelevant for current planetary ephemerides.

However, it has been realized during the last decade that for close solar conjunctions, additional enhanced $c^{-4}$ terms may no longer be negligeable, in particular for BepiColombo  MORE (Mercury Orbiter radio science Experiment) \citep{hees:2014pr,cappuccio:2021cq} measurements. 
As a consequence, for close conjunction events, one might have to go beyond Eq. (\ref{eq:shapirogeneral relativity}) and use a full second order time propagation time formula. This will be challenging in many different ways ---see Appendix \ref{sec:ShapiroO2}. 
For planetary ephemeris construction, Bepi-Colombo MORE experiment should provide range measurements up to close to 5$^{\circ}$ to the Sun, {that is,} close to conjunction \citep{fienga2022}. For such close-to-the-Sun observations, a specific analysis will have to be done in order to account for the neglected terms of Eq. (\ref{eq:shapirogeneral relativity}), but in the framework defined by the IAU (harmonic gauge). 



In any case, when one deals with an alternative theory of gravity, one has to take care of the consistency between the different levels of construction of the ephemeris: modifying the {Shapiro delay} or the equation of motion for massive bodies is not without consequences on the other.

\subsubsection{Radio science and navigation data}

This type of data represent more than 65$\%$ of the full data sample used for the construction of the planetary ephemerides.  It gathers measurements of distances in terms of time delays obtained during orbital or flyby phases of space missions, either during the regular navigation process{, or} during specific scientific sessions dedicated to radio science experiments. 
In this {latter} case, a detailed assessment of the capabilities of the radio transpondeur  used for the re-emission of the captured signal is {needed} for the analysis (see i.e. \cite{2020ITAES..56.4984C}). The measurement  is the time differences between the time of emission of the signal by a ground-based antenna, the time of capture by the {transponder} on board of a spacecraft, the time of re-emission by this {transponder} after amplification and the time of reception of the transmitted signal by the same or a different ground-based antenna. 
Using a known orbit of the spacecraft around or near the planet, it is then possible to deduce the planetary range measurement, by accounting for the distance between the spacecraft and the centre of mass of the planet estimated during the s/c orbit determination. The accuracy of the range constraints deduced from the radio science and navigation data is then strongly dependant on the uncertainties of the spacecraft orbit determination.

The distribution of the data per missions is given in Fig. \ref{fig:dataSC}. In numbers, the Mars orbiter observations dominate, followed by the Venus tracking data. In terms of accuracy, the Mars data also lead the fit together with the less numerous but important tracking data of the Juno mission orbiting Jupiter as well as the Cassini radio science and tracking data for Saturn. 
The predominance in numbers and accuracy of the Mars observations on the planetary ephemerides data sample explain why some tests of general relativity have been said {to be} driven by the Mars orbit \citep{will:2018cq}, whereas \cite{bernus2019} stressed also the importance of the outer planet radio science experiments.

\begin{figure}
\centering
\includegraphics[scale=0.35]{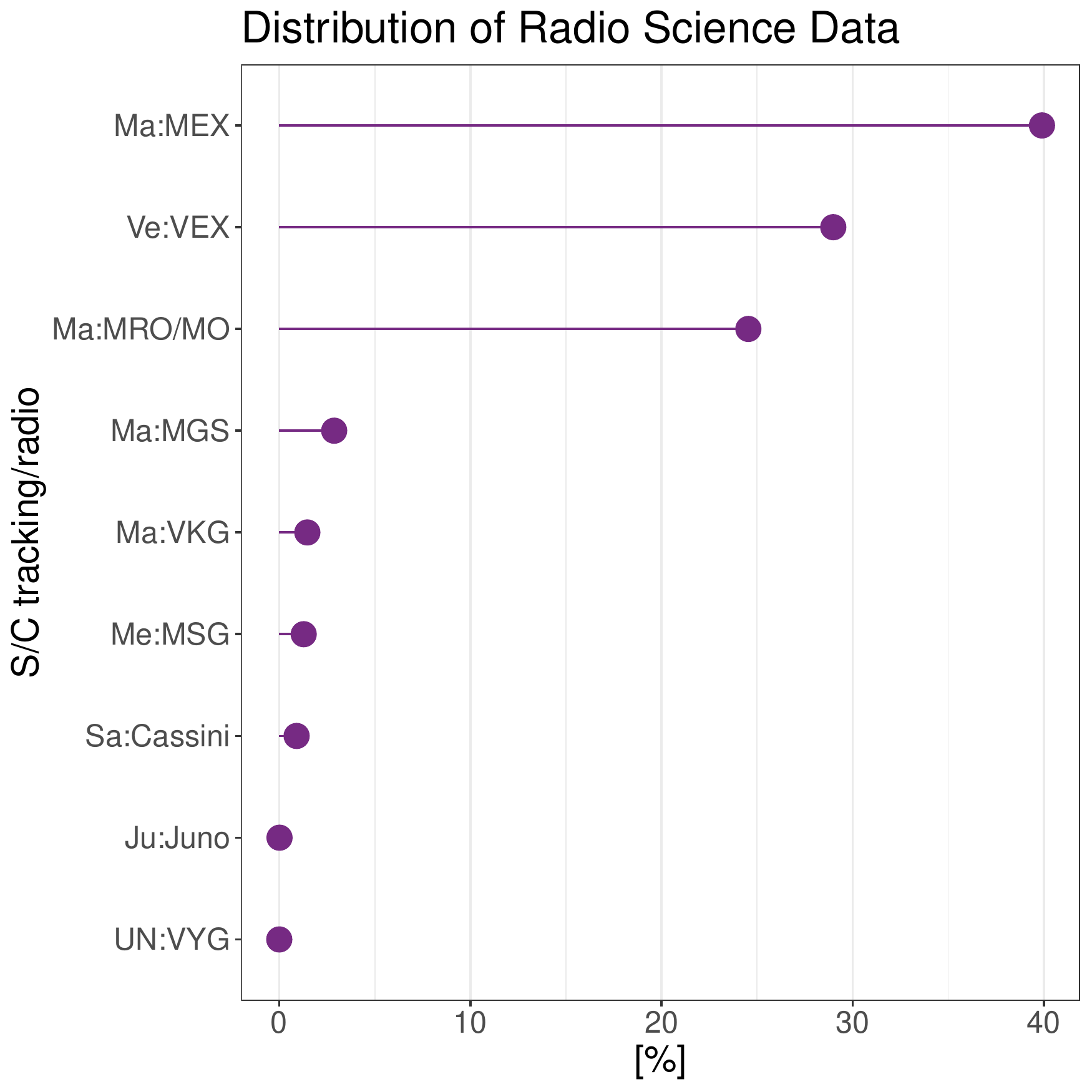}\includegraphics[scale=0.35]{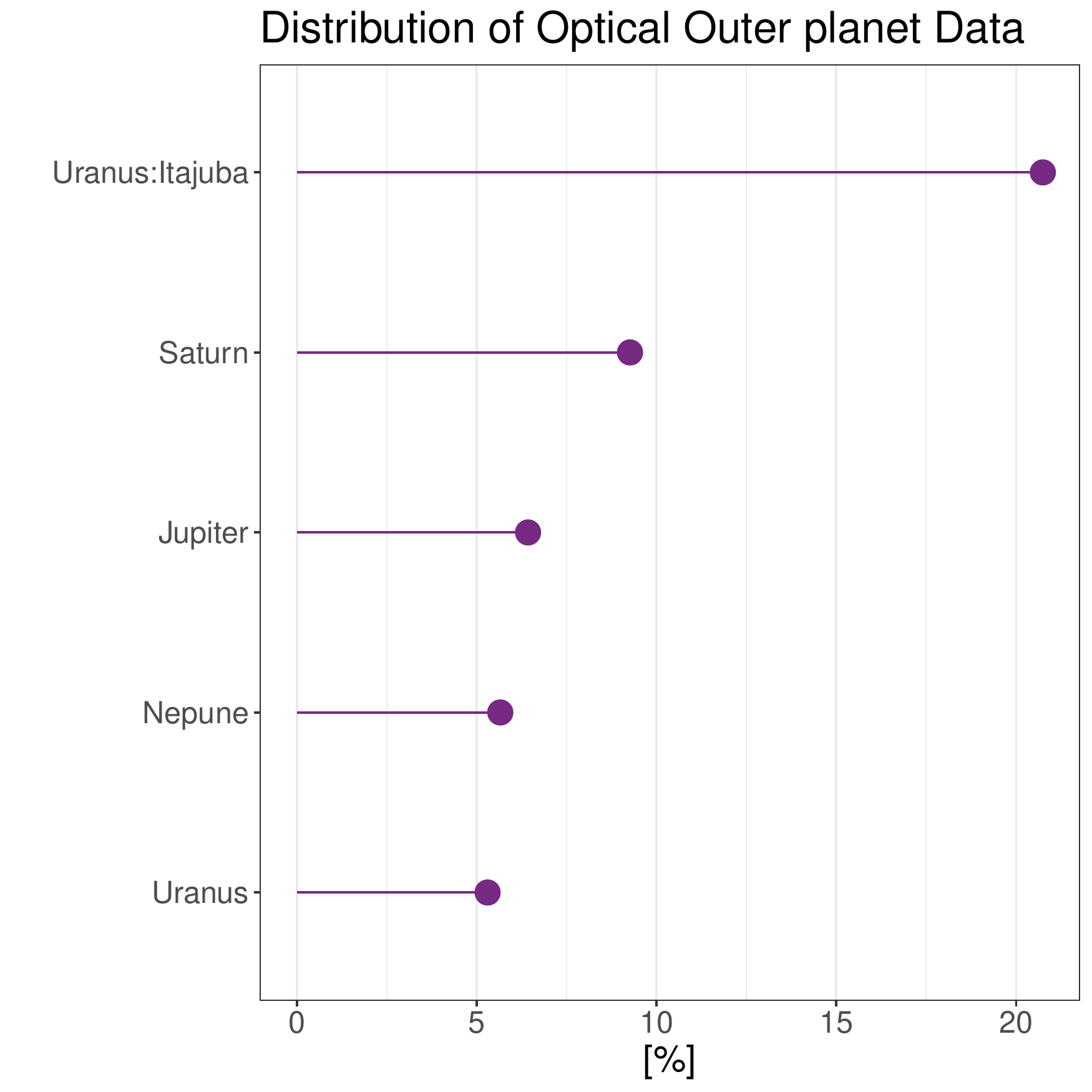}
\caption{Example of distribution in percentages of the radio science and tracking data and the optical observations. On the left-hand side, data samples are identified by planets (Ma stand for Mars, Me for Mercury, Ve for Venus, Ju for Jupiter, Sa Saturn and UN for Uranus and Neptune) and by missions. On the right-hand side, the data sample are identified according to the observed planet. A specific tag is made on the set of astrometric positions obtained by \cite{2022P&SS..21005376C}. The presented datasets correspond to the one of INPOP21a (see Table \ref{tab:res}). Roughly similar distribution is present for DE and EPM ephemerides.}
\label{fig:dataSC}
\end{figure}

Furthermore, in term of gravitational framework, these observations are also given in the framework used for studying the motion of the spacecraft around the planet. 
This framework is usually general relativity and discussions have been held regarding the consistency of testing alternative theories of gravity with observational constraints obtained in general relativity (i.e. \cite{2014A&A...561A.115V}).  
The most simple way to solve this problem is to operate the spacecraft orbit determination (OD) in the same framework as the one of the alternative theories to be tested. 
But as the s/c navigation is regularly interrupted{, even during orbital phases,} by engine boosts or attitude corrections, introducing significant changes in the orbit, OD is usually done over a short duration of time (from few hours to few days), in between these manoeuvres. 
Most of the time, up to now, alternative theories of gravity start to significantly differ from general relativity over longer time intervals.  
But it is clear that with the increase of the accuracy in the s/c orbit tracking, the modified general relativity {that one wants to test should also be implemented during} the OD process. See for example \cite{2020CQGra..37i5007D}.

In parallel to the radar tracking, navigation teams also acquire positions of the spacecraft relative to extra-galactic sources {by} using facilities also dedicated to VLBI observations. 
This type of observations, called Delta -DOR, represent less than 1 $\%$ of the full data sample  but they are crucial for linking the ephemeris reference frame to the ICRF (see sect. \ref{sec:ICRF}). The first were obtained in 1981 \citep{1981BAAS...13..555D} with the Viking Mars lander.
Technical description related to the production of this type of observations can be found in \cite{1994A&A...287..279F} and  \cite{Moyer2000}.  
The most recent planetary ephemerides account for a total of about 300 VLBI positions, mainly extracted from the tracking data of Mars, Venus, Jupiter and Saturn orbiters with  accuracies about less than 1 milliarcsecond of degree (mas) \citep{2011AJ....141...29J, 2019AAS...23330201J, 2021AJ....161..105P}. 


\subsubsection{Angular ground-based observations}

Finally, the observational data base is completed by the optical observations of outer planets. This set constitutes about 35$\%$ of the full data sample, and, despite their uncertainties, are crucial for Uranus and Neptune. 
For these two planets, only 1 flyby per planet has been acquired during the Voyager missions, leading to a weak constraint for several decade-long orbits.  
Direct transit observations of planets or satellites are part of the optical data samples as well as astrometric positions deduced from CCD observations and photographic plates. 
Both planet and satellite positions are included. In the case of satellite astrometry, the position of the planetary {barycenter} is deduced using satellite ephemerides (i.e. \cite{2022P&SS..21005376C}). 
The long-term (several decades) behaviour of outer planet orbits is then driven by the planet barycenter equation of motion as given in Eq. (\ref{eq_eih}), but also by the satellite dynamics.  
Additionally, depending on the reference frames used for the data publications, different algorithms are applied for transforming the astrometric positions  (for example given in FK3, FK4 or FK5) to ICRF. 
Such transformations guarantee, at the level of accuracy of the optical data (about 50 to 100 mas), the link between the INPOP outer planet frame and ICRF.  VLBI observations enforce this link at the VLBI accuracy which means a factor of at least 100 compared to the optical tie. On Fig. \ref{fig:dataSC} are given an example (for INPOP21a) of the percentages of optical observations per planet. A specific tag is made on the data sample provided by \cite{2022P&SS..21005376C} who did a new reduction of astrometric long-term observations of the Uranian main satellites obtained at the Pico dos Dias from 1982 to 2011, using the Gaia EDR3 as reference.

\subsection{Fitting procedures}
\label{sec:fit}

The main construction of planetary ephemerides uses a classic least squares approach. 
The total number of fitted parameters can be found in Table \ref{tab:param}. 
It changes from one ephemeris to an other, depending the perturbations included in the model and the observational parametrizations required for the data analysis. 
Thus, to the parameters presented in Table \ref{tab:param}, spacecraft bias accounting for additional transpondeur delays or additional ground-station delays \citep{Kuchynka12} have also to be added as well as  parameters related to solar plasma corrections \citep{2013A&A...550A.124V, pitjeva2021}. \\
Complementary methods have been proposed to make accuracy assessments, to overcome the problems of high correlated parameters, in particular in the frame of testing alternative theories of gravity (see sect \ref{sec:PPN}), or of multiple correlations between parameters, as in the case of the determination of asteroid masses \citep{2020MNRAS.492..589F}.
In the context of computing threshold values for the violation of general relativity, \cite{fienga2015cm} had tested genetic algorithm approaches for identifying sets of  parameters (PPN $\beta$, $\gamma$, the Sun oblateness   J$_{2}^{\odot}$ and secular variations of the gravitational mass of the Sun, $\dot{\mu}/{\mu}$) with which planetary ephemerides can be computed and fitted to the observations with an comparable accuracy {to} the ephemeris (in this case INPOP15a) built in general relativity. 
Different approaches based on $\chi^{2}$ analysis with fixed PPN or alternative theory parameters have been investigated as well, using random walk exploration methods associated with various cost functions \citep{bernus2019,2020PhRvD.102b1501B,bernus2022, fienga2022}. 
These methods have in common to be very conservative and to give larger constraints than those obtained with the least square procedures. See sect. \ref{sec:PPN} for the full discussion. 



\paragraph{A final but important comment is about the  masses used for the construction of the ephemerides.}
Most of the planetary masses are not obtained during the construction of the planetary ephemerides but during the orbit determination of one or several spacecraft flying in planetary systems (see Table \ref{tab:mass}). They are obtained by combination of data from different missions and techniques, usually in the general relativity framework. For example, in the case of Neptune and Uranus, it is even a combination of Voyager radio tracking data, ground-based optical observations and images acquired by the s/c that leads to the determination of the mass of the planetary system. The s/c orbit and gravity field (mass) determinations are complex as they depend on various sources of interactions from solar radiation pressure, atmospheric dragging, maneuver residual accelerations to non-stochastic accelerations.
For that reason, these s/c  orbit and gravity field determinations are usually done independently from the planetary ephemerides.\footnote{Even if iterations are sometimes necessary.} 
These masses are {therefore} obtained in the framework of the s/c orbit and of the gravity field determination algorithm, which assumes generally general relativity. 
Only, the gravitational mass of the Sun, the ratio between the mass of the Earth and the mass of the Moon (invariant to any change of the gravitational constant) as well as a variable number of asteroid gravitational masses (see sect. \ref{sec:aste}) are, in fact, estimated during the planetary ephemeris adjustment. 
In the case of a test of the violation of the Equivalence principle, the fact that some of the masses used in the ephemeris are obtained in the general relativity framework can introduce an inconsistency in the method. In principle, it should be necessary to consider a more global approach, including the  s/c orbit determination in the same framework than the one of the planetary ephemeris. So far, such an attempt was only operated for Parametrized post-Newtonian tests (see sect. \ref{sec:PPN}). 

\begin{table}
        \caption{Planetary masses used in DE440 \citep{2021AJ....161..105P} and INPOP19a \citep{inpop21a}. Are indicated from which missions the masses were extracted from with the corresponding references.}
    \begin{tabular}{c|l l}
    Planet & References & Mission \\
    \hline
    Mercury   & \cite{2020Icar..33513386K} & Messsenger \\
      Venus   & \cite{1999Icar..139....3K} & MAGELLAN \\
            Mars   & \cite{2016Icar..274..253K} & MRO, MGS, Odyssey, \\
        & &    Pathfinder, MER, Viking \\
            Jupiter barycenter   & \cite{2020GeoRL..4786572D} & Juno + 5 s/c flybys +  \\
            &\cite{Jacobson2021a} & ground-based and s/c optical data \\
            Saturn barycenter  & \cite{Jacobson2021b}& Cassini + ground-based + \\
            &\cite{Iess2020}& s/c optical data \\
 Uranus barycenter  & \cite{2014AJ....148...76J} & Voyager 2 + ground-based optical data\\
  Neptune barycenter  & \cite{2009AJ....137.4322J} & Voyager + ground-based and s/c optical data \\
  \hline
   \end{tabular}
    \label{tab:mass}
\end{table}

\section{Tests of alternative theoretical frameworks with planetary ephemeris}
\label{sec:fulltest}
Thanks to the accuracy of the most recent observations but also to the more than one century long data span, planetary ephemerides are excellent tools for testing general relativity and its alternatives. In principle, if the model of gravity is not accurate enough, or does not well represent the physics as seen by the observations,  it should then  lead to a deterioration of the differences (residuals) between observed and computed quantities{---deduced from the ephemeris}. When the deterioration of residuals built with an alternative theory  becomes statistically significant with respect to the residuals obtained in the general relativity framework, one can say that it disfavors the alternative theoretical framework---or favors general relativity over its alternative. On the opposite, if the residuals obtained with an alternative theory are significantly improved---that is, if there are significantly smaller---in comparison to the one obtained in the general relativity framework, it favors the alternative theory. However, so far, this has never occurred.

{In the following sections, we will explore various theories or phenomenologies that have been scrutinized using planetary ephemerides. The selection of certain theories or phenomenologies to test, rather than others, can be driven by several considerations. One of them could be the historical circumstances. For example, the Parametrized Post-Newtonian (PPN) framework was developed during the early stages of the general relativity testing era and has since been widely employed in tests involving planetary ephemerides.}

Another factor could be the complexity of the task, both in terms of numerical computations and statistical methodology. The larger the number of parameters to constrain a theory has, the larger the dimension of the theoretical space there is to examine. 
Additionally, there is an increasing risk of introducing multiple correlations between the parameters of the theory, as well as between these parameters and those of the solar system (i.e., masses).
Given that it can take up to eight hours of computation to adjust the solar system parameters of the ephemeris at a single point in the parameter space \citep{2023arXiv230607069M}, achieving a densely filled parameter space becomes highly computationally demanding if the parameter space dimension is greater than one. Moreover, the statistical methodology also grows more complex with an increase in the number of parameters to constrain.

There is also the risk of missing a genuine detection—that is, obtaining better residuals in an alternative theory than in general relativity for a specific parameter value—if the parameter space of alternative theories explored is too sparsely populated. Therefore, it is of paramount importance to have a densely filled parameter space, which consequently makes computational time requirements escalate with the number of parameters to constrain. 
In the case of a classic least square approach, the risk of high correlations between parameters associated with the observational uncertainties  also limit the possible theories to be tested.
As a result, tests are often conducted on theories or phenomenologies that hinge on few parameters only.

\subsection{Parametrized post-Newtonian framework}
\label{sec:PPNf}

For metric theories---that is, theories for which the additional gravitational fields\footnote{Sometimes called ``degrees of freedom'' in the theoretical literature.} do not couple to matter directly, such that the theories satisfy the weak equivalence principle by design---all the differences with respect to general relativity can be taken into account through parameters that appear in front of the gravitational potentials in the metric \citep{will2014lrr,will2018book}. Indeed, because the weak equivalence principle is enforced by hand by imposing that the additional gravitational fields do not couple to matter directly, particles still follow geodesics of the metric and proper time is still defined from the metric only.\footnote{Technically, a theory satisfies the weak equivalence principle if it exists a conformal (or disformal) representation for which the metric is the only gravitational field that appears in the matter Lagragian density {according to the \textit{comma-goes-to-semicolon rule} \citep{MTW}}. For instance, the additional scalar-field appears in the matter Lagrangian density in the Brans-Dicke theory when written in the Einstein representation (or frame), but not in the (original) Jordan representation. Hence, because of the latter, the Brans-Dicke theory satisfies the weak equivalence principle.}

Many theories then turn out to generate the same type of metrics, where the differences with respect to general relativity can be parametrized by a {set of a few} parameters in the metric only---which values depend upon the theory considered. This comes in handy, as it allows to derive all the equations necessary in planetary ephemerides---that are, the time-scale definitions, the motion of celestial bodies and the Shapiro delay---from a single original parametrized metric. This whole framework is therefore called the parametrized post-Newtonian (PPN) formalism.

Nevertheless, as it will be discussed in Sec \ref{sec:RefFrameThInAlts}, the coordinate system being used for planetary ephemerides matters, and one has to make sure that the coordinate system being used for alternative theories is compatible at the required level of accuracy with the IAU standards used  to convert observables (e.g. the roundtrip of an electromagnetic signal measured in terms of the proper time of the local clock) into spatial and temporal coordinate positions (e.g. the distance\footnote{Which is dependent upon the coordinate system being used.} of a spacecraft with respect to the ground-based station being used).

\label{sec:PPN}

\begin{table}
\caption{PPN parameters $\beta$ and $\gamma$ obtained with planetary ephemerides. Columns 1,2 and 3 give respectively the reference, the ephemeris involved and method used. As explained in the text (sect. \ref{sec:PPN}), we consider a classification based on 3 types of adjustment methods: the direct least square fit combining $\beta$, $\gamma$, all the ephemeris parameters and the oblateness of the Sun, J$_{2}^{\odot}$  (FF), the partial fit where some parameters are fixed such as $\gamma$ or J$_{2}^{\odot}$ and the other are fitted (PF), the random walk exploration (RW) where  $\beta$ and $\gamma$ are randomly sampled when the rest of the parameters (including  J$_{2}^{\odot}$) are fitted. Are given for information
the results obtained in considering only the Mercury orbit (MSG).}
\begin{tabular}{l l l l l l}
\hline
Reference & PE & Method & $\beta -1$ & $\gamma -1$ & J$_{2}^{\odot}$ \\
& & & $\times$ 10$^{5}$ & $\times$ 10$^{5}$  & $\times$ 10$^{7}$ \\
\hline
\citep{2001CeMDA..80..249P}  & EPM2000 & FF & 40 $\pm$ 20$^{*}$  & 10 $\pm$ 10$^{*}$ &  2.43 $\pm$ 0.67$^{*}$ \\
\citep{Standish2001}& DE405 & FF &  10 $\pm$ 10 $^{*}$ &   40 $\pm$ 10$^{*}$ &  2.46 $\pm$ 0.68$^{*}$ \\
\citep{2005AstL...31..340P} & EPM2004 & FF & 0 $\pm$ 10 $^{*}$ & -10 $\pm$ 20 $^{*}$ &  1.9 $\pm$  0.3\\
\citep{2009AA...507.1675F} & INPOP08a & PF &7.5 $\pm$ 12.5 & 0.0 & fixed to 1.82  \\
\citep{fienga2011cm} & INPOP10a  & PF &-4.1 $\pm$ 7.8 & -6.2 $\pm$ 8.1& fixed to 2.40 \\

\citep{konopliv:2011ic} & DE421 &  PF & {4 $\pm$ 24} &fixed to 2.1 & {fixed to 1.8} \\

\citep{2013MNRAS.432.3431P} & EPM2011 & FF &{-2 $\pm$ 3} &{4 $\pm$ 6} &{2.0 $\pm$ 0.20} \\
{\citep{2014A&A...561A.115V}} & INPOP13c & FF &{0.2 $\pm$ 2.5} &{-0.3 $\pm$ 2.5} & {2.40 $\pm$ 0.20}  \\
\citep{fienga2015cm} & INPOP15a & FF & -6.7 $\pm$ 6.9 & -0.8 $\pm$ 5.7  & 2.27 $\pm$ 0.25 \\
\citep{fienga2015cm} & INPOP15a & RW & 0.00 $\pm$ 6.90& -1.55 $\pm$ 5.01 & 2.22 $\pm$ 0.13\\
\citep{2017AJ....153..121P} & DE440 & PF &  -2.6 $\pm$ 3.9$^{*}$ & fixed to  2.1  & 2.25 $\pm$ 0.09$^{*}$ \\
\citep{genova2018nc} & DE438 & MSG &  -1.625 $\pm$ 1.8 & fixed to  2.1  & 2.246 $\pm$ 0.02 \\
\citep{fienga2022} & INPOP20a & PF &  1.9 $\pm$  6.28 &    2.64 $\pm$  3.44  &  2.165 $\pm$ 0.12\\\
\citep{fienga2022}& INPOP20a & RW &  -1.12 $\pm$ 7.16  &   -1.69 $\pm$ 7.49  & 2.206 $\pm$ 0.03 \\ 
\hline
\end{tabular}
\label{tab:ppn}
\end{table}

\begin{figure}
\centering
\includegraphics[scale=0.6]{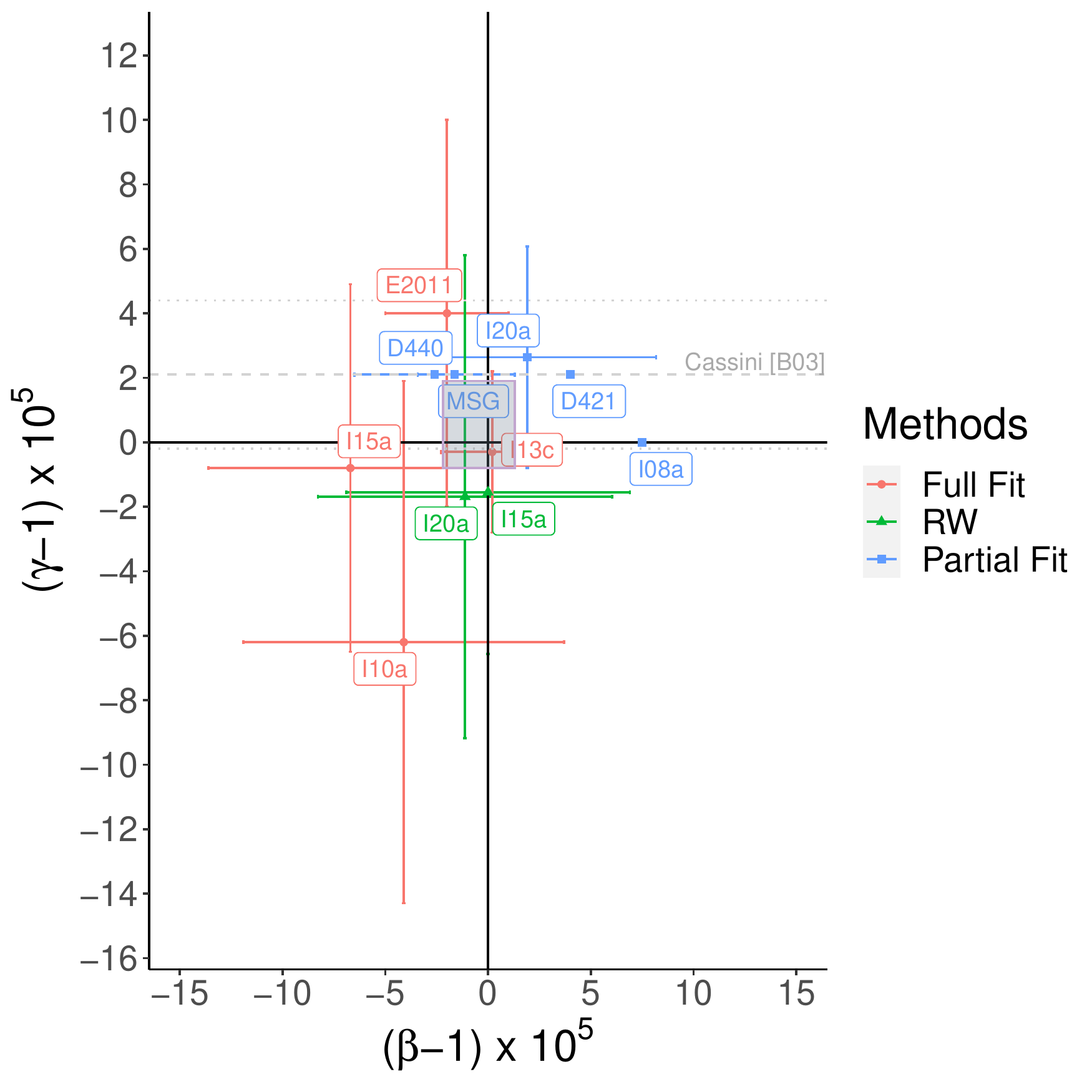}
\caption{Constraints on PPN parameters $\beta$ and $\gamma$ obtained from planetary ephemerides as presented in Table \ref{tab:ppn}. The shadded area is the zone common to all the constraints. The lightgrey lines give the constraint on $\gamma$ extracted from the Cassini experiment \citep{bertotti2003nat}. Three categories are considered: the values obtained from a full global fit, the one obtained using random walk exploration methods (RW) and finally, the one constrained either in fixing one of the two parameters (usually $\gamma$ like in \citep{konopliv:2011ic, 2017AJ....153..121P}) either in considering only one planetary orbit (like the Mercury orbit for \cite{genova2018nc} . These categories were labelled according to the Column 3 of  Table \ref{tab:ppn} and the annotated text refers to Column 2 of  Table \ref{tab:ppn}.}
\label{fig:ppn}
\end{figure}



Among the 10 PPN parameters {discussed} in \cite{will2014lrr} and \cite{will2018book}, we shall first only consider the Eddington-Robertson-Schiff parameters $\gamma$ and $\beta$, because most of the planetary ephemerides to this date focused on those parameters. A complete description of the PPN parameters will be discussed in Sec. \ref{sec:SPN}. Let us write the metric as follows \citep{klioner:2000pd}

\begin{widetext}
		\begin{align}
			g_{00}=&-1 + \frac{2 w}{c^{2}} -  \frac{2 \beta w^{2}}{c^{4}} + \Ol(c^{-5}) \nonumber\\
			g_{0i}=&- \frac{(4\gamma + 3 )}{2} \frac{ w_{i}}{c^{3}}  + \Ol(c^{-4}) \nonumber\\
g_{ij}=& \delta_{ij} \left(1 + \frac{2 \gamma w}{c^{2}} \right) + \Ol(c^{-3})  \, ,\label{eq:metric3}
		\end{align}
		\end{widetext}
In general relativity, the PPN parameters $\beta$ and $\gamma$ equal to one, but not for alternative theories in general. 

The form of the metric given in Eq. (\ref{eq:metric3}) is not sufficient in order to completely fix the coordinate system, and one has to impose an additional constraint (or gauge) in order to specify the metric field equations---see, e.g., \cite{klioner:2000pd}. Since the IAU recommends the coordinates (or gauge) to be harmonic for general relativity \citep{Soffel2003}, it seems relevant to consider harmonic coordinates for alternative theories as well, such that (at least) the coordinate systems of the alternative theory reduce to the IAU coordinate systems in the limit for which both theories cannot be distinguished at the level of accuracy of the observations.

\paragraph{Ambiguity of the definition of harmonic coordinates}

The very definition of what one may call \textit{harmonic coordinates} can be ambiguous for alternative theories of gravity. For instance, for scalar-tensor theories, one can impose the usual post-Newtonian harmonic conditions---$g^{\alpha \beta}\Gamma^0_{\alpha \beta}=0$---either in the Jordan frame, or in the conformal Einstein frame \citep{damour1992cqg,kopeikin:2004pr,minazzoli:2011cq,will2018book,kopeikin:2019pr}. But both choices reduce to the general relativity harmonic coordinate in the limit for which the difference between general relativity and the scalar-tensor theory becomes negligeable (e.g. for the Brans-Dicke parameter that is such that $\omega \rightarrow \infty$). Imposing the harmonic conditions in the Einstein frame has been shown to lead to more convenient properties \citep{damour1992cqg,kopeikin:2004pr,minazzoli:2011cq,kopeikin:2019pr}, notably with respect to the full theory of reference frames \citep{kopeikin:2004pr}. However, this definition of harmonic coordinates then becomes theory-dependent \citep{will2018book}. In order to alleviate the ambiguity, \cite{kopeikin:2004pr,kopeikin:2019pr} used the name ``\textit{Nutku gauge}''---or ``\textit{conformal harmonic gauge}''---when the harmonic conditions are imposed on the metric in the Einstein frame rather than in the Jordan frame. {In \citep{will2018book}, it corresponds to the ``\textit{generalized harmonic gauge}''}.

\subsubsection{Equations of motion, Lense-Thirring and Shapiro delay}

In harmonic coordinates, the EIHDL equation of motion in the barycentric frame (Eq. \ref{eq:accgeneral relativity}) reads as follows\footnote{However, the transformations between the barycentric and geocentric reference frames slightly depend on the specific type of harmonic coordinates being used \cite{kopeikin:2004pr}.}
\begin{widetext}
		\begin{align}
			\bm {a}^{{PPN}}_T=&-\sum_{A\neq T} \frac{\mu_A}{r_{AT}^3}\bm r_{AT}  \nonumber\\
			&-\sum_{A\neq T} \frac{\mu_A}{r_{AT}^3c^2}\bm r_{AT}\Bigg\{\gamma v_T^2+(\gamma+1)v_A^2-2(1+\gamma)\bm v_A.\bm v_T  -\frac{3}{2}\left(\frac{\bm r_{AT}.\bm v_A}{r_{AT}}\right)^2\nonumber \\
			&\hspace{3cm}-\frac{1}{2}\bm r_{AT}.\bm a_A -2(\gamma+\beta)\sum_{B\neq T}\frac{\mu_B}{r_{TB}}-(2\beta-1)\sum_{B\neq A}\frac{\mu_B}{r_{AB}}\Bigg\} \nonumber \\
			&+\sum_{A\neq T}\frac{\mu_A}{c^2r_{AT}^3}\left[2(1+\gamma)\bm r_{AT}.\bm v_T-(1+2\gamma)\bm r_{AT}.\bm v_A\right](\bm v_T-\bm v_A)  \nonumber \\
			&+ \frac{3+4 \gamma}{2}\sum_{A\neq T} \frac{\mu_A}{c^2r_{AT}}\bm a_A \, ,\label{eq_eih}
		\end{align}
		\end{widetext}		
with $\gamma$ and $\beta$, the Parametrized Post Newtonian (PPN) parameters. They are equal to 1 in general relativity, such that one recovers the usual EIHDL Eq. (\ref{eq:accgeneral relativity}) in that case. \\
Additionally, directly from the metric Eq. (\ref{eq:metric3}), one can deduce that the Shapiro delay becomes
	\begin{equation}
			c(t_r-t_e)^{{PPN}}=R+\sum_A(\gamma+1)\frac{\mu_A}{c^2}\ln\frac{\bm{n}\cdot\bm{r}_{rA}+r_{rA}+2\frac{(1+\gamma)\mu_A}{c^2}}{\bm{n}\cdot\bm{r}_{eA}+r_{eA}+2\frac{(1+\gamma)\mu_A}{c^2}} \, ,\label{eq:shapiroPPN}
		\end{equation}
with the same notations as Eq. (\ref{eq:shapirogeneral relativity}).

The Lense-Thirring acceleration is also modified and reads:
\begin{equation}
\bm{a}^{{PPN}}_{LT} =(1+\gamma) \frac{G S}{c^{2}r^{3}} \left[  \frac{3 \bm{k} . \bm{r}}{r^{2}} (\bm{r} \wedge \bm{v}) -  (\bm{k} \wedge \bm{v}) \right]
\label{eq:LTppn}
\end{equation}
with the notation of Eq. (\ref{eq:LT}).


It is also important to stress that only the PPN parameter $\gamma$ appears in the Shapiro delay formula that is used for the computation of the propagation time of light and doppler shift (see sect. \ref{sec:obs}). As a consequence, the impact of this parameter can be more easily disentangled from the effect of other parameters, in particular with conjunction events for which one directly probes the differential evolution of Eq. (\ref{eq:shapiroPPN}) with respect to evolution of the impact parameter during the conjunction \citep{bertotti2003nat}.
This disentangling of the $\gamma$ parameter is also enhanced by the LT contribution, {which depends} on $\gamma$.

Since the first radar echos obtained from the Mercury surface \citep{1964PhRvL..13..789S}, it has been possible to evaluate the departure from unity of the PPN parameters in the context of the planetary orbit computation.
Such estimations are done during the construction of the planetary ephemerides when the initial conditions of the planetary orbits are fitted to observations with a least square procedure \citep{1967AJ.....72..338A, 1976jden.book.....S}, together with other parameters such asteroid masses.

\begin{figure}
\centering
\includegraphics[scale=0.5]{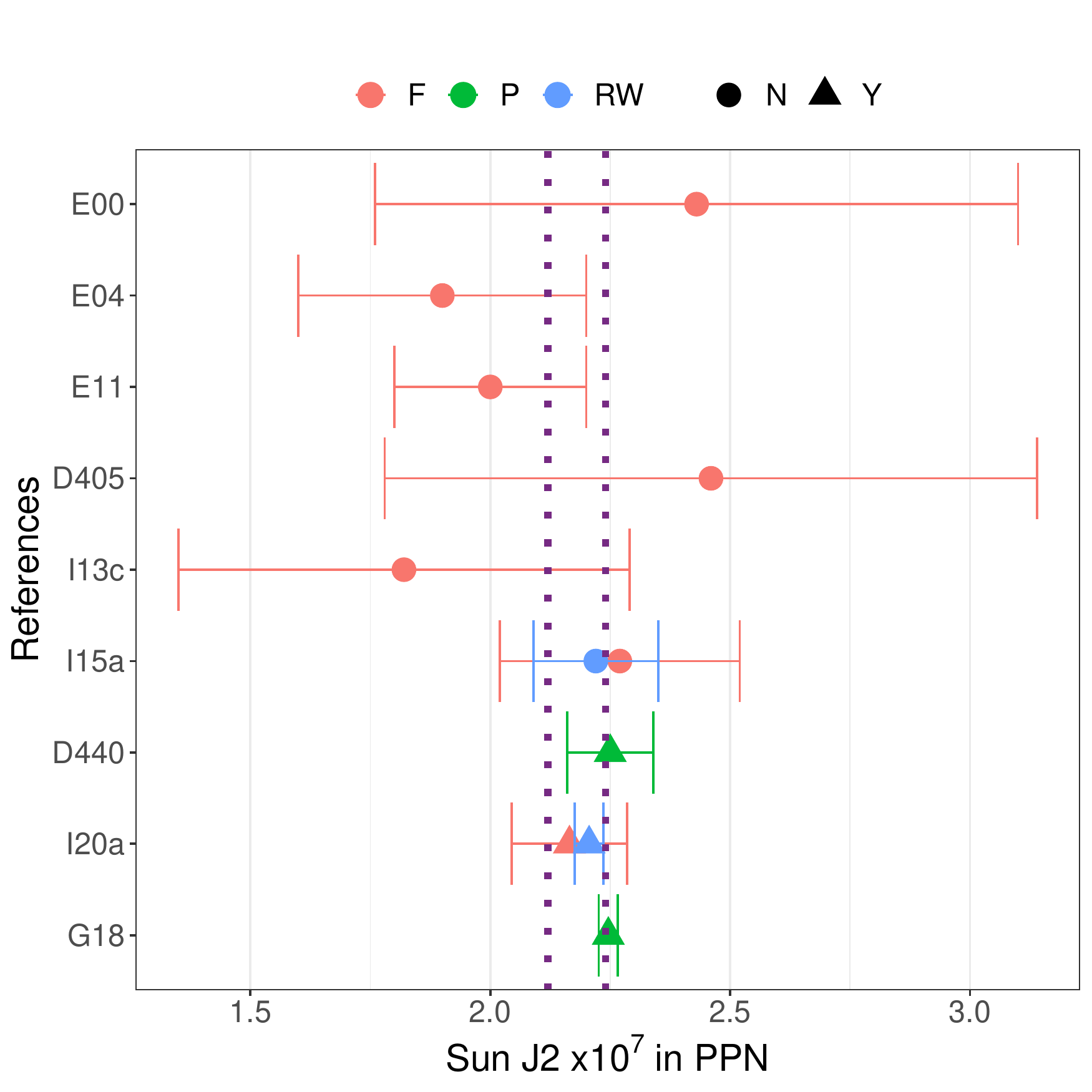}
\caption{Values of the Sun oblateness $J_{2}^{\odot}$ obtained {when} considering PPN parameters $\beta$ and $\gamma$ as varying parameters. The dotted lines give the limits of the less constraining helioseismologic value from \citep{1998MNRAS.297L..76P}. F stands for Full fit, RW for random walk exploration method and P for partially fitted ephemerides (some parameters are fixed). N (respectively Y)  indicates that the value has been obtained without (respectively with) LT. In y-axis, are given the references : G18 is \citep{genova2018nc}, I20a \citep{fienga2022}, D440 \citep{2017AJ....153..121P}, I13c \citep{INPOP13c}, E04 and E11 in \citep{2013MNRAS.432.3431P}. E00 is \citep{2001CeMDA..80..249P}, D405 \citep{Standish2001} and I15a \citep{fienga2015cm}. It is important to note that in \cite{genova2018nc}, only the orbit of Mercury is considered. }
\label{fig:j2ppn}
\end{figure}

\subsubsection{Time-scales}
\label{sec:RefFrameThInAlts}

As mentioned in Sec. \ref{sec:timescale}, planetary ephemerides produce the difference between the TT and the TDB. In the PPN framework, Eq. (\ref{eq_time2}) has to be modified as follows \citep{Manche2011}
\begin{widetext}
		\begin{align}
	a =& -\frac{1}{2} v_T ^{2} - \sum_{A\neq T} \frac{\mu_A}{r_{AT}} \, \nonumber \\
b =& -\frac{1}{8} v_T ^{4} +  \left(\beta - \frac{1}{2}\right) \left[ \sum_{A\neq T} \frac{\mu_A}{r_{AT}} \right]^{2} + \sum_{A\neq T} \frac{\mu_A}{r_{AT}} 
\Bigg\{ 2 (1+\gamma)\bm v_T.\bm v_A - (\gamma + \frac{1}{2}) v_T^{2}   \nonumber \\
 & - (1+\gamma) v_A^{2} + \frac{1}{2}\bm a_A.\bm r_{AT} + \frac{1}{2} \left(\frac{\bm v_A.\bm r_{AT}}{r_{AT}} \right)^{2} \Bigg\}  \nonumber \\
 & + (2\beta -1) \sum_{A\neq T} \frac{\mu_A}{r_{AT}} 
 \left(\sum_{B\neq A} \frac{\mu_B}{r_{BA}} \right).\label{eq_time2_PPN}
		\end{align}
		\end{widetext}

\paragraph{Comments on reference frame theory in alternative theories}

The difficult subject of reference frame systems in alternative theories has not been investigated systematically. In fact, it has been explored explicitly only for a class of massless scalar-tensor theories ``\`a la Brans-Dicke'' \citep{kopeikin:2004pr}, after some previous effort to characterize it in a general standard post-Newtonian approach \citep{klioner:2000pd}. Fortunately in practice, the differences between coordinate systems in general relativity versus PPN framework are numerically negligible---notably given the already tight constraints that one already has on the PPN parameters.

\subsubsection{Constraints and correlations}

An important discussion related to the PPN parameter determinations using least square methods is the correlations between $\beta$, $\gamma$ and other contributions to the planetary accelerations. 
This problem has been pointed out a long time ago \citep{1978AcAau...5...43A} but it is still vivid today.
%
In particular, if we consider the analytical expression of the advance of the perihelia for planetary orbits $\Delta \dot\varpi_{PLA}$, as given in \citep{will2014lrr}, the two main  contributions of this equation are the PPN term $(2+2\gamma -\beta)$ and a term giving the impact of the Sun oblateness $J_{2}^{\odot}$. But we can also add the main belt asteroid contribution and the LT effect defined in sect. \ref{sec:LT}, such as per orbit, one gets \citep{2003ASSL..293.....B}

	{\begin{widetext}
	\begin{align}
\Delta\dot\varpi_{PLA} =& \mu_{\odot}(t) \times \left( \frac{2}{3}\frac{{(2\gamma-\beta+2)} }{a \, (1-e^2)c^2} +  \Delta\dot\varpi_{LT}(S, \gamma) \right)  \nonumber\\
& + J_{2}^{\odot} \frac{R^2_{\odot}}{a^{2} \, (1-e^2)^2} + \Delta\dot\varpi_{AST}(\mu_{AST}, a_{AST}, e_{AST}) \,  \label{eq:ppnom}
		\end{align}
		\end{widetext}}
where   $a_{AST},e_{AST}, \mu_{AST}$ are respectively the asteroid orbit semi-major axis and eccentricity and its gravitational mass perturbing the planet orbits (main-belt and trans-neptunian objects), $a,e$ are the semi-major axis and eccentricity of the planet orbit and finally, $c$ and $\mu\odot$ are the speed of light and the gravitational parameter of the Sun. This latest quantity can be time-varying as discussed in sect. \ref{sec:mudot}.
$\Delta\dot\varpi_{LT}(S, \gamma)$ is the Lense-Thirring effect presented in sect. \ref{sec:LT}. A discussion about other contributing terms such as PN cross-terms, {de} Sitter precession and 2PN crossterms has been proposed by \cite{2018PhRvL.120s1101W}. The conclusion is that these 3 additional contributions could have some importances in the coming years with the Bepi-Colombo mission. {In any case, they are implicitely taken into account when numerically integrating the EIHDL equation of motion.}

Note that Eq. (\ref{eq:ppnom}) is an analytical approximation of the perihelia advance rate {that is useful in order to get a sense of some of the correlations that may arise between different quantities. The} actual value would have to be inferred from the numerical integration of the equations of motion in the solar system with the total acceleration, $\bm{a}_T^{\textrm{total}}$:
\bea
\bm{a}_T^{\textrm{total}} &=& \bm {a}_T(\mu_{SS},\gamma,\beta,IC_{SS})+ \bm {a}_{J_{2}^{\odot}}(\mu_{\odot},J_{2}^{\odot}) + \bm {a}_{LT}(S_{\odot}, \gamma),
\eea
where $\bm {a}_T, \bm {a}_{J_{2}^{\odot}},\bm {a}_{LT}$ are the point-mass interaction acceleration given in Eq. (\ref{eq_eih}), the acceleration induced by the oblateness of the Sun Eq. (\ref{eq:accJ2}) and the Lense-Thirring acceleration Eq. (\ref{eq:LTppn}) respectively, $IC_{SS}$ stands for the initial conditions of the numerical integration for the solar system bodies, including asteroids. 

It is rather understandable from Eq. (\ref{eq:ppnom}) that there can be some level of correlation between, e.g., the semi-major axis (which follows from the initial conditions in the numerical integration) and the post-Newtonian parameters $\gamma$ and $\beta$. One can also see that for disentangling the different contributions, it is more efficient to have several orbits with significantly different semi-major axis $a$ and eccentricities $e$ as the PPN acceleration depends on $a$ when the Sun oblateness acceleration depends on $a^{2}$ (see Eq. (\ref{eq:ppnom}) and (\ref{eq:J2})).
 It is also interesting to compare the Sun $J_{2}^{\odot}$ determinations obtained in general relativity---equivalent to consider $\beta$ and $\gamma$ fixed to one---presented on Fig. \ref{fig:j2} with the those obtained in adding the PPN parameters $\beta$ and $\gamma$ into the global planetary fit as on Fig \ref{fig:j2ppn}. It appears {first} that, besides the differences between ephemerides, the general relativity estimations (with $\beta$ and $\gamma$ fixed to unity) have globally  reduced error bars ({i.e. }less uncertainty induced by less correlation), as one can see, for example, with INPOP20a ($2.218 \pm 0.03$ in general relativity versus $2.165 \pm 0.12$ in PPN), or the partial fit obtained by \cite{genova2018nc} ($2.271 \pm  0.003$ in general relativity versus $2.246 \pm  0.02$ in PPN). 
 Secondly, biases affect also the $J_{2}^{\odot}$ determinations obtained with fitted PPN parameters as these latest can be significantly different from the  $J_{2}^{\odot}$ values obtained in general relativity.
 These bias are due to the correlations between the PPN parameters and $J_{2}^{\odot}$. In the case of the Earth-Moon system, the relation is even more complex as it also includes the tidal contributions depending on the internal structures of the two bodies.  
 
 In order to overcome the problem of the correlations between the parameters, other methods of inversion such as random walk exploration or Genetic Algorithm have been proposed (see sect.~\ref{sec:fit}). 
 They usually give more conservative constraints than the direct least square method as it is visible on Table \ref{tab:ppn} and Fig. \ref{fig:ppn}.
On these Table and Figure, one can find the estimations of PPN parameters  $\beta$ and $\gamma$ determinations based on planetary ephemerides. 
Three families of determinations are proposed. The first noted $FF$ gathers least square fitted evaluations of PPN parameters, $J_{2}^{\odot}$  and secular variation of the gravitational parameter of the Sun, $\mu_{\odot}$. 
As discussed previously on Sec. \ref{sec:j2} and \ref{sec:fit}, these parameters are strongly correlated and thus are affected by {biases} (see Fig. \ref{fig:j2} and \ref{fig:j2ppn}) and underestimated uncertainties. 
The second type of determinations given on Table \ref{tab:ppn} with the label $PF$ are the one deduced with one (at least) of these parameters fixed in the planetary ephemeris adjustment. These estimations are less correlated than the $FF$ ones and give more reliable assessments but are limited to only one value (usually either $\gamma$ or $J_{2}^{\odot}$ are fixed and $\beta$ is fitted). Finally, the latest category  is obtained by Genetic Algorithm \citep{fienga2015cm} or random walk exploration  \citep{fienga2022}. 
They investigate the space of parameters more largely than the two former methods, leading to usually larger intervals for possible general relativity violations.
On Table \ref{tab:ppn}, values obtained in the context of one single planet analysis (Mercury) with only one single spacecraft (MESSENGER) data analysis \citep{genova2018nc} {are also given for comparisons}. The obtained results show smaller uncertainties than the planetary determinations but might be affected by possible {biases}. As it was shown in the Eq. (\ref{eq:ppnom}), the use of multiple orbits favors the disentangling of the different planetary contributions. Moreover, the value of $\beta$ is obtained in introducing the Nordvedt relation Eq. (\ref{eq:eta}) in the equation of motion and in the fitting procedure which supposes to consider only metric theories and increases the correlations between the fitted parameters (see sect. \ref{sec:ep}).

Finally, as one can see on Fig. \ref{fig:ppn}, all the PPN estimations obtained with the three methods are consistent with general relativity, with no significant departure from unity for $\beta$ and $\gamma$. By considering the common overlaps of all the {recently} published intervals obtained with planetary ephemerides, the two following constraints are obtained:
	\begin{widetext}
	\begin{align}
(\beta - 1)  & =  (-0.45 \pm 1.75) \times 10^{-5} \nonumber\\
(\gamma -1) & =  (0.55 \pm 1.35) \times 10^{-5} \nonumber\\
\label{eq:gammabeta}		
    \end{align}
		\end{widetext}
These values represent an improvement of about 3 orders of magnitude for $\gamma$ and $\beta$ in comparison to the historic first determination of \citep{1978AcAau...5...43A}. We can expect the same type of improvements in the coming years, especially with the outcomes of the Bepi-Colombo mission \citep{2018Icar..301....9I, 2020CQGra..37i5007D}.

\subsubsection{ The extended PPN formalism}
\label{sec:SPN}

\paragraph{SPN gauge versus harmonic gauge}

Usually, the extended standard post-Newtonian formalism assumes a \textit{gauge} (or coordinate system) that is quite different from the one recommended by the IAU that is called the standard post-Newtonian (SPN) gauge. In general relativity already, while the harmonic condition\footnote{On top of the strong isotropy condition imposed on the metric, which fixes the space coordinate \citep{damour1990prd,Soffel2003}---see Sec. \ref{sec:pN}}  at the post-Newtonian level reads
\be
g^{\alpha \beta}\Gamma^0_{\alpha \beta}=\Ol(c^{-5}),
\ee
the SPN conditions read \citep{damour1990prd}
\bea
&&\partial_{j} g_{0 j}-\frac{1}{2} \partial_{0} g_{j j}=\Ol({c^{-5}}), \\
&&\partial_{j} g_{i j}-\frac{1}{2} \partial_{i}\left(g_{j j}-g_{00}\right)=\Ol(c^{-4}).
\eea
It implies that the field equations are modified, such that, instead of Eqs. (\ref{eq:poissongeneral relativity}-\ref{eq:poissongeneral relativityi}), one has \citep{damour1990prd}
\bea
&&\Delta w^{\textrm{SPN}}=-4 \pi G \sigma+\Ol(c^{-4}), \\
&&\Delta w_{i}^{\textrm{SPN}}-\frac{1}{4} \partial_{t} \partial_{i} w^{\textrm{SPN}}=-4 \pi G \sigma^{i}+\Ol(c^{-2}).
\eea
The difference between the potentials in the harmonic and the SPN gauges reads \citep{Soffel2003}
\be
w^{\textrm{SPN}} = w + c^{-2} \partial^2_{tt} \chi/2 + \Ol(c^{-3}), \qquad  w_{i}^{\textrm{SPN}} = w_i - \partial^2_{ti} \chi/8,
\ee
with 
\be
\chi :=G \int d^{3} x^{\prime} \sigma\left(t, \bm{x}'\right) \| \bm{x}-\bm{x}' \|.
\ee
In terms of coordinates, the transformation between the two coordinate systems reads {\citep{klioner:2000pd,will2018book}}
\bea
{t^{\textrm{SPN}} = t -\frac{1}{2} \partial_t \chi},\\
\bm{x}^{\textrm{SPN}} = \bm{x}.
\eea
While both gauges can be used in principle, the harmonic gauge has been found to be much more convenient for deriving the reference frame theory in the solar system \citep{damour1990prd}, such that harmonic coordinates have been used in the reference frame theory upon which the IAU recommendations have been built. {A thorough comparison between the two coordinate systems in the full PPN framework can be found in \citep{will2018book}}.

\paragraph{Full PPN metric in harmonic gauge}

The full PPN framework considers all the possible different potentials that could contribute to the metric, assuming that the Newtonian potential is the leading potential ---which may not be the case for a massive gravity theory for instance, see sect. \ref{sec:graviton}. Furthermore, the PPN framework can only deal with theories for which the Weak Equivalence Principle has been enforced by hand by requiring that the additional gravitational fields do not appear in the material part of the action. In the PPN framework, test particles {indeed} follow the geodesics of space-time, meaning that matter reacts to the space-time metric only. While this is necessary in order to have a theory for which all test particles fall alike by construction, one could question the theoretical relevance of such a restriction \citep{2012CQGra..29r4001D}---see Weak Equivalence Principle in sect. \ref{sec:ep}.

The full PPN metric in harmonic coordinates\footnote{{Defined here as a ``Nutku gauge'', that is $\partial_\nu [(1+(1+\gamma)U)\sqrt{-g} g^{\mu \nu}]=0$, and not $\partial_\nu [\sqrt{-g} g^{\mu \nu}]=0$---see discussion on the ambiguity of the definition of harmonic coordinates in Sec. \ref{sec:PPNf}.}} reads \citep{will2018book}:
\bea
g_{00}&=&-1+2 U+\left[2 {\psi_{\text {Harm }}}-2 \beta U^2\right]+\left(1-\frac{1}{2} \alpha_1+\alpha_2+2 \xi\right) \ddot{X}+ \nonumber\\
&& +\Phi_{\mathrm{Harm}}^{\mathrm{PF}},  \label{eq:g00SPN} \\
g_{0 j}&=&-\left[2(1+\gamma)+\frac{1}{2} \alpha_1\right] V_j-\frac{1}{4} \alpha_1 X_{, 0 j}+\Phi_{j \mathrm{Harm}}^{\mathrm{PF}}, \\
g_{j k}&=&(1+2 \gamma U) \delta_{j k}
\label{eq:g00willH}
\eea
with
\bea
{\psi_{\text {Harm }}}&=&\frac{1}{2}(2 \gamma+1-2 \xi) \Phi_1-(2 \beta-1-\xi) \Phi_2+\Phi_3+(3 \gamma-2 \xi) \Phi_4 + \nonumber\\
&& + \xi \Phi_6-\xi \Phi_W,\\
\Phi_{\text {Harm }}^{\mathrm{PF}} &=&-\alpha_1 w^2 U-\alpha_1 w^j V_j+\alpha_2 w^j w^k X_{j k}-\left(2 \alpha_2-\frac{1}{2} \alpha_1\right) w^j X_{, 0 j}, \\
\Phi_{j \text { Harm }}^{\mathrm{PF}} &=& -\frac{1}{2} \alpha_1 w^j U+\frac{1}{4} \alpha_1 w^k X_{j k},
\eea
	
where $\bm{w}$ is the coordinate velocity of the PPN coordinate system relative to the mean rest frame of the universe, and with the following metric potentials
\bea
&&U=\int \frac{\rho^{\prime *}}{\| \boldsymbol{x}-\boldsymbol{x}^{\prime} \|} d^3 x^{\prime},
V_j=\int \frac{\rho^{\prime *} v_j^{\prime}}{\|\boldsymbol{x}-\boldsymbol{x}^{\prime}\|} d^3 x^{\prime},
X=\int \rho^{\prime *}\|\boldsymbol{x}-\boldsymbol{x}^{\prime}\| d^3 x^{\prime},\\
&&\Phi_1=\int \frac{\rho^{\prime *} v^{\prime 2}}{\|\boldsymbol{x}-\boldsymbol{x}^{\prime}\|} d^3 x^{\prime}, \quad \Phi_6=\int \frac{\rho^{\prime *}\left[\boldsymbol{v}^{\prime} \cdot\left(\boldsymbol{x}-\boldsymbol{x}^{\prime}\right)\right]^2}{\|\boldsymbol{x}-\boldsymbol{x}^{\prime}\|^3} d^3 x^{\prime},\\
&& \Phi_2=\int \frac{\rho^{\prime *} U^{\prime}}{\|\boldsymbol{x}-\boldsymbol{x}^{\prime}\|} d^3 x^{\prime}, \quad \Phi_3=\int \frac{\rho^{\prime *} \Pi^{\prime}}{\|\boldsymbol{x}-\boldsymbol{x}^{\prime}\|} d^3 x^{\prime}, \\
&& \quad \Phi_4=\int \frac{p^{\prime}}{\|\boldsymbol{x}-\boldsymbol{x}^{\prime}\|} d^3 x^{\prime},\\
&& \Phi_W=\int \rho^{* \prime} \rho^{* \prime \prime} \frac{\left(\boldsymbol{x}-\boldsymbol{x}^{\prime}\right)}{\|\boldsymbol{x}-\boldsymbol{x}^{\prime}\|^3} \cdot\left[\frac{\left(\boldsymbol{x}^{\prime}-\boldsymbol{x}^{\prime \prime}\right)}{\|\boldsymbol{x}-\boldsymbol{x}^{\prime \prime}\|}-\frac{\left(\boldsymbol{x}-\boldsymbol{x}^{\prime \prime}\right)}{\|\boldsymbol{x}^{\prime}-\boldsymbol{x}^{\prime \prime}\|}\right] d^3 x^{\prime} d^3 x^{\prime \prime}\label{eq:PhiW},
\eea
where {$\rho^*:=\rho u^0 \sqrt{-g}$} is the ``conserved'' mass density\footnote{{$\rho$ is the density of rest mass of a fluid element as measured in a local, freely falling, momentarily comoving frame. One has $\rho = \sigma + \Ol(c^{-2})$, where $\sigma$ is defined in Eq. (\ref{eq:defsigma}).}}---in the sense that it satisfies the Newtonian conservation equation $\partial_t \rho^* + \partial_j (\rho^* v^j) = 0$. \footnote{Note that one therefore has $X=\chi+\Ol(c^{-2})$.} The whole set of PPN parameters is $\gamma, \beta, \xi, \alpha_1, \alpha_2, \alpha_3, \zeta_1, \zeta_2, \zeta_3, \zeta_4$.
The parameter $\xi$ is non-zero in any theory of gravity that predicts preferred-location effects such as a galaxy-induced anisotropy in the local gravitational constant GL (also called ``Whitehead'' effects); $\alpha_1, \alpha_2, \alpha_3$ measure whether or not the theory predicts post-Newtonian preferred-frame effects; $\alpha_3, \zeta_1, \zeta_2, \zeta_3, \zeta_4$ measure whether or not the theory predicts violations of global conservation laws for total momentum \citep{will2014lrr}. 


{Untill now, the full PPN framework has not been investigated with planetary ephemerides. First, one would have to derive the equation of motion in harmonic coordinates for {point-mass} interations that follows from the more general hydrodynamical situation depicted in Eqs. (\ref{eq:g00SPN}-\ref{eq:PhiW}). Also, let us stress that the issue of the definition of the SSB has to be carefully checked, notably for the parameters that correspond to a violation of the conservation laws of momentum. Finally, the more the parameters, the more computationally demanding the study is, and one therefore has to think about the best way(s) to explore and constrain the landscape of parameters. Hence, several studies still seem to be needed before being able to constrain the full PPN framework with planetary ephemerides.}

{Nevertheless, m}easurements of PPN parameters other than $\beta$ and $\gamma$ such as $\alpha_{1}, \alpha_{2}, \zeta_{1}$,in the weak field regime,  have been proposed early,  by \cite{1978AcAau...5...43A}, and in the context of the mission Bepi-Colombo by \cite{2002PhRvD..66h2001M, 2020CQGra..37i5007D}{, using Mercury orbit}. 
As explained in \citep{will2014lrr}, the contribution of the preferred-frame parameters are included in the ratio between the planet mass and the Sun mass, leading to contributions to a maximum of 1$\times 10^{-4}$ for Jupiter and  1$\times 10^{-7}$ for Mercury.  \cite{2020CQGra..37i5007D} expect a {constraint for $\alpha_1$} at the level of $10^{-7}$ and up to 7 $\times 10^{-8}$ for $\alpha_{2}$, with Bepi-Colombo and future Venus missions. These estimations are obtained with covariance analysis of tracking observations of s/c orbiting planets. They are interesting as they {estimate} the capabilities of future missions for disentangling of the PPN contributions on the planetary motions.


In the strong field regime, constraints have been obtained for the parameters $\alpha_{1},  \alpha_{3}$ and $\zeta_{2}$ using pulsar timing \citep{2003LRR....6....5S}. 
\subsection{Equivalence principle}
\label{sec:ep}

While a deviation for particles with different compositions falling equally in a given gravitational field would indicate a violation of the Weak Equivalence Principle (WEP), a deviation for extended bodies with different gravitational self-energies would be a sign of a violation of the Gravitational Weak Equivalence Principle (GWEP), which is part of the Strong Equivalence Principle (SEP) as defined in \citep{will2018book}. In all cases, with planetary ephemerides, one seeks to check the Universality of Free Fall (UFF)---that is, to check whether or not all bodies fall alike in a given gravitational potential.
Such violations naturally occur in various modifications of general relativity, and in particular in theories with more than four dimensions like in string theories \citep{1994GReGr..26.1171D,2002PhRvD..66d6007D}, for which the constants of the standard model of particles turn out to be dynamical entities \citep{damour2010prd,2012CQGra..29r4001D}.

As we will see, the violations of both the WEP and GWEP have in common the fact that the equation of motion at the Newtonian level reads as follows

\be
\boldsymbol{a}_{\mathrm{T}}=-\sum_{A \neq T} \mu^G_A \frac{\boldsymbol{r}_{A T}}{r_{A T}^{3}} \left(1+\delta_{\mathrm{T}}+\delta_{A T}\right), \label{eq:uffvgen}
\ee
where $\delta_T$ and $\delta_{AT}$ are coefficients that depend on the composition of the bodies $T$ and $A$, for the case of the violation of the WEP only. For the case of the violation of the GWEP only, $\delta_T$ depends on the ratio between the gravitational and inertial masses and $\delta_{AT}=0$---such that 
\be
\left(\frac{m^G}{m^I} \right)_T = 1 + \delta_T,
\ee
where $m^G$ and $m^I$ are the gravitational and inertial masses respectively. Note that $\mu^G_A$ is the gravitational parameter constructed with the gravitational mass $m^G_A$ such as $ \mu^G_A=(1+\delta_A) \mu_A^I$, where $\mu^I_A {:= G m^I_A}$ is the gravitational parameter obtained with the inertial mass $m^I_A$.
Eq. (\ref{eq:uffvgen}) is the most general equation that still satisfies Newton's third law of motion \citep{viswanathan2018mn}. As we will see in Sec. \ref{sec:dilaton}, one can have both a violation of the WEP and the GWEP at the same time---that is, $\delta_T$ can depend both on the composition of the body $T$ and on its internal gravitational energy.

The equivalence principle is currently tested at different scales: {on the one hand on the} laboratory scale for WEP with the torsion balance and MICROSCOPE, and {on} the other hand, the astronomical scale for GWEP with LLR \footnote{LLR tests also include a WEP contribution as the Earth and Moon {have} different compositions \citep{2001CQGra..18.2397A}.} and compact objects --- at the level of about 10$^{-13}$ with torsion balances \citep{2003PhRvD..68f2002A}, 10$^{-15}$ with the MICROSCOPE experiment \citep{2022CQGra..39t4009T} and  10$^{-14}$ with LLR \citep{williams:2012cq,viswanathan2018mn, 2021Univ....7...34B}  .

The latest LLR results \citep{2021Univ....7...34B} are more accurate that what can be achieved with the sensitivity of planetary ephemerides as discussed in \citep{viswanathan2018mn}.  We give however a brief overview of the equations at stake below, mainly in the perspective of the Bepi-Colombo mission that should lead to a significant improvement of the sensitivity of planetary ephemerides on the tests of the equivalence principle.

Finally, it is interesting to note that constraints on the violation of SEP have been obtained in strong field regime, mainly using the chronometry of millisecond pulsars. For a deeper dive into this captivating subject, we direct the reader to \citep{kramer:2021px,freire:2012mn,voisin:2020aa}.

\subsubsection{The case of metric theories}
\label{sec:SEPmt}

While metric theories satisfy the WEP by design (see Sec. \ref{sec:PPN})
, they do not satisfy the SEP in general---in the sense that bodies with different gravitational self-energies would not follow the same trajectories in general \citep{nordtvedt1968pr1,nordtvedt1968pr2}. This notably induces that for extended bodies with enough gravitational self-energy such as planets, the ratio between their gravitational and the inertial masses is no longer one as in general relativity, but instead depends on the body self-gravitational energy that can be approximated for uniformly distributed spherical bodies as follows
\be
\left(\frac{m^G}{m^I}\right)^{SEP}_T = 1 + \eta  \times \frac{3}{5} \frac{G m^{G}_T}{c^{2} R_T},
\label{eq:ratiomasses}
\ee
where $R_T$ is the radius of the astronomical body $T$ and $m^{G}_T$ is  its gravitational mass.
The parameter $\eta$, known at the Nordtvedt parameter,  quantifies the possible violation of the SEP and can be fitted together with the rest of the planetary and lunar ephemeris parameters. 
In the most generic case, $\eta$ can be seen as an independent parameter and be introduced in Eq. (\ref{eq:ratiomasses}), common to all the bodies. 
However, in the case of the metric theories, it can be related to the PPN parameters such as \citep{will2018book}:
\begin{align}
\eta = 4 (\beta -1) - (\gamma -1) - \alpha_{1} - \frac{2}{3} \alpha_2.
\label{eq:eta}
\end{align}
Although note that only the cases for which $\alpha_1=0$ and $\alpha_2=0$ have been considered in planetary ephemerides so far---see Sec. \ref{sec:SPN} for a discussion on the extended parametrized post-Newtonian framework.
In principle, the introduction of Eq. (\ref{eq:eta}) in the global adjustment of planetary ephemerides should increase the correlations between $\beta$, $\gamma$, $\eta$ and the other ephemeris parameters, including $J_{2}^{\odot}$ \citep{2007PhRvD..75b2001A}. However, 
simulations for the MORE Bepi-Colombo experiment indicate a significant improvement of the $\beta$, $J_{2}^{\odot}$ and $\eta$ correlation (from 0.9 to 0.3) thanks to the high accuracy of the Mercury-Earth range measurements that should be obtained (few tens of centimeters) by the mission \citep{2002PhRvD..66h2001M,2016PhRvD..93l3014D,2018Icar..301....9I, 2020CQGra..37i5007D}.
Finally, recent attempt using Messenger tracking data leads to a measurement of $\eta$ at the level of 7 $\times 10^{-5}$ but this estimation has been done in considering only the Mercury orbit \citep{genova2018nc}. Even in these very favorable conditions, it gives a slightly less restrictive limit than the latest LLR evaluation from \cite{2021Univ....7...34B} with a deduced $\eta$ value of 5 $\times 10^{-5}$. 

\subsubsection{The case of non-metric theories}
\label{sec:wep}

As discussed in Sec. \ref{sec:PPN}, metric theories satisfy the WEP because the additional gravitational fields (beyond the metric) do not couple directly to matter.
However, one could question the reason why additional gravitational fields would not couple to matter. Indeed, at the fundamental level, there does not seem to be any reason to expect that outcome \citep{2012CQGra..29r4001D}. Moreover, extra-gravitational fields with non-minimal couplings to matter are ubiquitous in various attempts to unify the whole set of fundamental interactions in physics---such as in Kaluza-Klein or superstring theories \citep{2012CQGra..29r4001D}, for instance. Finally, such non-minimal couplings can also appear from radiative quantum corrections of theories that satisfy the WEP at the classical level \citep{armendariz2012prd}---whereas  the WEP in general relativity is immune to quantum corrections due to the specific symmetries of general relativity \citep{armendariz2012prd}. Therefore, it seems that beyond general relativity, violations of the WEP should be expected at some level---either at the classical level already, or from quantum corrections to the classical action.

\paragraph{Decoupling mechanisms}
Since the  equivalence principle is observed to be satisfied with an ever increasing level of accuracy, non-metric theories need to possess mechanisms that are able to hide the effects of the additional gravitational fields in the solar system. They are often called ``decoupling'' or ``screening'' mechanisms.\footnote{Note that decoupling and screening mechanisms have been proposed first and foremost for theories that do satisfy the WEP, often in order to be able to explain the acceleration of the expansion of the universe without impacting with the well-constrained solar system phenomenology.} They can take their roots from the fact that the additional fields are simply massive enough to propagate over very short distances only{---such as what happens with flux compactification in string theory \citep{douglas:2007rm}---}or from the suppression of the coupling during the evolution of the universe---such as the Damour-Nordtvedt decoupling mechanism \citep{damour1993prd,damour1994npb}---or from the dependence of the effective mass, or coupling, of the additional gravitational field to the local density---such as the chameleon and symmetron mechanisms \citep{khoury:2004pr,hinterbichler:2010pl}---or from non-perturbative effects---such as the Vainstein mechanism \citep{vainshtein:1972pl,babichev:2013cq}---or simply from exact cancellations in the field equations---such as the intrinsic decoupling \citep{minazzoli:2013pr,minazzoli2016prd} that notably appears in entangled relativity \citep{arruga:2021pr}.

\subsubsection{The equivalence principle in a  planet-satellite system}

Considering the generic Eq. (\ref{eq:uffvgen}), we can deduce for the couple of the bodies $T$ and $A$ that the difference between their accelerations toward the Sun $S$ reads
\begin{eqnarray}
\Delta \bm{a} \equiv \bm{a}_A - \bm{a}_T &=& -  \frac{ G {\bar m_{AT}}}{r_{TA}^3}\bm r_{TA}+   G m^G_S\left[ \frac{\bm r_{ST}}{r_{ST}^3}-\frac{\bm r_{SA}}{r_{SA}^3} \right] \nonumber \\
&&+   G  m^G_S\left[ \frac{\bm r_{ST}}{r_{ST}^3} (\delta_T+\delta_{ST})- \frac{\bm r_{SA}}{r_{SA}^3} (\delta_A+\delta_{SA}) \right], \label{eq:EMacc}
\end{eqnarray}
with ${\bar m_{AT}}\equiv  m^G_A +  m^G_T+ (\delta_T+\delta_{TA})m^G_A+(\delta_A+\delta_{TA}) m^G_T$. 
For gravitationally bounded systems like the Earth-Moon system, one can further approximate both distances appearing in this last term as being approximately equal, such that the relevant term in order to check a potential violation of the EP through {a violation of} the universality of free fall (UFF)  is
\be
\Delta \boldsymbol{a}^{\overline{\mathrm{UFF}}} \approx \boldsymbol{a}_{\mathrm{E}} \Delta_{\mathrm{ESM}} \label{eq:approxWEP}
\ee
where $E$, $S$ and $M$ stand for the Earth, the Sun and the Moon respectively, and with
\be
\Delta_{\mathrm{ESM}}=\left[\left(\delta_{\mathrm{E}}+\delta_{\mathrm{SE}}\right)-\left(\delta_{\mathrm{M}}+\delta_{\mathrm{SM}}\right)\right].
\ee
Eq. (\ref{eq:approxWEP}) was introduced in \citep{williams:2012cq} for theories that violate the GWEP only, for which $\Delta_{\mathrm{ESM}} = \Delta_{\mathrm{EM}}$ such that
\begin{equation}
    \Delta_{EM} = \left[ \left(\frac{m^G}{m^I}\right)_E - \left(\frac{m^G}{m^I}\right)_M \right]. \label{eq:deltaSEP}
\end{equation}
In that case, Eq. (\ref{eq:approxWEP}) becomes
\begin{equation}
\Delta \boldsymbol{a}^{\overline{\mathrm{UFF}}} \approx \boldsymbol{a}_{\mathrm{E}} \Delta_{\mathrm{EM}}, \label{eq:approxSEP}
\end{equation}

\subsubsection{The equivalence principle at planetary scale}
\label{sec:EPNbody}

At planetary scale, Eqs. (\ref{eq:approxWEP}) or (\ref{eq:approxSEP}) are not valid anymore. {Indeed, p}lanets are at different distances from the Sun and EP tests cannot be limited to the differences between two accelerations. In that case, one has to use the more general Eq. (\ref{eq:uffvgen}) instead.

A confusion can occur when defining the SSB because the gravitational parameters that appear in the equations of motion are constructed with the gravitational masses; whereas the gravitational parameters that are used to define the SSB should be constructed with the inertial masses, as pointed out in \cite{2002PhRvD..66h2001M} and \cite{genova2018nc}---see also Sec. \ref{sec:dilaton}. It was an issue in particular for previous versions of planetary ephemerides, such as \cite{DE421}, where the SSB positions and velocities were estimated at each step of integration with the gravitational parameters appearing in the equations of motion.

However, as explained in Sec. \ref{sec:SSB}, in planetary ephemerides such as INPOP08 \citep{fienga:2008aa} and {the} followings, the SSB position and velocity are estimated once, before integrating the motion of planets, using planetary initial conditions, inertial masses and Eqs. (\ref{eq:SSB}) and (\ref{eq:SSB2}). As explained in Sec. \ref{sec:dilaton}, \cite{bernus2022} {have} checked with a specific theory that violates both the WEP and the GWEP that those equations indeed remain the ones that characterize the SSB at the required level of accuracy for planetary ephemerides---provided that inertial masses are used for the definitions and not gravitational masses---as originally suggested in \cite{2002PhRvD..66h2001M, genova2018nc}.

\subsubsection{Comment on the mass in the Shapiro delay equation}
\label{sec:mass_shapiro}

{As soon as the SEP is violated, there is a potential ambiguity about the gravitational parameter that appears in the Shapiro delay Eq. (\ref{eq:shapiroPPN}). Indeed, the gravitational parameter in the Shapiro delay could either be constructed on the inertial mass, or on the gravitational mass. As we will see in Sec. \ref{sec:dilaton} Eqs. (\ref{eq:shapiro_dilaton}-\ref{eq:grav_params}), one can derive from first principles that the gravitational parameter that appears in the Shapiro delay is the one built upon the inertial mass, that is $\mu^I_A = G m^I_A = (1-\delta_A) G m^G_A$, with :
\bea
c(t_r-t_e)&=&R+\sum_A(\gamma+1)\frac{\mu^I_A}{c^2}\ln\frac{\bm{n}\cdot\bm{r}_{rA}+r_{rA}+2\frac{(1+\gamma)\mu_A}{c^2}}{\bm{n}\cdot\bm{r}_{eA}+r_{eA}+2\frac{(1+\gamma)\mu_A}{c^2}} \, , \\
&=&R+\sum_A(\gamma+1)\frac{\mu^G_A}{c^2} \left(1 - \eta  \times \frac{3}{5} \frac{\mu^{G}_A}{c^{2} R_A} \right) \times \\
&&\qquad \qquad \ln\frac{\bm{n}\cdot\bm{r}_{rA}+r_{rA}+2\frac{(1+\gamma)\mu_A}{c^2}}{\bm{n}\cdot\bm{r}_{eA}+r_{eA}+2\frac{(1+\gamma)\mu_A}{c^2}} \, . \nonumber
\eea}
{There is a somewhat simple explanation to that. One must recall that the original Nordtvedt effect was derived for the sum of point particles of (inertial) mass $m$ that are gravitationally bounded together. The metric is bent by those masses $m$, and the sum of all the masses correponds to the inertial mass of the overall body that is composed from all thoses masses, to lowest order \citep{nordtvedt1968pr2}. Since the Shapido delay simply derives from the null condition $ds^2=0$ for light, the Shapiro delay must depend on this sum of inertial masses. What the Nordtvedt effect tells additionally, is that the center of mass of this collection of masses does not follow the same trajectory as would an inertial mass, but follows the trajectory given by the following equation instead
\be
\boldsymbol{a}_{\mathrm{T}}=-\sum_{A \neq T} \mu^G_A \frac{\boldsymbol{r}_{A T}}{r_{A T}^{3}} \left(1 + \eta  \times \frac{3}{5} \frac{\mu^{G}_T}{c^{2} R_T} \right). \label{eq:Nordt}
\ee}

\subsubsection{The issue of time in non-metric theories}

The issue with theories that violate the WEP is that they usually also violate the local position invariance \citep{uzan2011lrr}, such that the behaviour of clocks not only may be modified with respect to general relativity depending on the position in the gravitational field, but also may depend on the clock composition.
Therefore, one may expect that this should complicate the tests of such theories with planetary ephemerides, given that positions of astronomical bodies are given in one of the coordinate times recommended by the IAU, and which are perfectly independent of the composition of the clocks that has been used.
Fortunately, the local position invariance is tested to a very high level of accuracy on Earth with various atomic clocks \citep{rosenband:2008fk,guena:2012ys,leefer:2013xy,godun:2014sf,huntemann:2014nr}. Indeed, thanks to the eccentricity of the Earth's orbit, one can probe a potential variation of the ratio between the frequencies of atomic clock with different compositions depending on the variation of the local value of the Sun's gravitational field.
Because the local position invariance is tested with atomic clocks at a level of accuracy that is far beyond what is used in space navigation, one can neglect this potential effect in planetary ephemerides.

\subsection{Variation of the gravitational constant G}
\label{sec:mudot}
{With respect to the subject of the potential variation of constants \citep{uzan2011lrr}, planetary and lunar ephemerides have been mostly used for testing an hypothetical variation of Newton's constant $G$, or a related value, the Astronomical Unit \citep{1983PhRvL..51.1609H, 1987Icar...71..337A, 1989AdSpR...9i..71A,1993CeMDA..55..313P}.}

Before 2012 and the decision by the International Astronomical Union  \citep{2012IAUJD...7E..40C} to redefine the Astronomical Unit (AU)  as a constant {with a} fixed value, the AU was indeed part of the parameters estimated during the construction of planetary ephemerides. The Gauss constant $k$  was fixed \citep{Standish2001} and the AU was estimated from this equation 
\begin{equation}
GM_{\odot} = (AU)^{3}k^{2}/D^{2} 
\end{equation}
where D is the length of the day  and $GM_{\odot}$ is the gravitational parameter of the Sun. 
In this context, AU values and its {hypothetical} time variations have been estimated with planetary fit  by \cite{2004CeMDA..90..267K}. 
In the same manner,  {an estimation of} $\frac{\dot{G}}{G}$ was also obtained, together with the value of AU. A first test of such a {variation} has been initiated by \cite{1983PhRvL..51.1609H}, using the data obtained with the Viking mission on Mars.  On a regular basis, the  planetary ephemerides  were then published with updates for $\frac{\dot{G}}{G}$ based on a global adjustment including  also AU and PPN parameters \citep{1993CeMDA..55..313P, Standish2001, 2005AstL...31..340P}. Results are gathered in Table \ref{tab:GM}.

After 2012, the AU has been fixed to 149597870700~m, closing the door to the estimation of $d{AU}/dt$, but leading to a more consistent determination of   the gravitational parameter of the Sun, {$GM_{\odot}$} noted {$\mu_{\odot}$}, now fitted in planetary ephemerides.
As it is reviewed in \cite{uzan2011lrr}, some theories propose that the gravitational constant can vary with time. Thus, { an hypothetical variation} of $G$ in time {has} been introduced in the planetary ephemerides, but it is not straightforward since it affects both the planetary and the Sun gravitational parameters $\mu$.
. 


For the planets, {because} they do not {lose or gain mass}, the time variation of G induces a direct variation of the gravitational parameter $\mu_P$ such $\dot{\mu}_{P}/\mu_{P} = \dot{G}/G$. For the Sun, its intrinsic mass loss $\dot{M_{\odot}}$ has also to be considered. {As a consequence,} the following {equations are} added to the system of equations integrated numerically during the planetary ephemerides construction
\bea
\frac{\dot{\mu}_\odot}{\mu_\odot} &=& \frac{\dot{G}}{G} + \frac{\dot{M_{\odot}}}{M_{\odot}},
\label{mudotdefSun}
\eea
where ${\mu}_\odot$ is the gravitational paramater of the Sun and
\bea
\frac{\dot{\mu}_{P}}{\mu_{P}} &=& \frac{\dot{G}}{G},
 \label{mudotdef}
\eea
with $\mu_{P}$ is the gravitational parameter of bodies other than the Sun.
\\

At each step $t$ of the numerical integration  of the planetary equations of motion, the following quantities are estimated :
{\begin{eqnarray}
&&M_{\odot}(t) = M_{\odot}(t_0) + (t-t_0) \times \dot{M_{\odot}}, \\
&&M_{P}(t) = M_{P},\\
&&G(t) = G(t_0) + (t-t_0) \times \dot{G},\\
&&\mu(t)= G(t)  \times M(t),
\label{mudot}
\end{eqnarray}}
where $t_0$ is the date of the origin of the planetary ephemeris.
{Let us note that the gravitational parameter $\mu(t)$ also appears} in the computation of the Shapiro delay (see sect \ref{sec:shapiro}). 
In this case, the value of $\mu(t)$, corresponding to the date of the observation, is computed with Eq. (\ref{mudot}) and re-introduced in the {Shapiro} Eq. (\ref{eq:shapirogeneral relativity}).

$M_{\odot}(t_0)$ is the mass of the Sun fitted during the construction of the ephemeris,
and $G(t_0)$ {is} the Newtonian gravitation constant as defined by the IAU \citep{2011CeMDA.110..293L}. 
The effect of the time variation of G being largely induced by the gravitational parameter of the Sun more that the ones of the planets, one then deduces {the value} of $\dot{G}/G$ by considering its impact of the Sun contribution (Eq. \ref{mudotdefSun}) and a fixed value for the Sun total mass loss, $\frac{\dot{M_{\odot}}}{M_{\odot}}$. 
\cite{2012SoSyR..46...78P} have proposed an interval of values for the total solar mass loss  of  
\be \label{eq:PPmest}
\frac{\dot{M_{\odot}}}{M_{\odot}} = (-0.67 \pm 0.31) \times 10^{-13} { {\quad (2\sigma)}} \, \textrm{yr}^{-1},
\ee

This estimation considers the mass {loss} by the 
Sun, but also the mass {gained} by falling materials (comets, asteroids etc...). 
 \cite{2011ApJ...737...72P}  have estimated a mean mass loss from wind emission of charged particles during the 11-year solar cycle :
 \be
\frac{\dot{M_{\odot}}}{M_{\odot}} = (-0.55 \pm 0.15) \times 10^{-13} \, { {\quad (3\sigma)}} \, \textrm{yr}^{-1} ,
\ee
instead of {Eq. (\ref{eq:PPmest})}.
In 2021, a detailed evaluation of  \cite{pitjeva2021} {gives} 
\begin{equation}
\frac{\dot{M_{\odot}}}{M_{\odot}}= { {(-1.105 \pm 0.235) \times 10^{-13} \quad (3\sigma)}}\, \textrm{yr}^{-1} ,
\end{equation}
This value accounts for the solar wind radiation as well as the accumulation of interplanetary dust falling on the Sun together with comets. The authors conclude {that the value of $\frac{\dot{M_{\odot}}}{M_{\odot}}$ in \cite{2012SoSyR..46...78P} is overestimated}

On Table \ref{tab:GM}, are gathered {the values of }$\dot{\mu_\odot}/{\mu\odot}$ obtained by different authors  jointly with PPN parameter estimations as well as the $\dot{G}/G$ deduced {using the} value of $\frac{\dot{M_{\odot}}}{M_{\odot}}$ specified in the same Table. 
It is important to stress that the correlations between the PPN parameters, $\dot{\mu_{\odot}}/{\mu_{\odot}}$ and the other fitted parameters of the planetary ephemerides being non zero (see i.e. Table 6 from \cite{fienga2015cm}), the values given in Table \ref{tab:GM} have to be considered as part of a global fit, and consequently {they are planetary ephemeris dependent}. 
Direct adjustments with all parameters fitted together \citep{2013MNRAS.432.3431P,pitjeva2021},  partial fits \citep{konopliv:2011ic, fienga2022} but also  random walk exploration algorithms have been used for obtaining constraints on $\dot{\mu_{\odot}}/{\mu_{\odot}}$, given in Table  \ref{tab:GM}. 
As explained in Sect \ref{sec:fit} and \ref{sec:PPN}, the partial fit consists {to fix} one PPN parameter and {then fit} $\dot{\mu_{\odot}}/{\mu_{\odot}}$ together with the rest of the planetary parameters. 
The random walk exploration algorithm fixes the PPN parameters  and $\dot{\mu_{\odot}}/{\mu_{\odot}}$ according to random values, and performs a regular fit for the rest of the planetary parameters. The obtained ephemerides are then selected according to different statistical criteria \citep{fienga2015cm, fienga2022}.
The {latest two} approaches (partial fit and random walk exploration algorithms) give larger intervals of possible values than the direct fit of all parameters, as one can see {in} Table \ref{tab:GM}.

One can also mention $\dot{G}/G$ estimations deduced, not directly from planetary ephemerides, but from the adjustment of one single planet orbit during the analysis of radar tracking data analysis of one given space mission. This is the case for example with Mercury and the MESSENGER data operated by \cite{genova2018nc}, noted {\it{MSG}} in Table \ref{tab:GM}. In this work, it has to be stressed that only Mercury orbit was considered in the analysis.
Nevertheless, in all the cases produced in Table \ref{tab:GM}, no clear indication of {a} time variation of G is visible, despite the improvements {of} the planetary ephemerides and {of} the estimation of the solar mass loss \citep{pitjeva2021}.
Finally, other technics or methods have been used for measuring possible variations of the gravitational constant, G. One can cite for example, the determination of  $\dot{G}/G$ deduced from pulsar timing \citep{2003LRR....6....5S,2015ApJ...809...41Z, Zhu2018, kramer:2021px}. The obtained limit is then one order of magnitude greater (i.e. {$\lvert \dot G/G \rvert < 0.9 \times 10^{-12}$ yr$^{-1}$} from \cite{Zhu2018}) than the one obtained with planetary ephemerides although the gravitational regimes are quite different (strong field regime for the pulsar measurements and weak field for planetary ephemerides). 
\begin{table}
\caption{Values of  $\dot{\mu_{\odot}}/{\mu_{\odot}}$ found in the literature deduced from planetary ephemerides. The Column 3 indicates the type of method used for the estimation: either the full fit (FF), the partial fit (PF) in  or the random walk exploration (RW). See text for details. For \cite{fienga2022},  PF$^{**}$ indicates that the oblateness of the Sun is strongly constrained by helio-seismological value. So this solution cannot be seen as a free fit but as a partial fit. Analysis of the MESSENGER mission is labeled MSG in Column 3 using only the Mercury orbit. The $\dot{G}/G$  values are deduced with $\dot{\mu}/{\mu}$ and  ${\dot{M_{\odot}}}/{M_{\odot}}$ given in Column 5. The value of $\dot{G}/G$ deduced from \cite{2013MNRAS.432.3431P} is not the one published by the authors but was obtained in using the same value of ${\dot{M_{\odot}}}/{M_{\odot}}$  as for \cite{konopliv:2011ic}.}
\begin{tabular}{l l l l l l}
\hline
Reference & PE & Method & $\dot{\mu_{\odot}}/{\mu_{\odot}}$  & $\frac{\dot{M_{\odot}}}{M_{\odot}}$ &  $\dot{G}/G$  \\
& & & $\times$ 10$^{14}$ yr$^{-1}$ & $\times$ 10$^{14}$ yr$^{-1}$& $\times$ 10$^{14}$ yr$^{-1}$\\
\hline
Before AU fixed & & & \\
\cite{1993CeMDA..55..313P} & EPM1988 & FF & NA & NA & 470 $\pm$ 470 \\
\cite{2001CeMDA..80..249P} & EPM2000 & FF & NA & NA & 4 $\pm$ 8\\
\cite{Standish2001}& DE405 & FF &  NA & NA & 1 $\pm$ 8\\
\cite{2005AstL...31..340P} & EPM2004 & FF & NA & NA & -2  $\pm$ 5 \\
\hline
After AU fixed & & & \\
\cite{konopliv:2011ic} & DE421 & PF & 1 $\pm$ 16 & -9.2 $\pm$ 6.1 & {10.2 $\pm$ 22.1}$^{*}$ \\

\cite{2012SoSyR..46...78P} & EPM2010 & FF & -5 $\pm$  4 & -6.7 $\pm$ 3.1 & {1.65 $\pm$ 8.77} \\

\cite{2013MNRAS.432.3431P} & EPM2011 &  FF & -6.3 $\pm$ 6.4 &  -6.7 $\pm$ 3.1 & 0.4 $\pm$ 11.1 \\ 
\cite{fienga2015cm} & INPOP15a & FF & -5.0 $\pm$ 2.9 & -9.2 $\pm$ 6.1 & 4.2 $\pm$ 9.0  \\
 &  & RW & -4.3 $\pm$ 7.4 &  -9.2 $\pm$ 6.1 & 4.9 $\pm$ 13.5  \\
\cite{genova2018nc} & & MSG & -6.130  $\pm$ 1.47 & -10  $\pm$ 1 & 4 $\pm$ 7.5 \\
 \cite{pitjeva2021} & EPM2021 & FF & -10.2 $\pm$ 1.4 & -11.05  $\pm$ 2.35 & 0.85 $\pm$ 3.75 \\
\cite{fienga2022} & INPOP20a & PF$^{**}$ & -8.8 $\pm$ 2.9 & -9.2  $\pm$  6.1 &  0.4 $\pm$ 9.0\\
& & & & -11.05  $\pm$ 2.35 & 2.25 $\pm$ 5.25 \\
& & RW & -10.3 $\pm$ 22.8 &  -9.2  $\pm$ 6.1 & -0.8 $\pm$ 28.4  \\ 

\hline
\end{tabular}
\label{tab:GM}
\end{table}

\subsection{Massive gravity}
\label{sec:graviton}

Unlike in electromagnetism---- like with the Procca theory \citep{derham:2014lr,1936JPhyR...7..347P}----, there is not a unique definition for what massive gravity might entail \citep{derham:2014lr}. In field theory, massive interactions typically result in a Yukawa suppression of these interactions at the scale of the Compton wavelength. However, due to its tensorial rather than vectorial nature, this may not necessarily be the case for a fully consistent theory of massive gravity \citep{derham:2014lr}. Nonetheless, from a phenomenological perspective, it is possible to test whether a Yukawa suppression of gravitational potentials occurs within the solar system.
Formally, this would lead to the following modification of the Newtonian potential \citep{will:2018cq}
\begin{equation}
w=w_{\textrm{Newton}}\mathrm{exp}(-r/\lg),
\label{eq:massive}
\end{equation}
which can be developped as \cite{bernus2019}
\be
w=w_{\textrm{Newton}} \left(1+\frac{1}{2} \frac{r^2}{\lg^2} \right)+\Ol{(\lg^{-3})},\label{eq:massivepert}
\ee
after a convenient change of coordinate system that absorbs the constant term in the gravitational potential---which has no impact on the observables.

As discussed in Sect~\ref{sec:ISL}, this modification is different from a \textit{fifth force}, for which the new potential is an affine function of the Yukawa suppression instead of a linear function \citep{fischbach:1992me,will2014lrr}. Fifth forces usually originate from the existence of an additional gravitational field that is massive---e.g. a massive scalar field {\citep{wagoner:1970pr,hees:2018pr}}---rather than considering that the field equation on the metric perturbation itself has a mass term. Indeed, Eq. (\ref{eq:massive}) is solution of a massive gravitational potential equation that reads
\be
\Delta w - \frac{w}{\lg^2}=-4\pi G \sigma. \label{eq:Ysuppr}
\ee

Obviously, as long as $\lg$ is big enough, the gravitational phenomenology in the Newtonian regime can reduce to the one of general relativity to any given level of accuracy. Also, if $\lg$ is large enough that only the leading order correction in Eq. (\ref{eq:massivepert}) has a significant contribution to the metric, one can assume that only the Newtonian part of the post-Newtonian expansion is modified with respect to the equations of motion in general relativity Eq. (\ref{eq:accgeneral relativity}). \\
In that situation, the equation of motion only has one extra term with respect to the usual EIHDL equation Eq. (\ref{eq_eih}) that reads {\cite{bernus2019}}
 \bea
 \delta \bm a_{A}^{\lg} = \frac{1}{2\lg^2} \sum_{A\neq T} \frac{\mu_T}{r_{AT}}\bm r_{AT} + \Ol(\lg^{-3}), \label{eq:accgravmas}
\eea
and further assuming that light still propagates along null geodesics, the Shapiro delay reads {\citep{bernus2019}}

	\bea\label{eq:shapiro_lg}
			c(t_r-t_e)&=&c(t_r-t_e)_{GRT} \\
&&+ \sum_{A}\frac{\mu_A}{c^2}\frac{1}{2 \lg^2} \ln \left[ b^{2}\frac{\bm{n}\cdot\bm{r}_{rA}+r_{rA}}{\bm{n}\cdot\bm{r}_{eA}+r_{eA}} + \bm{n}\cdot (r_{rA}\bm{r}_{rA}-r_{eA}\bm{r}_{eA}) \right] \, ,\nonumber
		\eea
where $c(t_r-t_e)_{GRT}$ corresponds to the general relativity light time given in Eq. (\ref{eq:shapirogeneral relativity}), $b$ is the minimal distance between the light path and the central body (here the Sun). This expression is an approximation at {c$^{-2}$} level, the additional terms induced by gravitational field mass being negligible relative to the present day accuracy for commonly admitted $\lg$ (with $\lg > 2.8 \times 10^{12}$ km \citep{will:2018cq, will:1998pd}).

Likewise, from Eq. (\ref{eq:massivepert}), the difference between a clock $A$ and a BCRS time $t$ still is
\begin{equation}
\frac{d \tau_{A}}{d t}=1-\frac{1}{c^{2}}\left[v_{A}^{2} / 2+U \left(\mathbf{x}_{A}\right)\right], \label{dtauAdt}
\end{equation}
up to terms of order $ \Ol{(\lg^{-2} c^{-2})}$. For instance, assuming a conservative bound of $\lg > 2 \times 10^{12}$ km, the correction to Eq. (\ref{dtauAdt}) at the surface of the Earth would be less than $10^{-17} (\sim R_\oplus^2/\lg^2\textrm{, where } R_\oplus\textrm{ is the radius of the Earth})$ time less than the contribution of the last term in the bracket of Eq. (\ref{dtauAdt})---that is, far beyond what can be achieved with present clocks stability \citep{guena:2012ie}.

{By analogy with standard quantum physics, the} Compton length can also be interpreted in terms of {a} mass of the graviton $m_g$ following the relation:
\bea\label{eq:mglg}
\lg=\frac{\hbar}{c m_g},
\eea
with $\hbar$ the Planck constant, and $c$ the speed of light.
In 2018, \cite{will:2018cq} proposed to use solar system ephemerides to improve the constraints on $\lg$ in the Newtonian limit.  The starting point was that a massive gravitational field should lead to a modification of the perihelion advance of solar system bodies. Hence, based on current constraints on the perihelion advance of Mars derived from Mars Reconnaissance Orbiter (MRO) data, Will estimates that the Compton wavelength should be bigger than $(1.4 - 2.7) \times 10^{14}$km (resp. $m_g < (4-8) \times 10^{-24}$ eV/$c^2$), depending on the specific analysis. However, making an estimation from quantities derived from ephemerides that assumed other theorerical frameworks---here, the perihela advances per orbit and their uncertainty in either general relativity or PPN frameworks---cannot account for the fact that the mass of the graviton is correlated to the various parameters of the ephemeris (eg. masses, semi-major axes etc.). While a graviton with a non-zero mass may impact the solar system dynamics, so also does a change of the various other parameters of the ephemerides. Because of the correlation between $m_g$ (or $\lg$) and other parameters, any modification induced by a non-null value of $m_g$ may---at least in part---be reabsorbed by the modification of other parameters of the ephemerides.

In order to overcome this issue, \cite{bernus2019,2020PhRvD.102b1501B} and \cite{2023arXiv230607069M} have built planetary ephemerides fully developed in the massive gravity framework of Eq. (\ref{eq:massive}) and fitted over the data sample of INPOP17a , INPOP19a and INPOP21a respectively. The results of these investigations are given on Table \ref{tab:graviton}. 

\cite{bernus2019,2020PhRvD.102b1501B} had used a method of random walk exploration that is more conservative that the Monte Carlo Markov Chain (MCMC) algorithm used by \cite{2023arXiv230607069M}. 
With the same random walk exploration method, but using the updated INPOP21a ephemerides, \cite{2023arXiv230607069M}  obtained a constraint that is 3 times smaller than \cite{2020PhRvD.102b1501B}. This improvement is induced by the use in INPOP21a of the latest Juno and Mars orbiter tracking data up to 2020 as well as a fit of the Moon-Earth system to LLR observations also up to 2020. By improving the procedure with MCMC, \cite{2023arXiv230607069M} was able to push the limit of detection of the mass of the graviton at a new level, with a constraint at $1.01 \times 10^{-24} \; eV c^{-2}$ (resp. $\lambda_g \geq 122.48 \times 10^{13} \; km$) with a $99.7\%$ confidence level. 



It is somewhat interesting to compare these constraints to the ones deduced from the observation of gravitational waves. Indeed, it is assumed that a massive gravitational field that leads to Eq. (\ref{eq:Ysuppr}) might also modify the dispersion relation of gravitational waves as follows \citep{GWTC3b,will:1998pd}
\begin{equation}
E^{2}=p^{2} c^{2}+m_g c^2, \label{eq:dispm}
\end{equation}
where
$E$ and $p$ are the energy and momentum of the wave.

Such a modified dispersion relation causes gravitational waves frequency modes to propagate at different speeds, leading to an overall modification of the phase morphology of gravitational waves with respect to the general relativity predictions. Since the morphology of gravitational wave phase has been consistent with general relativity so far, it {led} to severe constraints on the value of $m_g$ that are reproduced on Table \ref{tab:graviton}.

Even if a massive gravity theory {actually} leads to both the phenomenological consequences represented in Eqs. (\ref{eq:Ysuppr}) and (\ref{eq:dispm}), there is absolutely no reason for the constraints from ephemerides on the one hand, and {from} gravitational waves on the other hand, to be at the same level accuracy. 

Each type of constraints is relevant on its own right given that they test different phenomenologies---that is, Eq. (\ref{eq:massive}) versus Eq. (\ref{eq:dispm})---which may (or may not) be related, depending on the underlying massive gravity theory that one is considering. For instance, screening mechanisms that kick-in for high density environments---such as the Vainshtein mechanism \citep{Babichev_2013}---may impact Eq. (\ref{eq:massive}) and not Eq. (\ref{eq:dispm}). This notably seems to be the case for ghost-free massive gravity \citep{derham:2014lr}.

\begin{table}
\footnotesize
\caption{Limits obtained for the Compton length $\lg$ in km as defined in Eqs.  (\ref{eq:accgravmas}) and (\ref{eq:shapiro_lg}). Are also given the corresponding values in term graviton mass $m_g$ in eV$/c^2$.  Are also indicated, for comparisons, the values obtained with INPOP17a \citep{bernus2019}, INPOP19a \citep{2020PhRvD.102b1501B},  and INPOP21a  \citep{2023arXiv230607069M} as well as the estimations for the dynamical mode from Virgo-Ligo  GWTC-1 and GWTC-3 \citep{GWTC1b,GWTC3b}. For INPOP21a, two values at 90$\%$ confidence level (CL) are given : the one indicated in the column {\it{RW}} corresponds to value obtained with the same method (random walk exploration) as \cite{bernus2019} and \cite{2020PhRvD.102b1501B} and the one given in Column {\it{MC}} corresponds to MCMC results. }

     \begin{tabular}{c | l l | c c c c}
         \hline
 & GWTC-1  & GWTC-3 & INPOP17a & INPOP19a  & \multicolumn{2}{c}{INPOP21a}\\
& & & & & RW & MCMC\\
CL & {0.90}    & {0.90}                  & 0.90   & {0.90}  & 0.9 & 0.9 \\
    \hline
       Graviton mass             &     &                   &          & && \\
         $\lg \times10^{-13}$   [km]            &                {2.6} &  {9.77}  & 1.83  &3.93    & 12.01  & 209.67 \\
         $m_g \times10^{23}$   [eV/c$^2$]     & {4.7} & {1.27}  & 6.76 & 3.16   &   1.03  & 0.059 \\ 
    \hline
         Fifth Force                           &                   &           &                                        &&& \\
    $ \frac{\lambda}{\sqrt{\mid \alpha \mid}} \times10^{-13}  $[km], $\alpha > 0$ &         &                            & 1.83 & 3.93 &{} &\\
    $\frac{\lambda}{\sqrt{\mid \alpha \mid}} \times10^{-13} $[km], $\alpha < 0$  &         &                         & & 3.77  & &\\
    \hline
    \hline
    \end{tabular}
        \label{tab:graviton}

\end{table}

\subsection{Yukawa potential and Fifth Force}
    \label{sec:ISL}
    
    Whereas one can imagine that the metric field itself has a mass---see Sec. \ref{sec:graviton}---it is also possible to imagine the existence of additional gravitational fields that are massive, while the metric field would remain massless. In such situations, the potential often\footnote{{But not always \citep{hees:2018pr}}.} becomes an affine function of the Yukawa suppression as follows \citep{wagoner:1970pr,fischbach:1992me,will2014lrr}
    \begin{equation}
    w=w_{\textrm{Newton}}(1+\alpha\mathrm{exp}(-r/\lambda)),
    \label{eq:yukawa}
    \end{equation}
    where $\alpha$ is the strength (relative to gravity) and $\lambda$ the range of the force. {This type of modifications of the Newtonian potential is often referred to as a \textit{fifth force} \citep{will2014lrr}.\\}
    Depending on whether the additional massive field couples universally to matter or not, the fifth force can be either composition-dependent or independent \citep{will2014lrr}. 
    {Solar system tests---such as the ones realised with planetary ephemerides---usually focus on composition independent models}.
In \cite{konopliv:2011ic},  constraints on the Yukawa potential were deduced from the analysis of Mars orbiter tracking data and from the construction of a Mars updated ephemeris. Fig. \ref{fig:yukawa} shows that already in 2011, {the constraints from planetary ephemerides were at the level of the constraints from lunar ephemerides.}
%
 {As explained in \cite{bernus2019}, in the limit $\lambda \gg r$ and with $\alpha > 0$, one can use the constraints}
given in Sect~\ref{sec:graviton} for the test of massive graviton, {in order to deduce constraints on the Yukawa potential.}
These constraints are given in Table \ref{tab:graviton} and Fig. \ref{fig:yukawa}. 
The green lines on this Figure show the improvements of the new generation of planetary ephemerides relative to the one use in \cite{konopliv:2011ic}{, as well as} the new limits obtained at solar system scale. 
{A new limit (labelled {\it{IBC}}) obtained by simulating the introduction of Bepi-Colombo MORE experiments as predicted by \cite{2020CQGra..37i5007D} and \cite{moriond2022} is also indicated.}
{Let note that, because fifth force models in Eq. (\ref{eq:yukawa}) depend on two parameters ($\alpha$ and $\lambda$), whereas the massive graviton models  in Eq. (\ref{eq:massive}) depend on only {one} ($\lambda_g$), the mapping between the two breaks down for small Compton wavelength. This explains why the green lines in Fig. \ref{fig:yukawa} are restricted to the right part of the plot.} 

\begin{figure}
\centering
\includegraphics[scale=0.5]{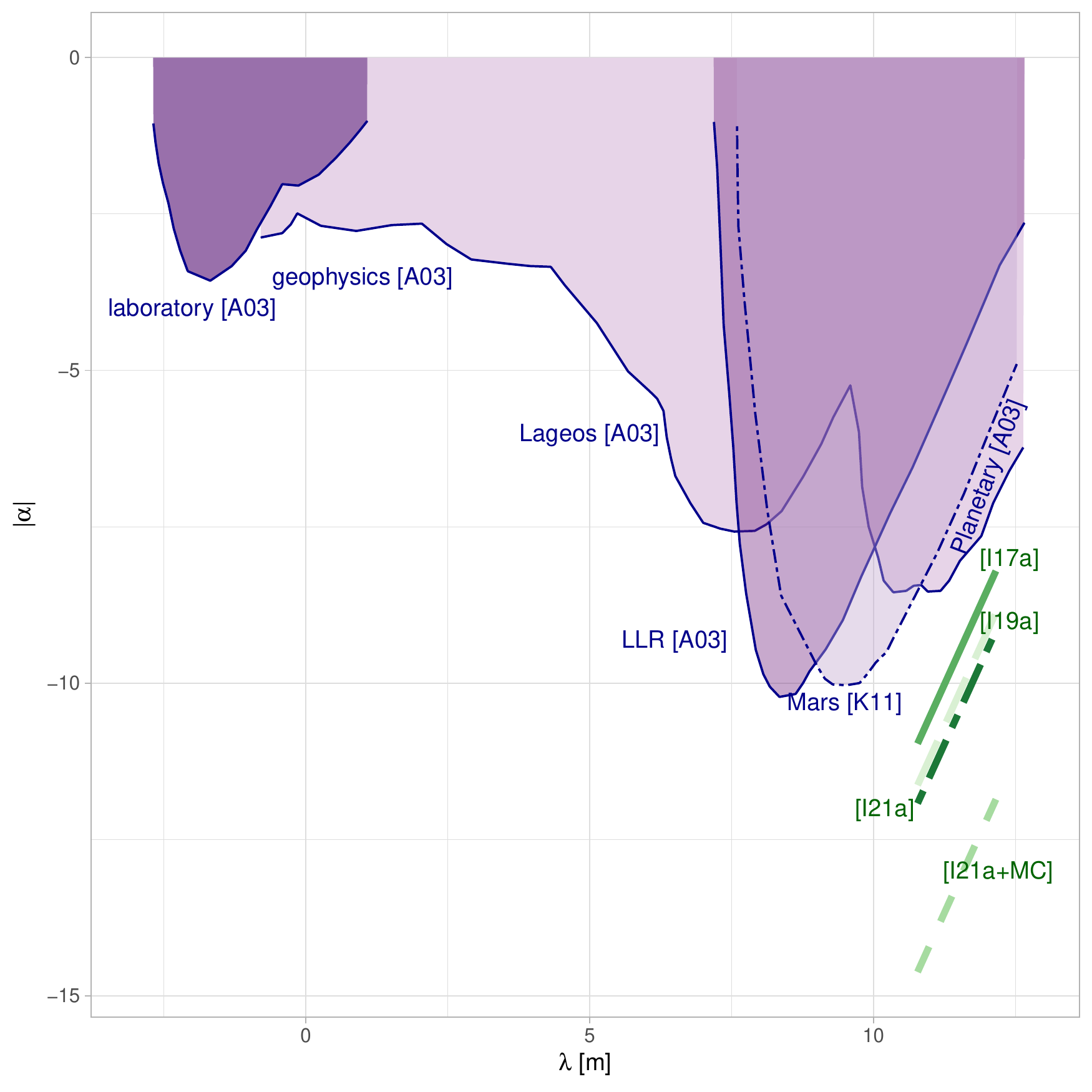}
\caption{Observational constraints obtained for the Yukawa potential extracted from A03, standing for \cite{adelberger2003}. The green lines give the constraints deduced from the INPOP planetary ephemerides graviton tests presented in Table \ref{tab:graviton} : I17a stands for \cite{bernus2019}, I19a for \cite{2020PhRvD.102b1501B}, I21a and I21a+MC for \cite{2023arXiv230607069M}. K11 indicates the limits deduced from Mars tracking data analysis by \cite{konopliv:2011ic}.}
\label{fig:yukawa}
\end{figure}

\subsection{Einstein-dilaton theories}
\label{sec:dilaton}

The case of Einstein-dilaton theories is somewhat very interesting for the phenomology of alternative theories in the solar system, because it allows one to consistently derive the equations of motion, Shapiro delay and conserved quantities in a framework that leads to a violation of both the WEP and the GWEP.

Einstein-dilaton theories are scalar-tensor theories that violate the WEP at the fundamental level, because they possess at least one massless scalar-field that couples non-minimally to matter fields. They are somewhat expected as being the low energy effective gravitational action of string theories \citep{1994GReGr..26.1171D,2002PhRvD..66d6007D}\footnote{Let us note, however, that the current consensus among string theorists is that all the scalar fields acquire a potential through the moduli stabilization mechanism \citep{douglas:2007rm}, such that they mediate gravity with a very small range instead of being of infinite range---{unlike} the part of gravity that is mediated by the metric field.}---and other higher than four dimensional theories \citep{overduin:1997pr}---which generically predict the existence of one (or several) additional scalar-field(s) that mediates gravity: the dilaton field (and the moduli fields that come from the compactification of the extra-dimensions in string theories).
Tests of Einstein-dilaton theories with planetary motions have been proposed by {\cite{damour2011cqg}} based on the equations of \cite{damour2010prd}.  {The first results for the massless dilaton with planetary ephemerides} have been published in \cite{bernus2022}.

\subsubsection{Equations of motion and Shapiro delay}

The general action for an Einstein-dilaton theory\footnote{{With a massless dilaton field. Note that \cite{damour2010prd} assume that the phenomenology they derive is valid for light dilaton fields as well. However, even a minute mass can lead to an entirely different phenomenology in the solar system, as one can check in \citep{hees:2018pr}.}} may be written as \citep{minazzoli2016prd,bernus2022}
\begin{equation}
\begin{aligned}
S\left[\boldsymbol{g}, \psi_{i}, \varphi\right]=& \frac{1}{2 \kappa c} \int\left(f(\varphi) R-\frac{\omega(\varphi)}{\varphi} \varphi^{\mu} \varphi_{, \mu}\right) \sqrt{-g} \mathrm{~d}^{4} x \\
&+\frac{1}{c} \int\left(\mathcal{L}_{\mathrm{SM}}\left[\boldsymbol{g}, \psi_{i}\right]+\mathcal{L}_{\mathrm{int}}\left[\boldsymbol{g}, \psi_{i}, \varphi\right]\right) \sqrt{-g} \mathrm{~d}^{4} x
\end{aligned} \label{eq:actEdilaton}
\end{equation}
where $\mathcal{L}_{\mathrm{SM}}$ is the Lagrangian density of matter described by the standard model of particle physics and $\mathcal{L}_{\mathrm{int}}$ the Lagrangian density of the interactions between the dilaton field $\varphi$ and matter. Such interactions can be parametrized by arbitrary functions of the scalar-field as follows

{
        	\begin{align}
			\mathcal{L}_{int}&=-\frac{D_e(\varphi) \beta_e (e)}{2e}F_{\mu\nu}F^{\mu\nu}-\frac{D_g(\varphi)\beta_3(g_3)}{2g_3}G^a_{\mu\nu}G_a^{\mu\nu} \nonumber\\  
			&-\sum_{i=e,u,d}(D_{m_i}(\varphi)+\gamma_{m_i}D_g(\varphi)) m_i\bar{\psi}_i\psi_i \label{eq_lagrdilaton}
		\end{align}}
where $F_{\mu\nu}$ is the Faraday tensor, $G^a_{\mu\nu}$ is the gluons tensor, {$e$ and $g_3$ are respectively the photons and the gluons coupling constants, $\beta_e(e)=\lambda\partial\ln e/\partial\lambda$ and $\beta_3(g_3)=\lambda\partial\ln g_3/\partial\lambda$ are their respective beta functions relative to the quantum scale invariance violation, where $\lambda$ is the energy scale of the considered physical processes}, $m_i$ is the fermion mass, $\psi_i$ their spinor, and $\gamma_{m_i}=-\lambda\partial\ln m/\partial \lambda$ is the beta function relative to the dimensional anomaly of the fermion masses coupled to the gluons. The $D_i(\varphi)$ functions describe the different couplings between the matter fields and the dilaton. This Lagrangian density is a straightforward non-linear generalisation of the action considered by  \cite{damour2010prd}.    \footnote{{Although note that in \cite{damour2010prd}, the dilaton fields does not couple to all the trace terms---that is, the classical part of the trace in addition to all the relevant quantum trace anomalies. However, it has been shown in \cite{nitti2012prd} that it is much more convenient to consider the parametrization in Eq.(\ref{eq_lagrdilaton}) because it recovers the fact that in the limit of metric theories, the dilaton field couples to the total trace, as it should. Indeed, with this parametrization, metric theories corresponds to $D_i=D_j \forall i,j$. The fact that in metric theories, any gravitational scalar degree of freedom must couple to the total trace is a property of conformal couplings. {This is consistent to the fact that the mass of a composite object equals the total trace of the fields that compose the particle due to the constraint that the internal stresses all vanish (this is true even if some of the internal forces do not contribute to the trace, such as classical electromagtetism)  \citep{nitti2022x}.} 
}}

{Assuming a linear coupling---such as in \cite{damour2010prd}---and at leading order in the composition dependent effects, the acceleration reads \citep{bernus2022}}
		  
		  \begin{align}
			\bm {a}_T=&-\sum_{A\neq T} \frac{\mu_A}{r_{AT}^3}\bm r_{AT}\left(1+\delta_T+\delta_{AT}\right)  
			 -\sum_{A\neq T} \frac{\mu_A}{r_{AT}^3c^2}\bm r_{AT}\Bigg\{\gamma v_T^2 +(\gamma+1)v_A^2 \nonumber\\  
			 &-2(1+\gamma)\bm v_A.\bm v_T  -\frac{3}{2}\left(\frac{\bm r_{AT}.\bm v_A}{r_{AT}}\right)^2-\frac{1}{2}\bm r_{AT}.\bm a_A -2\gamma\sum_{B\neq T}\frac{\mu_B}{r_{TB}} +\sum_{B\neq A}\frac{\mu_B}{r_{AB}}\Bigg\} \nonumber \\
			&+\sum_{A\neq T}\frac{\mu_A}{c^2r_{AT}^3}\left[2(1+\gamma)\bm r_{AT}.\bm v_T-
			(1+2\gamma)\bm r_{AT}.\bm v_A\right](\bm v_T-\bm v_A)  \nonumber\\  
			&+ \frac{3+4 \gamma}{2}\sum_{A\neq T} \frac{\mu_A}{c^2r_{AT}}\bm a_A 
			\label{eq_eihmod}
		\end{align}
Eq. (\ref{eq_eihmod}) depends on  $\gamma=(1-\alpha_0^2)/(1+\alpha_0^2)$  and $\delta_A=\delta_A^d+\delta_A^N$ with $\delta_A^d=\alpha_0\tilde{\alpha}_A/(1+\alpha_0^2)$ and $\delta_A^N=(\gamma-1)\omega_A\mid/\tilde{m}_Ac^2$ where $\omega_A$ is the self-gravitational energy of the body A. 
    The fundamental parameters on top of which they are built are $\alpha_0$, the universal coupling constant, and $\tilde{\alpha}_A=d_{\hat{m}}Q_{\hat{m}}^A+d_{\delta m}Q_{\delta m}^A+d_{m_e}Q_{m_e}^A+d_eQ_e^A$, where  $Q_{\hat{m}}^A$, $Q_{\delta m}^A$, $Q_{m_e}^A$, and $Q_e^A$ are the dilatonic charges, estimated according to the composition of the considered bodies. In these equations, $A$ stands for the planet to consider, $\mu_A$ being its gravitational parameter.
        \cite{bernus2022} simplify the problem by considering two average charges for the telluric and gaseous bodies only, because of {the} similar dilatonic charges for these two classes of objects. This approximation leads to the reduction of the number of the tested parameters from 10 to 3 for the linear coupling case: $\alpha_{0}$, {$\tilde{\alpha}_{T}$, $\tilde{\alpha}_{G}$, for which $T$ and $G$ stand for telluric and gaseous bodies respectively. Let note that $\mu_A$ is here the product of $G$ with the gravitational mass $m^G_A$. \\}
From there, one has to also account for different effects, such as the Nordvedt effect that (in the case of the linear  coupling)  reads $\delta_A^N=- (1-\gamma)\frac{3\mu_A}{5R_Ac^2}$ with $R_A$ the planet radius, or the modification of the time travel that, at the required level of accuracy, reads \citep{bernus2022}
	\begin{equation}
{c(t_r-t_e)=R+\sum_A (1+\gamma-\delta_A)\frac{\mu_A}{c^2}\ln\frac{\bm{n}\cdot\bm{r}_{rA}+r_{rA}+\frac{4\mu_A}{c^2}}{\bm{n}\cdot\bm{r}_{eA}+r_{eA}+\frac{4\mu_A}{c^2}} \, .}
\label{eq:shapiro_dilaton}
		\end{equation}
Let us note that
\be
(1+\gamma-\delta_A) \mu_A = (1+\gamma) \mu^I_A, \label{eq:grav_params}
\ee
at the required level of accuracy, where $\mu_A^I$ is the gravitational parameter constructed with the inertial mass, such that $\mu^I_A = (1-\delta_A) \mu_A$. This means that the mass involved in the Shapiro delay is the inertial mass and not the gravitational mass{---as already discussed in Sec. \ref{sec:mass_shapiro}}. 

\subsubsection{Conserved quantities and the definition of the SSB}

From the Lagrangian formulation of the equations of motion, \cite{bernus2022} show that the following barycenter constant vector is a first integral of the equations of motion 
\begin{equation}
			\bm{q}=\bm{G}-\bm{V}t ,\label{const_lag_dil}
\end{equation}
where
		\begin{equation}
			\bm{G}=\frac{c^2}{h}\sum_A\mu_A\bm{z}_A\left(1-\delta_A+\frac{v_A^2}{2c^2}-\frac{1}{2c^2}\sum_{B\ne A}\frac{\mu_B}{r_{AB}}\right) \label{bary_lag_dil}
		\end{equation}
		are the coordinates of the relativistic barycenter of the system and
		\begin{equation}
			\bm{V}=\frac{c^2\bm{P}}{h}
		\end{equation}
		is the velocity of the barycenter motion. $h$ is the conserved energy---whose value does not affect what follows but can be found in \cite{bernus2022}---and $\bm{P}$ is the conserved linear momentum that reads
		\begin{align}
			\bm{P}&=\sum_A\bm{p}_A \nonumber\\
			&=\sum_A\mu_A\bm{v}_A\left[ 1-\delta_A + \frac{1}{2c^2}\left( v_A^2 - \sum_{B\ne A}\frac{\mu_B}{r_{AB}} \right) \right] \nonumber\\
			&\quad- \frac{1}{2c^2}\sum_A\sum_{B\ne A}\frac{\mu_A\mu_B}{r_{AB}}(\bm{n}_{AB}\cdot\bm{v}_A)\bm{n}_{AB}. \label{eq_momlin}
		\end{align}
From Eq. (\ref{eq_momlin}), one can see that the sum at leading order is over the gravitational parameter based on the inertial mass $\mu_A^I$, because one has $\mu_A^I = \mu_A (1-\delta_A)$ at leading order. This can be used in order to check what has been discussed in Sec. \ref{sec:EPNbody}. {Let us note that it is consistent with \cite{klioner2016parametrized}, and with \cite{1995PhRvD..52.4455D} in the $\delta_A=0$ limit.}

\subsubsection{Results}
    
		        In \cite{bernus2022}, the planetary ephemerides used were  INPOP19a \citep{fienga2019inpop} and, as in \cite{fienga2020aa}, after integrating Eq. (\ref{eq_eihmod}) for all bodies, the orbits are adjusted to planetary observations and  statistical criteria (cost functions) are applied for selecting the distribution of the tested parameters for which deduced ephemerides are compatible with instrumental uncertainties. 
The obtained results for the linear coupling parameters are given in Table \ref{tab:dilatonres}.
 	
	\begin{table}
	\caption{Intervals of possible values for the 3 dilaton parameters as defined in \cite{bernus2022}: $\alpha_0$, the universal coupling,  $\alpha_T$ the telluric planet coupling and $\alpha_G$ the gazeous planet coupling. 
	}
	\label{tab:dilatonres}
	\centering
		\begin{tabular}{lrr }
		\hline
		&   \multicolumn{2}{c}{INPOP19a}  \\
  & \multicolumn{2}{c}{ \citep{bernus2022}} \\
			 Confidence:      & 90\%            & 99.5\%   \\
			\hline
			$\alpha_0(\times 10^{5})$  & $-0.94\pm 5.35$ &$1.01\pm23.7 $  \\
			$\alpha_T(\times 10^{6})$  & $0.24\pm1.62$    & $0.00\pm24.5 $ \\
			$\alpha_G(\times 10^{5})$  & $0.01\pm4.38$   & $-1.46\pm12.0 $  \\
			\hline
			$(\gamma-1) \times 10^{{8}}$ & 0.2$ \pm$ 6  & $0.2 \pm 11.2$ \\
			\hline
					\end{tabular}
		\end{table}
	
	At 3-$\sigma$, \cite{bernus2022} obtained constraints at the level of 10$^{-4}$ for $\alpha_0$ and $\alpha_G$ and 10$^{-5}$ for $\alpha_T$. These results reflect the better accuracy reached for the telluric planets in planetary ephemerides thanks mainly {to} Mars orbiters. They should be improved by the future Bepi-Colombo measurements of Mercury orbit.

\subsection{MOND}
\label{sec:mond}

The modified Newtonian dynamics (MOND) 
{is a framework that modifies Newtonian dynamics $\bm{a}=\bm{g}_N$ (where $\bm{a}$ is
the acceleration of a test particle and $\bm{g}_N$ is the Newtonian gravitational field) by  $\bm{a}=\bm{g}$ with \citep{1983ApJ...270..365M,Milgrom:2014}
\be
\mu ~\bm{g} = \bm{g}_N,
\ee
where the interpolating function $\mu$ is a function of the ratio $g/a_0$ between the norm of the gravitational field $g$ and the MOND acceleration scale $a_0$. 
It was developed as an alternative explanation of the galactic rotation curves and the empirical Tully–Fisher relation without relying on dark  matter haloes \citep{1983ApJ...270..365M}.} 
Several attempts to verify this framework at scales different from the galactic one have been made in the past years \citep{ 10.1111/j.1365-2966.2009.15302.x,2011MNRAS.412.2530B, 10.1142/S021827180701153X,skordis:2021pl}.
In the solar system, three main consequences of the MOND phenomenology have been studied \citep{1983ApJ...270..365M,hees:2014pr}. Two of them have been shown to be negligible considering uncertainties of planetary mean motions \citep{2006MNRAS.371..626S}. {The last one is not negligible however, and is known as} the External Field Effect (EFE) \citep{10.1111/j.1365-2966.2009.15302.x,milgrom:2010mn,2011MNRAS.412.2530B}. The effect stems from the non-linearity of MOND equations, and---at leading order in the multipole expansion \citep{2011MNRAS.412.2530B}---it induces an anomalous quadrupolar correction to the Newtonian potential $\delta \Phi(Q_2)$ that reads \citep{10.1111/j.1365-2966.2009.15302.x}
\be
\delta \Phi(Q_2) = - \frac{Q_2}{2}r^{2} \left( cos^{2} \theta - \frac{1}{3}\right),
\ee
where $\theta$ could be the angle pointing either toward the Galactic center, or toward the Newtonian galactic field---depending on the model considered \citep{2014PhRvD..89j2002H}.
 This factor Q$_{2}$ changes with the shape of the MOND coupling functions and, in the solar system, \cite{2011MNRAS.412.2530B} estimated that the values of Q$_{2}$ can vary from 3.8 $\times$ 10$^{-26}$ to 4.1 $\times$ 10$^{-26}$ s$^{-2}$, {and then evaluated that it would therefore lead} to additional advances of perihelia up to 5.81 mas.cy$^{-1}$ for Saturn and even bigger values for Uranus (-10.94  mas.cy$^{-1}$). \cite{10.1111/j.1365-2966.2009.15302.x} had also proposed additional Saturn perihelion precession rate of about 1.8 mas.cy$^{-1}$.
 Nevertheless---as notably emphasized in \citep{2011MNRAS.412.2530B}---it is not consistent to take into account only a particular MOND effect, like perihelion precession, and to compare it with constraints obtained in other frameworks, such as the PPN framework. To tackle this issue, \cite{2014PhRvD..89j2002H} re-adjusted the parameters of planetary ephemeris when the EFE is taken into account.
In \citep{2014PhRvD..89j2002H}, {the anomalous quadrupolar correction enters directly as a modification of the space-time metric that reads:}
  	\begin{widetext}
	\begin{align}
ds^2 =& \left( -1 + \frac{2\mu}{c^2 r} -2 \delta \Phi(Q_2) + 2 \left(\frac{\mu}{c^2 r}\right)^{2} \right) c^2 dt^2 \nonumber\\
& + \left( 1 + \frac{2\mu}{c^2 r} 
+2 \delta \Phi(Q_2) \right) dl^2, \label{eq:MOND}
		\end{align}
		\end{widetext}
with $dl^2 \equiv dx^2+dy^2+dz^2${, because this is what one ought to expect from a relativistic realisation of MOND dynamics}.
Using this new definition of the metric,  \cite{2014PhRvD..89j2002H} introduced modifications into the EIHDL equations of motion and in the {Shapiro delay}. 
An alternative planetary ephemeris has been built in this new framework and {has been }adjusted to observations. 
A specific focus has been brought on the Saturn orbit as it is supposed to be the most affected by the MOND modification (as also estimated by \cite{2014PhRvD..89j2002H},) but also the most constrained by accurate observations (in this case Cassini tracking data). 
By fitting directly  $Q_2$ with the rest of the planetary parameters, \cite{2014PhRvD..89j2002H} obtained a constraint on $Q_2$ of about $(3 \pm 3 ) \times 10^{-27}$ s$^{-2}${, which excludes a important interval of {$Q_2$} values computed theoretically in \cite{2011MNRAS.412.2530B} for various MOND interpolating functions $\mu$.}
Considering other coupling functions, it is still possible to maintain a MOND formalism in the solar system but at the cost of very limited possibilities of couplings.  
This result is consistent with \cite{10.1111/j.1365-2966.2009.15302.x} for which the predicted perihelion precession for Saturn, estimated for coupling functions $\mu_{\alpha}$ with $\alpha > 1/3$,  was significantly smaller than {in} \cite{2011MNRAS.412.2530B}.

\section{Inconsistent tests with ephemeris outputs}
\label{sec:derivedQ}

In the literature, it is possible to find a wide range of {\it{constraints}} supposedly deduced from planetary ephemerides. There are of two types: 


The first type of tests are those considering residuals obtained after the modifications of the equations of motion of the planetary systems--- by, for example, introducing additional terms produced by general relativity-like linearization  of an alternative theory--- but without re-adjusting the newly modified dynamical system to observations. This approach is equivalent to not considering any attempt to improve the ephemerides as described in Fig. \ref{fig:ephem} in Sec. \ref{sec:basisephem}. It is clearly an issue because any modification of the dynamical {model} must be re-adjusted to observations before any interpretation of the obtained residuals. In essence, any comparison of a simulation, which presupposed the parameters of the solar system bodies inferred in a different theoretical framework, with the observations, will likely overestimate the deviations in the new theoretical framework. This would primarily be because the simulation did not use the most appropriate set of parameters for the solar system bodies—which are, those that minimize the residuals, and which are obtained through readjustment in a given framework. We can refer the reader to the extensive discussion of this very basic concept applied to the massive gravity problem and notably presented in the supplementary materials of \cite{bernus2019}. 


The second category of indirect tests involves what we will term as \textit{derived quantities}, as explained in Sec. \ref{sec:defder}. We will discuss these indirect tests in detail in the upcoming section.

\subsection{Definitions}
\label{sec:defder}

Several quantities can be provided as outputs of a given planetary ephemeris, notably in order to give an idea of the accuracy of that ephemeris with respect to specific aspects of the solar system phenomenology---such as, for instance, the values of the perihelion and node advances per orbit for a given astronomical body, and their uncertainties; or the secular variation of the gravitational constant $\dot G/G$ and its uncertainty. 
We shall call those quantities \textit{derived quantities} in what follows.

Derived quantities could also be the values and uncertainties of parameters such as the post-Newtonian parameters $\gamma$ and $\beta$, or of the Compton wavelength in massive gravity, or in the context of a \textit{fifth force}. 
Table \ref{tab:alter} gives a non-exhaustive series of examples of possible interpretations of planetary ephemeris derived quantities for testing alternative theories. 
It falls outside the scope of this review to detail theories that have not been confronted and re-adjusted against planetary observations but have only considered the derived quantities defined above. The  reason for limiting this review to fully tested theories, as presented in Sec. \ref{sec:fulltest}, is our belief that the published constraints listed in Table \ref{tab:alter}  yield to unrealistic constraints, as we will explain in Secs. \ref{sec:crap} and \ref{sec:dont}. In particular, considering only derived quantities or un-fitted residuals is equivalent to neglecting that these latest were obtained in a given framework (usually general relativity), with instrumental uncertainties and correlations between parameters.


\begin{table}
\centering

\caption{Examples of interpretation of ephemeris derived parameters. $d \dot{\varpi}_{supp}$ and $d \dot{\Omega}_{supp}$ indicate the supplementary advances in perihelia and nodes respectively.}
\begin{tabular}{l l | c | l}
\hline
Theories & Section & Impact on orbits & References \\
\hline
{MOND} & \ref{sec:mond} &{$d \dot{\varpi}_{supp}$, $d \dot{\Omega}_{supp}$} &\citep{2011MNRAS.412.2530B}  \\
{AWE/chameleons} && {variation of PPN parameters} & \citep{2007PhRvD..75l3007F} \\
&&{EP} & \citep{burrage2018} \\
{Scalar field theories} && {$\dot{G}/G$} & \citep{2003AnHP....4..347U} \\
&& {variation of of PPN parameters} & \\
{dark  Energy} & &{$\dot{G}/G$} & \citep{2009PhRvD..79j4026S} \\
{dark  Matter}  & \ref{sec:peri} &{linear drift of AU} & \citep{2010AdSpR..45.1007A}\\
&& {a$_{supp}$} & \citep{1994ApJ...437..529N}\\
&& {$d \dot{\varpi}_{supp}$, $d \dot{\Omega}_{supp}$} & \citep{2008PhRvD..77h3005F}\\
{Yukawa, fifth force} & \ref{sec:ISL} & {$d \dot{\varpi}_{supp}$} & \citep{2010LRR....13....7M} \\
{f(r)}  && {a$_{supp}$} & \citep{2010LRR....13....3D}\\
&&{variation of of PPN parameters} & \\
Massive graviton & \ref{sec:graviton}& $d \dot{\varpi}_{supp}$, $d \dot{\Omega}_{supp}$ & \citep{will:2018cq} \\
\hline
\end{tabular}
\label{tab:alter}
\end{table}

\subsection{The case of the advance of the perihelia and nodes}
\label{sec:peri}

It is traditional when one investigates possible laws of gravitation in the solar system to follow the Einstein's steps and to consider supplementary advances in the planetary orbital angles (mainly perihelion and node).  Several methods have been developed in the past fifty years in order to estimate possible remaining advances in planetary perihelia and nodes that can be fully explained by general relativity. 
\\
The most direct method  (i.e.  presented in \cite{2013MNRAS.432.3431P}) is the adjustment of a quantity $\Delta \varpi$ or $\Delta \Omega$ (respectively supplementary advances of perihelia and node) for all of the planetary orbits or only for some of them, together with the rest of the planetary and relativistic parameters (masses, Sun oblateness, initial conditions, PPN parameters...). 
One can get an intuition from Eq. (\ref{eq:ppnom})  that the fit of such derived quantities is affected by strong correlations between various parameters of the ephemeris, such that the results are plagued with {biases} and underestimated uncertainties.\\
The second approach is to introduce possible rotations of the planetary planes {while} considering  $\Delta \varpi$ or $\Delta \Omega$ {fixed}, and to build new planetary ephemerides integrated with these fixed additional rotations and  fitted to observations. The result  is then the limit of possible rotations that one can add without degrading the planetary residuals (i.e. \cite{2018mgm..conf.3694F}). The advantage with the method is the uncorrelated estimation of maximum value for $\Delta \varpi$ or $\Delta \Omega$, the drawback is that what is obtained is only an upper bound. \\
Finally, a third method consists in averaging  planetary orbits \citep{2017AJ....153..121P} for obtaining residual precession of the perihelia. In this case, as for the two former methods, the deduced residual precession could be induced  by a violation of general relativity{, but  also---and more likely---}by some other sources of uncertainties (i.e unmodeled asteroid perturbations).  \\
Table \ref{tab:bounds1} gathers some of the recently obtained values for $\Delta \varpi$, following the three methods described above.
However, based on the arguments presented in this sect. \ref{sec:derivedQ}, the direct interpretation of these quantities, $\Delta \varpi$ or $\Delta \Omega$, in terms of possible violation of general relativity is strongly discouraged. \\

\paragraph{Estimation of the dark  matter density in the solar system}

An example of misleading conclusion can be taken from the dark  matter density estimation. 
In \cite{2013MNRAS.432.3431P}, direct estimations of the density of the dark  matter inside the orbit of Saturn have been tested  using two different implementations. The first one consists in adding an additional acceleration to the equations of motion of the EPM planetary ephemerides such as:
\bea
 \bm a_{A}^{DM} =  \frac{\mu_{M(R)}}{r_{AT}}\bm r_{AT}, \label{eq:accDM}
\eea
where $\mu_{M(R)}$ is the gravitational parameter of an additional matter in a sphere of radius $R$ around the Sun. It turned out that such a direct modeling is highly affected by the uncertainties induced by the asteroid masses and no conclusive measurement of the mass of dark  matter inside the solar system has been obtained in using Eq. (\ref{eq:accDM}). 
The second attempt {in \cite{2013MNRAS.432.3431P}} was from the secular advance of perihelia $d \varpi_{A}^{DM}$ following the equation from \cite{2006IJMPD..15..615K}
\begin{eqnarray}
 \Delta \varpi_{A}^{DM} =  - 3 \pi \frac{{\rho_{DM}}}{M_{sun}}\sqrt{1-e_{A}^{2}}, 
 \label{eq:accDM2}
\end{eqnarray}
where $\Delta \varpi_{A}^{DM}$ is the supplementary advance of perihelia of the planet $A$ induced by dark  matter of density $\rho_{DM}$, supposed uniformly distributed at the planetary distances, $e_{A}$ being the eccentricity of the  orbit. The most stringent constraint gives a density for the dark  matter up to the Saturn orbit of about 
$\rho_{DM} < 1.1 \times 10^{-20}$ g.cm$^{-3}$,  leading to a dark  matter mass smaller than 7.1 $\times 10^{-11}$ solar mass.
It is interesting to note that this estimation is 5 orders of magnitude higher than the density for local dark matter halo proposed by \cite{ 10.1111/j.1365-2966.2011.18564.x, weber10,Wardana20}  and estimated either by galactic simulations or by fit to observations (including Gaia DR2 in \citep{Wardana20}). 
The third approach is in the interpretation of the secular variation of the Sun gravitational mass as a combination of several phenomena including the fall of dark  matter towards the Sun \citep{1985ApJ...296..679P, 2004PhRvD..69l3505L, Blennow18} . In the scenario of dark  matter falling into the Sun, the mass of the Sun should decrease less rapidly.   However one can discuss the difficulty of disentangling the different contributions from dust and comets falling into the Sun and plasma ejecta that have to be accounted for in the Sun mass equation (see sect. \ref{sec:mudot} for discussion). An attempt has been proposed by \cite{2005ARep...49..134K} but leading to a constraint of few percents of the sun mass that can be assimilated to falling dark  matter. This {value} is even bigger than the one proposed by \cite{2013MNRAS.432.3431P} and is {therefore} not in agreement with the expected estimations for local dark matter density.

Recently, \cite{2022MNRAS.510.5154B} simulated the impact of dark  matter located in the galactic halo on the motion of objects in the solar system. Their conclusions are that only objects located in the outer solar system (after 80 AU){, or objects situated in saddle points,} can allow a detection. This is consistent with what has been already discussed, for example, by \cite{1993PhLA..184...41K}.
Finally, in sect. \ref{sec:mond}, the determination of $Q_2$ obtained by  \cite{2014PhRvD..89j2002H} can also be seen as a measurement of the Galactic potential acting on the solar system, either induced by the stellar population or induced by dark  matter. This value presented in sect. \ref{sec:mond}, shows a clear lack of sensitivity of planetary ephemeris for the detection of the tidal interactions coming from our galaxy. 

All these results favour the fact that planetary ephemerides are not yet accurate enough to measure local dark  matter influences in the solar system.


\begin{table}[t] 
\centering
\caption{1-$\sigma$ uncertainties (mas/yr) on the perihelion advance per orbit as extracted from \citep{2013MNRAS.432.3431P} using a full fit of the advances and of the associated ephemeris, from \citep{2018mgm..conf.3694F} and  \citep{Fienga2011}  with fixed values of the maximum advances and from \citep{2017AJ....153..121P}.} 
\begin{tabular}{l c c} 
\hline 
Planet& \citep{2013MNRAS.432.3431P} & \citep{Fienga2011} \\
Mercury&$0.03$ & 0.006   \\
\noalign{\smallskip}
Venus&$0.016$ & 0.015  \\
\noalign{\smallskip}
Earth&$0.0019$ & 0.009  \\
\noalign{\smallskip}
Mars&$0.00037$& 0.0015 \\
\noalign{\smallskip}
Jupiter&$0.28$ & 0.42  \\ 
\noalign{\smallskip}
Saturn&$0.0047$ & 0.0065  \\ 
\hline
\hline 
& \citep{2018mgm..conf.3694F} & \citep{2017AJ....153..121P} \\ 
Mercury & $0.02$ & $0.015$ \\
\hline 
\end{tabular}
\label{tab:bounds1}
\end{table}


\subsection{What can often be found in the literature}
\label{sec:crap}
Most of the bounds in the literature do not come from planetary ephemerides developed in a given theoretical or phenomenological framework, but instead they use the uncertainty on derived quantities as an input of the maximum tolerated departure from general relativity that would be compatible with observations. For instance, if the uncertainty on the perihelion advance per orbit for a given astronomical body and a given ephemeris is less than a specific value, it is often assumed that any modification of gravity has to induce an effect that is less than this former limit. Therefore, if one computes that a specific theory should induce an effect on, e.g., the perihelion advance per orbit for an astronomical body that is bigger than the uncertainty on this derived quantity, then it is often claimed that this constraint rules out this specific theory.

At first sight, it seems like a reasonable thing to do. Unfortunately, doing this is problematic for the simple reason that derived quantities are obtained assuming a specific theoretical framework, and that there is no guarantee that the adjusted parameters of the ephemeris (masses, initial conditions etc.) would be the same in another theoretical framework. Even more, it is not impossible, a priori, that another theoretical framework actually leads to smaller residuals than general relativity after adjusting the parameters of the ephemeris in this framework---which would mean that this alternative to general relativity is favoured by the data for the considered model of the solar system.

In other words, while one would claim to have derived a constraint on an alternative theory from a derived quantity, actually one could very well have missed a signal in favour of the alternative theory instead---which is precisely the opposite of giving a constraint. This shows that using derived quantities as an input to constrain alternatives to general relativity is not trustworthy in general.

The only way to compare the merit of two different theoretical frameworks is to compare statistically the accuracy of the ephemerides adjusted to the data in each theoretical framework---that is, the statistical amplitude of their residuals. If the ephemeris in a given theoretical framework has significantly better (i.e. smaller) residuals than in another theoretical framework, then it means that the former is favoured by the data within the planetary model considered. Different planetary models---see e.g. Sects. \ref{sec:aste} and \ref{sec:j2}---might also lead to different answers with this respect.

%

\subsection{Why consistency matters}
\label{sec:dont}
The reason why the adjusted parameters of an ephemeris (masses, initial conditions etc.) are in general different when one considers an alternative theory to general relativity is because all the parameters of the ephemeris, the ones describing gravity (e.g. PPN parameters, Compton wavelength of the Yukawa suppression etc.) and the ones describing the bodies themselves and their orbits (masses, initial conditions, shapes etc.), are more or less correlated to one another. The high degree of correlation is also somewhat accentuated by the specific symmetry of planetary orbits, because most of the motions are close to the ecliptic plane (and with relatively low eccentricities), which limits the disentanglement of the effects of different parameters during the fit of the ephemeris parameters---as one can get an intuition from the approximated analytical expression, for instance, of the advance of the node Eq. (\ref{eq:ppnom}).

This is exemplified notably with the oblateness of the Sun $J_{2}^{\odot}$ in Figs. \ref{fig:j2} and \ref{fig:j2ppn}.\footnote{{See also Fig. 6 in \citep{2002PhRvD..66h2001M} for an illustration of the correlation between $J_{2}^{\odot}$ and $\beta-1$ in simulations.}} Indeed, due to the high level of correlations between the oblateness of the Sun and the PPN parameters $\gamma$ and $\beta$, both the adjusted value of the $J_{2}^{\odot}$ and its uncertainty are affected when $\gamma$ and $\beta$ are not set to 1 a priori (see sec \ref{sec:PPN} for a full discussion) .


From a formal perspective, Eq. (\ref{eq:ppnom}) gives a good illustration of the problem in terms of perihelion advance per planetary orbit.
Because many different parameters contribute to the advance, if an advance is, say, induced by a parameter related to the description of gravity beyond what is acceptable in terms of deviation with respect to the data, the fit will often lead to a modification of other parameters in order to compensate for this unacceptable contribution---such that, in the end, the final solution remains as close as possible to the data (i.e. the residuals are minimised). 


As long as the statistical properties of the residuals of the ephemerides in distinct theoretical frameworks are not significantly different, one cannot say which of the two theoretical frameworks better explains the observations. This tells why one cannot simply estimate the modification of the perihelion advance per orbit in an alternative theory and compare it with the output value obtained while assuming general relativity.

Another example is given by the Compton wavelength of a Yukawa suppression of the Newtonian potential in a massive gravity framework---see sect. \ref{sec:graviton}. {An illustration of the high degree of correlations between parameters} is given by \cite{bernus2019}  {in} a table{---reproduced here in Table \ref{table:corr}--- that gathers}  the correlations between the Compton wavelength and some of the solar system parameters for the INPOP17a planetary model. \\ 
Furthermore,  statistics of the residuals for several planets were also displayed by \cite{bernus2019} in order to show their evolution  if one assumes that the solar system parameters are given by fits obtained when assuming general relativity instead of re-adjusting them in the massive gravity framework. The conclusion of this investigation is that not adjusting planetary ephemeris parameters within the framework of massive gravity would have led to an overestimation of the constraint on the value of the Compton wavelength by about one order of magnitude.

\paragraph{The Pioneer anomaly}

Finally, one can discuss the case of the Pioneer anomaly. For some years, the unexplained supplementary acceleration detected during the navigation of Pioneer 10 and 11 escaping the solar system, keeps the community active in looking for some possible violations of GR that could produce such a phenomena. The reader can see \citep{2010LRR....13....4T} for a complete review.  Among the alternative theories {that were} proposed {to} explain the Pioneer anomaly, some also impact the orbits of outer planets. Most of the authors just consider the effect of the induced modification of the planetary equations of motion without considering the new adjustment that one should do for adapting the initial conditions to this new model \citep{2009CQGra..26d5014L, 2009AJ....137.3615I,2010OAJ.....3....1I}. In 2009 and 2010, \cite{2009ApJ...697.1226P, 2010IAUS..261..179S} and \cite{2010IAUS..261..159F} show that the modification required for explaining the Pioneer acceleration anomaly induced, after fit,  residuals marginally compatible with the  observational accuracies reached at this epoch (before the inputs of the Juno mission). These results restricted severely the possibility of such modifications impacting also planetary orbits. {After that, only remained the} alternative theories {that were} affecting the s/c orbit but not the planetary bodies.
{Another attempt to explain the Pioneer anomaly within the boundaries of conventional physics that does not affect the motion of the planetary bodies was put forward by \cite{kopeikin:2012pr}. In their proposal, the Pioneer effect was perceived as the cosmological consequence of a quadratic divergence between the time scales of electromagnetic wave propagation within the Doppler tracking system and the atomic clocks on Earth.}
{However, \cite{2010SSRv..151...75B} and \cite{2012PhRvL.108x1101T} conclusively demonstrated that the Pioneer acceleration can be explained by considering the distinct thermal properties of each spacecraft face. Either way, the Pioneer anomaly thus serves as a compelling illustration of the importance of developing a fully consistent model when testing alternative theories to General Relativity, and ensuring that this model is fitted to observations—a step that is undeniably crucial.}
\begin{table}
	\caption{Examples of correlations from \citep{bernus2019} between various INPOP17b parameters and the Compton wavelength $\lg$. $a$, EMB and $M_{\odot}$ state for semi-major axes, the Earth-Moon barycenter and the mass of the Sun respectively.}

\resizebox{\columnwidth}{!}{
\begin{tabular}{c|ccccccc}
		~ &$\lambda_g$& $a$ Mercury  &$a$ Mars & $a$ Saturn& $a$ Venus &$a$ EMB & $GM_{\odot}$ \\
		\hline
		$\lambda_g$ &1 &0.50&0.49&0.04&0.39&0.05&0.66\\
        $a$ Mercury &$\cdots$& 1& 0.21& 0.001& 0.97& 0.82& 0.96 \\
		$a$ Mars  &$\cdots$& $\cdots$& 1 &0.03&0.29& 0.53 &0.06 \\
		$a$ Saturn &$\cdots$& $\cdots$ &$\cdots$& 1& 0.003 & 0.02& 0.01 \\
		$a$ Venus &$\cdots$&$\cdots$&$\cdots$&$\cdots$&1& 0.86& 0.94\\
		$a$ EMB &$\cdots$&$\cdots$& $\cdots$& $\cdots$ &$\cdots$&1 &0.73\\
		$GM_{\odot}$  &$\cdots$&$\cdots$ &$\cdots$& $\cdots$&$\cdots$&$\cdots$ &1\\
	\end{tabular}
}
	\label{table:corr}
\end{table}

\section{Future directions}

\subsection{{Theory}}

On the theory side, {many aspects of alternative theories remain to be studied and implemented in planetary ephemerides.}
Theorists do not lack of new ideas, and therefore one should not fall short of new theories to investigate with planetary ephemerides. However, not all existing theories that can lead to significant variations in the solar system have been constrained with planetary ephemerides yet. As an example, not even the full {PPN framework} Eqs. (\ref{eq:g00willH}-\ref{eq:PhiW}) have been completely investigated so far {with} planetary ephemerides. This is not a surprise given the many parameters involved, and given that the more parameters to test, the more difficult, and computationally demanding, the study is{---and also the worse the constraints on each parameter are}.\\ 

Another example can also be given with the case of Brans-Dicke-like scalar-tensor theories. A considerable portion of Brans-Dicke-like scalar-tensor theories---that is, theories defined by $\mathcal{L}_{int}=0$ in Eq. (\ref{eq:actEdilaton})\footnote{{Indeed, strictly speaking, Brans-Dicke theories are an even more restricted group, which is such that $\mathcal{L}_{int}=0$ and $\omega(\varphi)=\omega_{BD}$ in Eq. (\ref{eq:actEdilaton}).}}---have been exquisitely constrained by observations of binary pulsars, owing to a strong field effect that cannot occur within the solar system. Specifically, with certain selections for the function $\omega(\phi)$, the scalar-field within compact objects like neutron stars can be amplified through a nonlinear effect known as \textit{scalarization} \citep{damour1993pl,damour:1996pr}. This effect should lead to a significant violation of the Strong Equivalence Principle\footnote{{Even, occasionally, with large values of $\omega(\varphi_0)$.}}---see Sec. \ref{sec:SEPmt}---that is not seen in binary pulsars. For a deeper dive into this captivating subject, we direct the reader to \cite{kramer:2021px,freire:2012mn,voisin:2020aa}.

The PPN framework in Sec. \ref{sec:PPNf} encompasses this class of theories with $\gamma = (1+\omega(\varphi_0))/(2+\omega(\varphi_0))$, where $\varphi_0$ is the asymptotic value of $\varphi_0$ at the edge of the solar system,\footnote{{Which could vary over cosmological times.}} and $\beta-1 = \omega'(\varphi_0) (3+2\omega(\varphi_0))^{-2} (4+2\omega(\varphi_0))^{-1}$. The constraints on $\gamma$ and $\beta$ obtained with planetary ephemerides in the PPN framework are given in Fig. \ref{fig:ppn}.

However, those constraints assume that $\gamma$ and $\beta$ are independent parameters in the  field equations{---which lead to the definition of the time and space coordinate system, the equation of motion (see Eq. \ref{eq:accgeneral relativity}) and of the {Shapiro delay} (see Eq. \ref{eq:shapiroPPN})---}while they are not independent in those scalar-tensor theories whenever $\beta \neq 1$---that is $\omega'(\varphi_0) \neq 0$---given that $\omega(\varphi_0)$ appears in both $\gamma$ and $\beta$. 

As a consequence, one cannot directly convert the constraints in Fig. \ref{fig:ppn} in terms of constraints on $\omega(\varphi_0)$ and $\omega'(\varphi_0)$. Therefore, it would be interesting to test those specific theories with planetary ephemerides in order to compare with the constraints with binary pulsars, in the regions of the theory space where one does not have the non-linear strong-field scalarization effect. For instance, assuming $\omega(\varphi)=\omega_{BD}$---that is, $\omega'(\varphi_0)=0$ and $\beta = 1$---tests involving pulsars lead to $\omega_{BD}>130\times 10^3$ \citep{voisin:2020aa}---although note that this value depends on the unknown equations of state of neutron stars. 

{Considering} the constraint on $\gamma$ obtained with the Cassini experiment \citep{bertotti2003nat}, the limit on $\omega(\varphi_0)$ {instead is} $\omega(\varphi_0) > 40\times 10^3$. This threshold should also be better than what is currently possible with planetary ephemerides---but perhaps not too far from what will be possible with the additional data from Bepi-Colombo.




A general reminder of Sec. \ref{sec:derivedQ}  is that, while the PPN framework serves as a highly convenient phenomenological apparatus for testing alternative theories, its limitation lies in its inability to accommodate potential dependencies between PPN parameters (such as the discussed above or the one of Sec. \ref{sec:dilaton}) that could emerge within a given theory. For example, in Brans-Dicke-like scalar-tensor theories, there is a certain degree of interdependence at the level of the field equations between $\gamma$ and $\beta$ since {they both depend on the parameter $\omega(\varphi_0)$ of the theory.}

Consequently, one has to keep in mind that it is not generally possible to directly translate constraints obtained on PPN parameters to potential underlying theories—unless the theory predicts the parameters to be independent at the {level of the field equations of the considered theory (e.g. Scalar-Tensor, massive gravity, etc.)}. Therefore, to establish constraints on the parameters of a specific theory—--like $\omega(\varphi_0)$ and $\omega'(\varphi_0)$—-- one {should} directly test that particular theory.

\subsection{{Observation}}

{In terms of observations, much progress is anticipated, particularly in light of active or future planetary missions such as Bepi-Colombo.}
This mission will orbit Mercury for more than a year and will provide unprecedented accurate measures of the Mercury-Earth distance, and consequently, {will produce} stringent new limits {on deviations from} general relativity. 
A lot of publications propose {to include} Bepi-Colombo range simulations {in order to provide} possible new constraints on classic general relativity tests, such as advance of {the} Mercury {perihelion}, PPN parameter estimations, SEP or alternative tests \citep{2002PhRvD..66h2001M,2007PhRvD..75b2001A,2016PhRvD..93l3014D,2018Icar..301....9I,2020CQGra..37i5007D,2022RemS...14.4139V,fienga2022,moriond2022}. As previously explained, a specific care should be taken on the consistency between the definition of the considered framework (i.e. harmonics versus non-harmonics gauges, definition of the solar system barycenter ...), the tests performed and the claimed accuracy.
The question of consistencies and correlations between astronomical constraints and general relativity tests in the {full} PPN context will then become even more urgent to address. \\
On the other side of the solar system, missions towards gas giants and outer solar system will also be interesting for testing another types of general relativity violation such as dark sector violations or dark matter clumps \citep{2021ExA....51.1737B, 2018cosp...42E.292B}.\\
Finally, among more exploratory projects,  LISA-like configurations for  interplanetary laser distance measurements between telluric planets (Earth, Mars and Venus) have been proposed as a way to gain accuracy in planetary ephemerides and sensitivity to general relativity violations such as the secular variations of the gravitational constant \citep{2018P&SS..153..127S}. Despite the technical challenges of such project \citep{2022P&SS..21405415B, 2022P&SS..21505423B}, the outcome of these measurements would indeed impact the global accuracy of the ephemerides, improve significantly the Bepi-Colombo results but also allow {for} better constraints on the distribution of mass in the solar system.


\section{{Summary}}

This paper describes how the planetary ephemerides are built in the framework of General Relativity and how they can be used to test alternative theories. {It focuses} specifically on the dependencies that exist behind the definition of the reference frame (space and time) in which the planetary ephemeris is described, the equations of motion that govern the orbits of solar system bodies and {electromagnetic waves}. 
This paper then summarizes the results obtained considering {consistent} modifications of the ephemeris framework with direct comparisons with the observations of planetary systems. 
The PPN formalism is the one that has been the most heavily tested, and {the} results of {its} confrontation with planetary astrometry constitutes the most developed part. The paper then moves on {to} {specific alternatives to general relativity} such as Einstein-dilaton theories, {a massive} graviton {phenomenology} and MOND. The paper finally concludes on some comments and recommendations regarding misinterpreted {estimations} of the advance of perihelia, giving examples such as the Pioneer anomaly interpretation or some attempts to measure dark matter in the solar system.\\
As we hope this paper demonstrates, the consistency of the planetary ephemeris framework is a crucial aspect in the field of testing alternative theories. Misinterpretations of results obtained in the general relativity framework can lead to significant errors ({e.g.} Pioneer anomaly) or over optimistic constraints ({e.g.} mass of the graviton).\\

\section*{Acknowledgements}
{The authors thank  Aurélien Hees and Vincenzo Mariani for their valuable inputs on the initial versions of the paper. OM expresses also gratitude to Francesco Nitti, Federico Piazza, and Aurélien Hees for stimulating discussions on the relationship between composite particle mass and quantum trace anomalies. AF thanks  L. Blanchet, A. Hees, E. Gourgoulond, S. Reynaud, P. Wolf and C. Will for their supports, comments, discussions all these years. AF is also debtfull to J. Laskar and M. Gastineau for the development of the INPOP planetary ephemerides. Finally, AF is grateful to the French Space Agency and Observatoire de la Côte d'azur for the financial supports.}


\bibliographystyle{spbasic}
\bibliography{global}   

\appendix
\section{Derivation of the Shapiro delay at the $c^{-4}$ level, and related issues}
\label{sec:ShapiroO2}

In the following lines, we indicate some arguments stressing the complexity of keeping a consistent framework for the derivation of the Shapiro delay at the $c^{-4}$ level in the general data analysis framework of planetary ephemerides.

Indeed, although there are several $c^{-4}$ formulae given in the literature already, e.g. \cite{ashby:2010cq,deng:2012pr,linet:2013cq,hees:2014pr,linet:2016pr,cappuccio:2021cq,zschocke:2022pr}, most of them are not consistent with the coordinate system recommanded by the IAU---the harmonic gauge---that is used in planetary ephemerides in order to describe the motion of celestial bodies. Hence, one still has to convert those propagation time formulae in the harmonic coordinate system.
For instance, many $c^{-4}$ Shapiro equations are derived from an isotropic metric \cite{ashby:2010cq,linet:2013cq,hees:2014pr,linet:2016pr,cappuccio:2021cq}, whereas the metric in harmonic coordinates is not isotropic at the full $c^{-4}$ level, even for a spherical object at the center of the coordinate system---see, e.g., \cite{minazzoli:2012cq} for the difference between harmonic and isotropic coordinate metrics at the $c^{-4}$ level in the framework of general relativity and scalar-tensor theories. A second order ($c^{-4}$) propagation time formula with harmonic coordinates has recently been derived in \cite{zschocke:2022pr}, but it assumes general relativity and a single body at rest. In general, most of the propagation time formulae in the literature indeed simplify the problem by assuming staticity of the celestial bodies during the propagation of light, whereas one may have to go beyond this approximation at the required level of accuracy \cite{bertone:2014cq,zschocke:2016pr}. A $c^{-4}$ propagation formula in harmonic coordinates\footnote{Note that there can be two distinct versions of harmonic coordinates in scalar-tensor theories, based on whether the harmonic gauge condition $g^{\alpha \beta} \Gamma^\sigma_{\alpha \beta}=0$ is imposed on the metric in the Jordan frame or in the Einstein frame \cite{minazzoli:2011cq,deng:2012pr}. This further complicates the discussion on the coordinate systems to be used in alternative theories.
} for light in the solar system---that is, with many moving bodies---has been derived in \cite{minazzoli:2011cq,deng:2012pr} in the framework of general relativity and scalar-tensor theories, but the actual propagation time that results from it remains to be derived. {Otherwise, a $c^{-4}$ propagation time formula for light rays in harmonic coordinates, but restricted to one arbitrarely moving pointlike body, has been derived in \cite{zschocke:2016pr}.}

\end{document}